\pdfoutput=1
\documentclass[cits]{JINST}
\usepackage{xspace}
\usepackage{caption}
\usepackage[skip=0cm,list=true,labelfont=it]{subcaption}

\newcommand{\unit}[1]{\ensuremath{\mathrm{\,#1}}\xspace}

\DeclareFontFamily{U}{euc}{}
\DeclareFontShape{U}{euc}{m}{n}{<-6>eurm5<6-8>eurm7<8->eurm10}{}%
\DeclareSymbolFont{AMSc}{U}{euc}{m}{n} 
\DeclareMathSymbol{\umu}{\mathord}{AMSc}{"16}

\newcommand{\cpp}{\texttt{C++}\xspace} 
 
\newcommand{\fref}[1]{Figure~\ref{#1}} 

\usepackage[outermarks]{titlesec}

\titleformat{\section}
  {\normalfont\Large\bfseries}{\thesection}{1em}{}
\titleformat{\subsection}
  {\normalfont\large\bfseries}{\thesubsection}{1em}{}
\titleformat{\subsubsection}
  {\normalfont\normalsize\bfseries}{\thesubsubsection}{1em}{}
 
\titleformat{\paragraph}
  {\normalfont\normalsize\bfseries}{\theparagraph}{1em}{}
\titleformat{\subparagraph}
  {\normalfont\normalsize\bfseries}{\thesubparagraph}{1em}{}
\titlespacing*{\section}      {0pt}{3.5ex plus 1ex minus .2ex} {2.3ex plus .2ex}
\titlespacing*{\subsection}   {0pt}{3.25ex plus 1ex minus .2ex}{1.5ex plus .2ex}
\titlespacing*{\subsubsection}{0pt}{3.25ex plus 1ex minus .2ex}{1.5ex plus .2ex}

\titlespacing*{\paragraph}    {0pt}{3.25ex plus 1ex minus .2ex}{1.5ex plus .2ex}
\titlespacing*{\subparagraph} {0pt}{3.25ex plus 1ex minus .2ex}{1.5ex plus .2ex}
\title{Construction and commissioning of a technological prototype of a high-granularity semi-digital hadronic calorimeter}

\author{
  G.~Baulieu$^a$, M. Bedjidian$^a$, K.~Belkadhi$^b$, J.~Berenguer$^f$, V.~Boudry$^b$,  P.~Calabria$^a$, S.~Callier$^d$, E.~Calvo~Almillo$^f$, S.~Cap$^i$,  L.~Caponetto$^a$, C.~Combaret$^a$, R.~Cornat$^b$, E.~Cortina Gil$^c$, B.~de Callatay$^c$, F.~Davin$^c$,  C.~de la Taille$^d$,  R.~Dellanegra$^a$, D.~ Delaunay$^a$, F.~Doizon$^a$, F.~Dulucq$^d$,  A.~Eynard$^a$, M-C.~Fouz$^f$,  F. Gastaldi$^b$, L.~Germani$^a$, G.~Grenier$^a$, Y.~Haddad$^b$,  R.~Han$^a$, J-C.~Ianigro$^a$, R.~Kieffer$^a$,  I.~Laktineh$^a$\thanks{Corresponding author.}, N.~Lumb$^a$, K.~Manai$^g$\footnote{On leave for College of Sciences and Arts. Bisha University, Bisha, Kingdom of Saudi Arabia.},
  S.~Mannai$^c$, H.~Mathez$^a$, L.~Mirabito$^a$, J.~Prast$^i$, J.~Puerta Pelayo$^f$, M.~Ruan$^b$,
  F.~Schirra$^a$, N.~Seguin-Moreau$^d$, A.Steen$^a$, W.~Tromeur$^a$, M.~Tytgat$^h$,
  M.~Vander~Donckt$^a$, G.~Vouters$^i$, N.~Zaganidis$^h$.\\ 
  \llap{$^a$} Universit\'e de Lyon, Universit\'e Lyon 1, CNRS/IN2P3, 
  IPNL, 4 Rue E.~Fermi, 69622 Villeurbanne Cedex, France.\\
  \llap{$^b$}Laboratoire Leprince-Ringuet -- \'Ecole polytechnique, CNRS/IN2P3,
  Palaiseau, F-91128 France.\\
  \llap{$^c$}Center for Cosmology Particle Physics and Phenomenology (CP3),
Universit\'{e} Catholique de Louvain, Belgium.\\
  \llap{$^d$}Organisation de Micro-Electronique G\'en\'erale Avanc\'ee, \'Ecole Polytechnique, CNRS/IN2P3.\\
  \llap{$^f$}Centro de Investigaciones Energeticas, Medioambientales y Tecnologicas, Madrid, Spain. \\
   \llap{$^g$} Tunis El Manar University, Rommana 1068, Tunis BP 94, Tunisia. \\
   \llap{$^h$}Ghent University, Dept. Physics and Astronomy, Proeftuinstraat
   86, B-9000 Gent, Belgium.\\
    \llap{$^i$}LAPP, Universit\'e Savoie Mont Blanc, CNRS/IN2P3, Annecy-le-Vieux, France.\\
 E-mail: \email{laktineh@in2p3.fr}
}

\abstract{ 

  A large prototype of $1.3\unit{m^3}$ was designed and built as a
  demonstrator of the semi-digital hadronic calorimeter (SDHCAL)
  concept proposed for the future ILC experiments.  The prototype is a
  sampling hadronic calorimeter of 48 units. Each unit is built of an
  active layer made of $1\unit{m^2}$ Glass Resistive Plate Chamber (GRPC)
  detector placed inside a cassette whose walls are made of stainless
  steel. The cassette contains also the electronics used to read out
  the GRPC detector. The lateral granularity of the active layer is
  provided by the electronics pick-up pads of $1\unit{cm^2}$ each. The
  cassettes are inserted into a self-supporting mechanical structure
  built also of stainless steel plates which, with the cassettes walls, play the role of the absorber.  The prototype was designed to be very compact and important efforts were made to minimize the number of services cables to optimize the efficiency of the Particle Flow Algorithm techniques to be used in the future ILC experiments.  The different components of the SDHCAL prototype were studied individually and strict criteria were applied for the final selection of these components.  Basic calibration procedures were performed after the prototype assembling.
  The prototype is the first of a series of new-generation detectors equipped with a power-pulsing mode intended to reduce the  power consumption of this highly granular detector.  A dedicated
  acquisition system was developed to deal with the output of more
  than 440000 electronics channels in both trigger and triggerless modes.  
  After its completion in 2011, the prototype was commissioned using cosmic rays and particles beams at CERN.

    }
\begin{document}


\section{Introduction}

The success of future high-energy experiments intended to investigate
physics phenomena in the TeV range will be determined by their ability
to measure precisely the energy of jets associated with the production
of bosons such as $W^\pm, Z^0, H^0$.  %
Among the different proposed methods to obtain high-precision jet
energy measurements, one of the most attractive techniques is based on
the Particle Flow Algorithm (PFA) approach~\cite{PFA_1,PFA_2}. %
In this approach, particles are to be followed in the different sub-detectors
and their energy or momentum to be estimated in the sub-detector which
provides the best resolution of such a measurement.  For the PFA techniques to be successfully and efficiently applied, electromagnetic and hadronic calorimeters  need to have tracking capabilities,
in addition to their usual role of energy measurement.  This could be  achieved  by
designing a new generation of sampling electromagnetic and hadronic
calorimeters with high-granularity in both the transverse and the
longitudinal directions %

The challenge of this new generation of calorimeters is to cope with
the millions of electronic channels needed to read out such a
high-granularity and yet still compact and hermetic detector.  %
To address the compactness and power consumption, genuine designs are
needed. The SDHCAL prototype, whose development is also a part of the CALICE collaboration activities to study highly granular calorimeters, has successfully achieved these contradictory requirements. 

After a general description of the SDHCAL prototype we describe
the GRPC detector developed for this prototype in the third
section. The new electronic readout is detailed in the fourth
section. In the fifth and the sixth sections the hardware and the software  developed for the acquisition system are respectively described. The seventh section depicts the cassette structure used to
assemble the different components of the active layer as well as the self-supporting absorber mechanical structure to host the cassettes. 
In section eight a description of  event building, data quality control  and  detectors performance is given. Finally the technical performance of the SDHCAL prototype are summarized in section nine.

\section{General description}

The basic unit of the SDHCAL is made of a cassette containing the
active layer. 
The latter  is composed of a GRPC detector and its embedded readout electronics.  
The cassettes are inserted into a self-supporting mechanical structure made of $1.5\unit{cm}$ thick plates of stainless steel.
This structure forms  a robust frame to host the cassettes on the one hand and constitutes the essential part of the hadronic calorimeter's absorber on the other hand. 
The cassette walls and the structure plates are made of the same material and thus complete the calorimeter's absorber. In total $2\unit{cm}$ of stainless steel is separating two consecutive active layers. 

A simple cooling system made of two thin boxes ($2\unit{cm}$ width) each covering one lateral side of the prototype is used. A water circuit inside the two boxes allows to absorb part of the heating produced by the electronic readout system. This cooling system will not be necessary when the power-pulsing scheme will be operated in ILC power cycle schemes. However in beam test conditions (longer data taking cycles) this was found helpful to stabilize the temperature of the GRPC detectors during data taking.
A dedicated system to produce the gas mixture needed to operate the GRPC was conceived to feed in parallel the different GRPC of the prototype. 
The prototype is also equipped with a high voltage system to polarise the GRPC individually. 
Finally, the data collected by the active layers are treated by an acquisition system that was developed for this purpose.

\section{GRPC construction}

\subsection{GRPC description}
\begin{figure}[h!]
  \centerline{\includegraphics[width=0.7\textwidth]{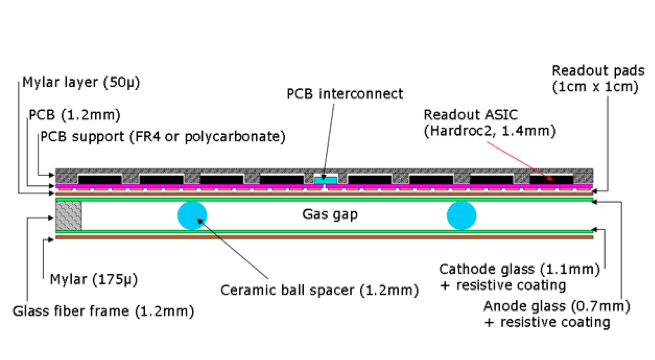}}
  \caption{Cross-section through a $1\unit{m^2}$ chamber.}
  \label{chamber}
\end{figure}

The GRPC developed for the SDHCAL is made of two glass plates with a thickness
of $0.7\unit{mm}$ for the one associated to the anode and $1.1\unit{mm}$ for that associated to the
cathode. The two plates are separated by a $1.2\unit{mm}$ space which is
maintained constant thanks to special spacers (Figure \ref
{chamber}). The distance and the size of those spacers were optimized
to eliminate the detector dead zones while providing a uniform
electric field between the two plates. The two glass plates are
covered on their outer face by a conductive painting.  This allows to
apply high voltage polarization to create an electric field in between. The gap
between the two plates is filled with a gas mixture of TetraFluoroEthane(93\%), CO2(5\%) , SF6(2\%). The first gas provides the primary electrons-ions when
ionized by the crossing charged particle while the second and the
third are UV photons and electrons quencher respectively. Their role is to
limit the size of the avalanche that follows the creation of primary
electrons. It also reduces the probability of producing additional avalanches away from the charged particle impact in the GRPC.  It is worth mentioning here that both TetraFluoroEthane and SF6 are not eco-friendly. Their use in future detectors could be questioned. Therefore, attempts to replace them by new gases with similar functionalities but with much lower Global Warming Potential index are being currently made. To replace the TetraFluoroEthane for instance, new refrigerant gases proposed by car industry like the HFO-1234yf are being investigated.

The gas tightness is provided by a robust insulating frame made of glass reinforced epoxy laminate (Veronite$\circledR{}$ FR-4). This frame is only $3\unit{mm}$ wide leading to a dead area of the detector of 
less than 1.3\%. A gas distribution system was also developed. It allows to
renew the gas content of the chamber in an efficient way taking into
consideration the fact that gas inlets and outlets are to be on one
side of the chamber. The system is designed to reduce the gas
consumption. This and the recycling progress achieved by the RPC-gas
group at CERN, are important elements to reduce the cost of the gas
consumption of such hadronic calorimeters in the future experiments.


\subsection{Chamber design}
\label{chamber_design}
The $1\unit{m^2}$ chambers consist of the elements shown in
Figure~\ref{chamber}.  Precision ceramic (Zr O$_2$) balls of diameter
$1.2\unit{mm}$ are used as spacers to separate the two electrodes: the anode ($0.7\unit{mm}$ thickness) and the cathode ($1.1\unit{mm}$ thickness).  
Readout pads of area $1\unit{cm} \times 1\unit{cm}$ are
isolated from the anode glass by a thin Mylar foil$\unit{50 \mu m}$.  These pads are
etched on one side of a PCB; on the other side are located the
front-end readout chips.  Finally, a
polycarbonate spacer (`PCB support' in Figure~\ref{chamber}) is used to
`fill the gaps' between the readout chips and to improve the overall
rigidity of the cassette (detector/electronics `sandwich') as well as the lateral homogeneity of the active layer.  The total
theoretical thickness of the assembly is $5.825\unit{mm}$.  Taking into
account air gaps and engineering tolerances, the true thickness was  found to be slightly bigger $5.98\pm 0.04\unit{mm}$ but not 
exceeding $6.15\unit{mm}$. 
\subsection{Resistive coating}
\label{coating}
To identify the best resistive coating for the chamber glass different products were tested. All of them providing surface resistivity  in the range $1-20\unit{M\Omega/\square}$
%

\begin{figure}
\centerline{\includegraphics[width=0.95\columnwidth]{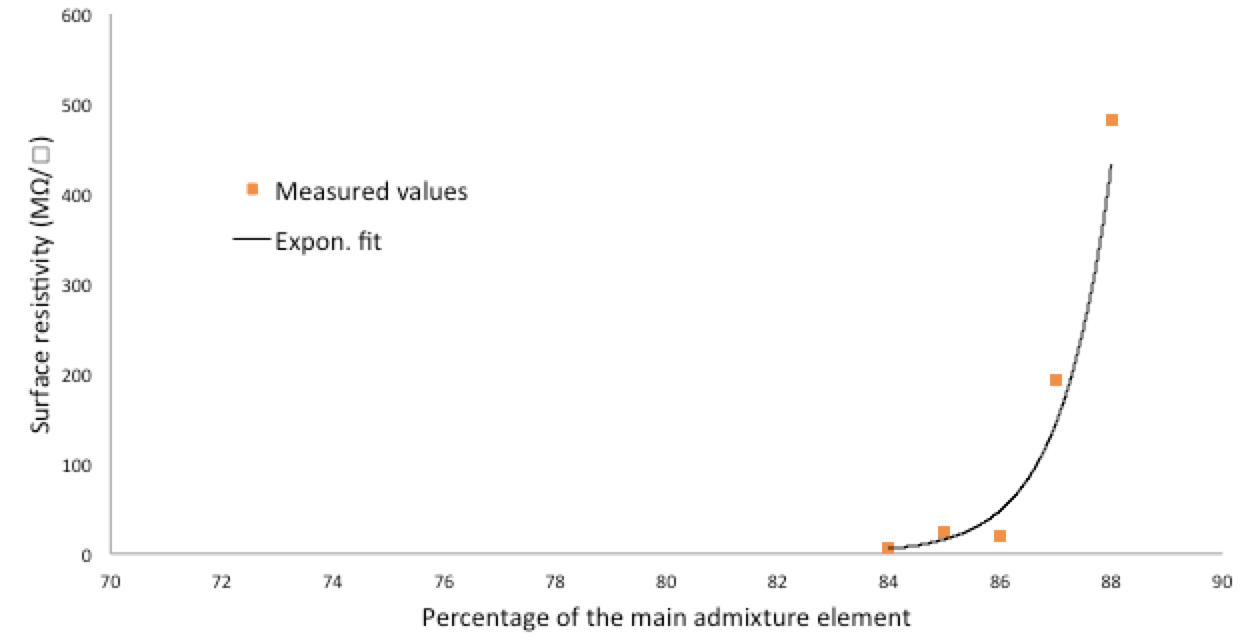}}
\caption{Surface resistivity as a function of mix ratio for bi-component colloidal graphite.}\label{resistivity_graph}
\end{figure}

One of the commercial products we tried, the Licron$\circledR{}$, was found to be problematic in that it has the
tendency to `migrate' away from the high voltage  strip glued to the cathode glass, resulting in loss of the HV contact after a very short
time (a few days of continuous use).  We have constructed several
large-area GRPC using another commercial product  (Statguard$\circledR$), an inexpensive floor
paint used in electrostatic discharge (ESD) applications.  These chambers have been successfully
operated \cite{tipp09}; however the paint was applied to the glass
using a paint brush, which is found not suitable for mass
production.  Attempts to coat large areas with Statguard using the
silk screen technique produced unsatisfactory results.  In addition, this product
also has a long time constant to reach a stable surface resistivity
(typically two weeks).  Research was therefore carried out to find a
product specifically designed for silk screen printing with the
correct surface resistivity.  Two products were identified, both of
which are based on colloids containing graphite.  One of these
products is a single component paint with a dry surface resistivity of
$1-10\unit{M\Omega/\square}$ after being deposited using the silk screen
method.  The second product comes as two components which must be
mixed by the user.  The surface resistivity may be adjusted over a
wide range by changing the mix ratio (Figure~\ref{resistivity_graph}).  Both products require baking at around
170$^o$C to attain a stable surface resistivity.

\begin{figure}
\centerline{\includegraphics[width=0.75\columnwidth]{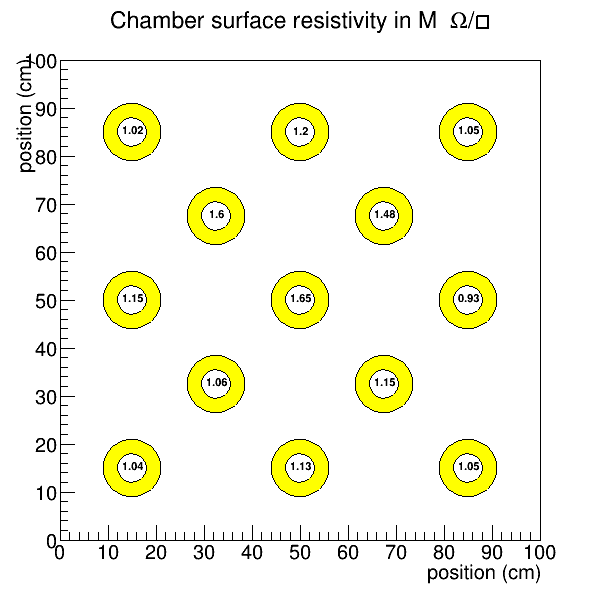}}
\caption{Surface resistivity for $1\unit{m^2}$ glass coated with bi-component colloidal graphite. 
For a painted surface, resistivity has been measured in 13 locations. 
The surface used for the measurement is displayed as a yellow ring. The corresponding measured value is shown inside the ring.}\label{colloid_resistivity}
\end{figure}

\begin{figure}
\centerline{\includegraphics[width=0.95\columnwidth]{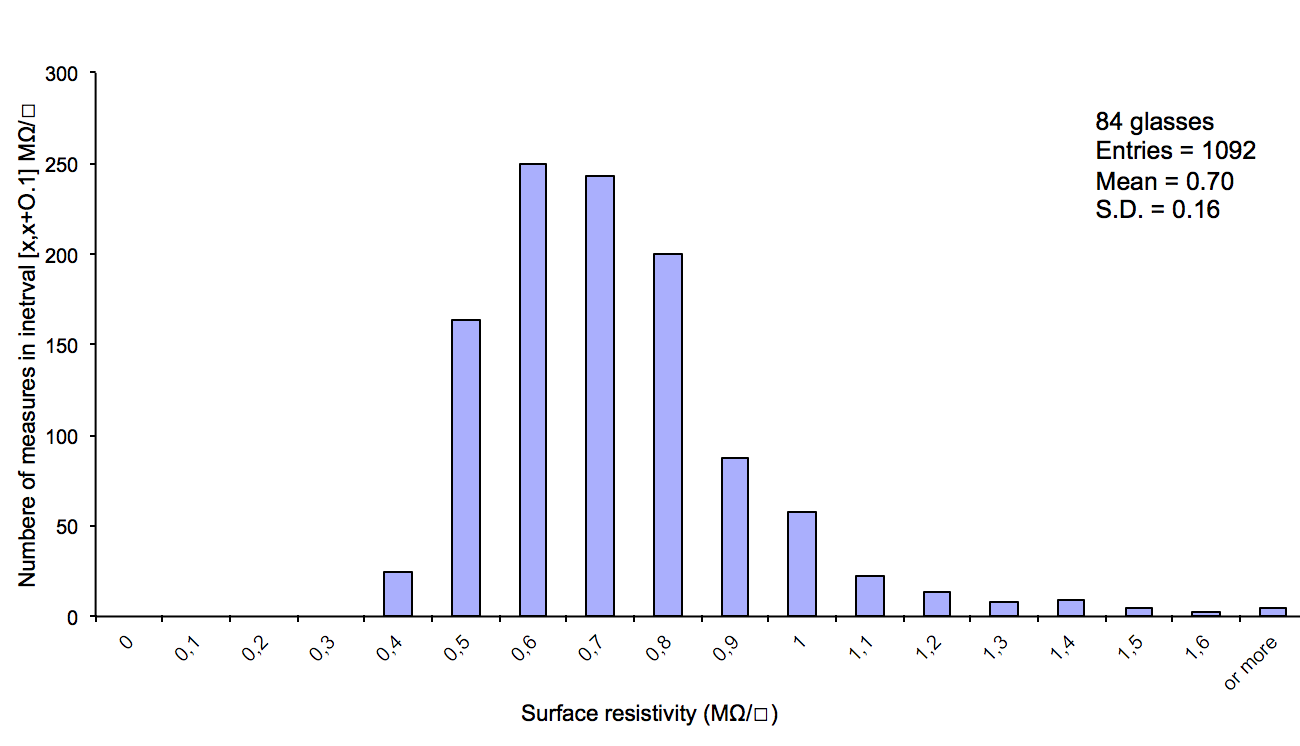}}
\caption{Distribution of average surface resistivity for 84 of $1\unit{m^2}$ glasses coated with bi-component colloidal graphite.
Each coated surface has been measured at 13 locations as shown on figure~\protect\ref{colloid_resistivity}. 
Three surfaces are responsible for most of the values above $1.2\unit{M\Omega/\square}$ and are not used for the computation of the mean and standard deviation.}\label{resistivity_distr}
\end{figure}

The measured surface resistivity at various points over an early
example of a $1\unit{m^2}$ glass coated with the bi-component paint are
shown in Figure~\ref{colloid_resistivity}.  The mean value is
$1.2\unit{M\Omega/\square}$ and the ratio of the maximum to minimum values
is 1.8.  A study was also made of the repeatability of the surface
resistivity between different mix batches.  It was found that surface
resistivity in the range $1-2\unit{M\Omega/\square}$ could be reliably
reproduced, again with a factor of approximately two between minimum
and maximum values.  These results are considered entirely
satisfactory, since it is known from previous tests~\cite{tipp09} that
significant impact on chamber performance begins to be measurable only
for resistivity variations of the order of a factor 10.

Indeed for the majority of the production glasses, a slightly lower
surface resistivity was chosen; Figure~\ref{resistivity_distr} shows the
distribution of average values for 84 glasses.  The repeatability of
the coating process is well demonstrated, the standard deviation of
the distribution being $0.16\unit{M\Omega/\square}$ for a mean of $0.70\unit{M\Omega/\square}$.

Electrical contact to the paint layer is made by means of copper tape with conductive adhesive. The adhesive was chosen so it has no chemical interaction with the coating components in order to avoid the migration of the painting molecules away from the high voltage contact and to ensure a long term contact stability\footnote{After three years of running no loss of high voltage connection was observed in the 48 GRPCs of the prototype.}.

\subsection{Spacer distribution}

\begin{figure}
\centerline{\includegraphics[width=1.0\columnwidth]{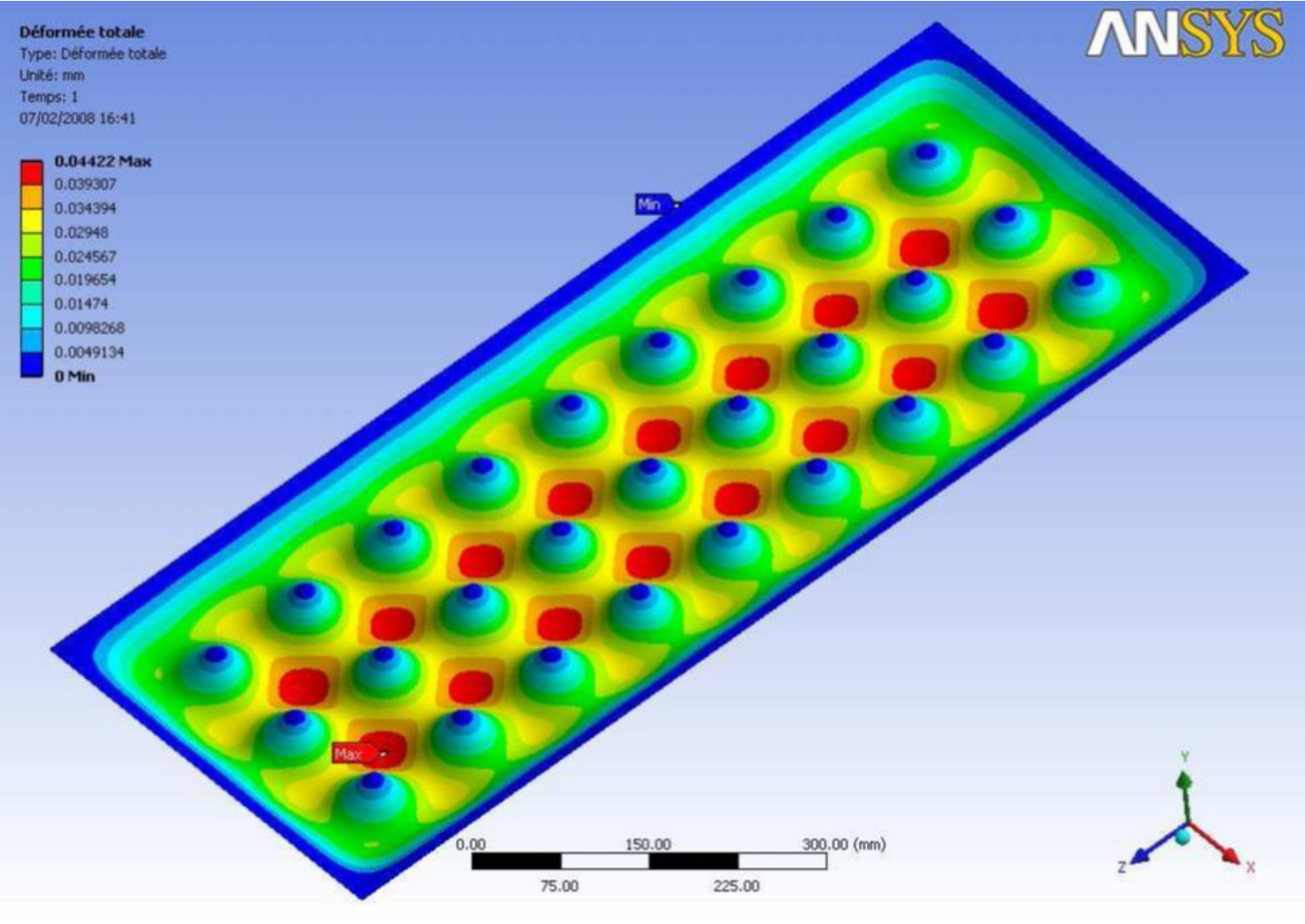}}
\caption{Finite element analysis to optimize the gas gap spacer distribution. Spacers are placed to keep surface bending within acceptable value when High Voltage is on.}\label{fea}
\end{figure}

A finite element calculation was made to determine the optimum
distance between the gas gap spacers (ceramic balls). The goal of this study was to maintain a constant distance between the two electrodes ensuring the same electric field and hence the same gain in all the GRPC chamber while introducing the smallest dead zone.
Two static forces acting on the glass plates were considered: the weight of the
anode plate (chamber horizontal) and the electrostatic force between
the plates for a potential difference of $8\unit{kV}$ (maximum likely chamber
voltage).  The results of the simulation (Figure~\ref{fea}) indicate
that a maximum glass deformation of 44 microns occurs for a ball
spacing of $100\unit{mm}$.  This maximal deformation\footnote {The opposite force produced by the slight excess of the pressure inside the chamber with respect to the outside atmosphere was not taken into account.}  was considered acceptable in terms of the computed gain variation due to the non-uniformity of the
gas gap height.

\subsection{Distribution of gas within the chamber}

\begin{figure}[!ht]
\includegraphics[width=1.0\columnwidth]{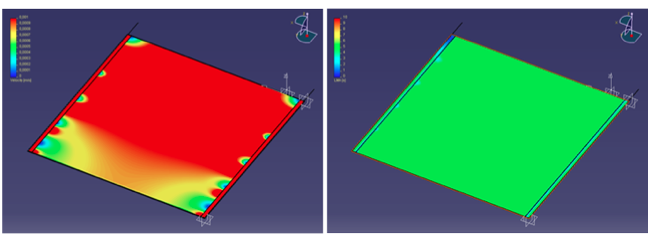}
\caption{ Left  gas speed profile in the range 0-1 mm/s;   Right:  Least mean age profile in the range 0-10 s. Gas flow has been modeled using finite element analysis.}
\label{gas_distribution}
\end{figure}

Gas distribution within the RPC is improved by channeling the gas
along one side of the chamber and releasing it into the main gas
volume at regular intervals.  A similar system is used to collect the
gas at the other side of the chamber.  A finite element model has been
established to verify the gas distribution (see \cite{tipp09} for
details).  Results from this model are shown in
Figure~\ref{gas_distribution}.  The simulation confirms that the gas
speed is reasonably uniform over most of the chamber area and that
this design significantly improves the distribution of gas with
respect to a chamber with no gas channels.  A profile of the quantity
known as the `Least Mean Age' (LMA) was also generated by the model.
This is defined as the time for gas to reach a given point in the
chamber after entering the volume, including the effects of diffusion.
At most points in the chamber the LMA was calculated to be around $5\unit{s}$. When compared with the estimated gas speed  shown in Figure~\ref{gas_distribution}, the LMA value indicates clearly 
that diffusion plays an important role in the distribution of the gas.

\subsection{Mechanical assembly}
\begin{figure}
\centerline{\includegraphics[width=.50\columnwidth]{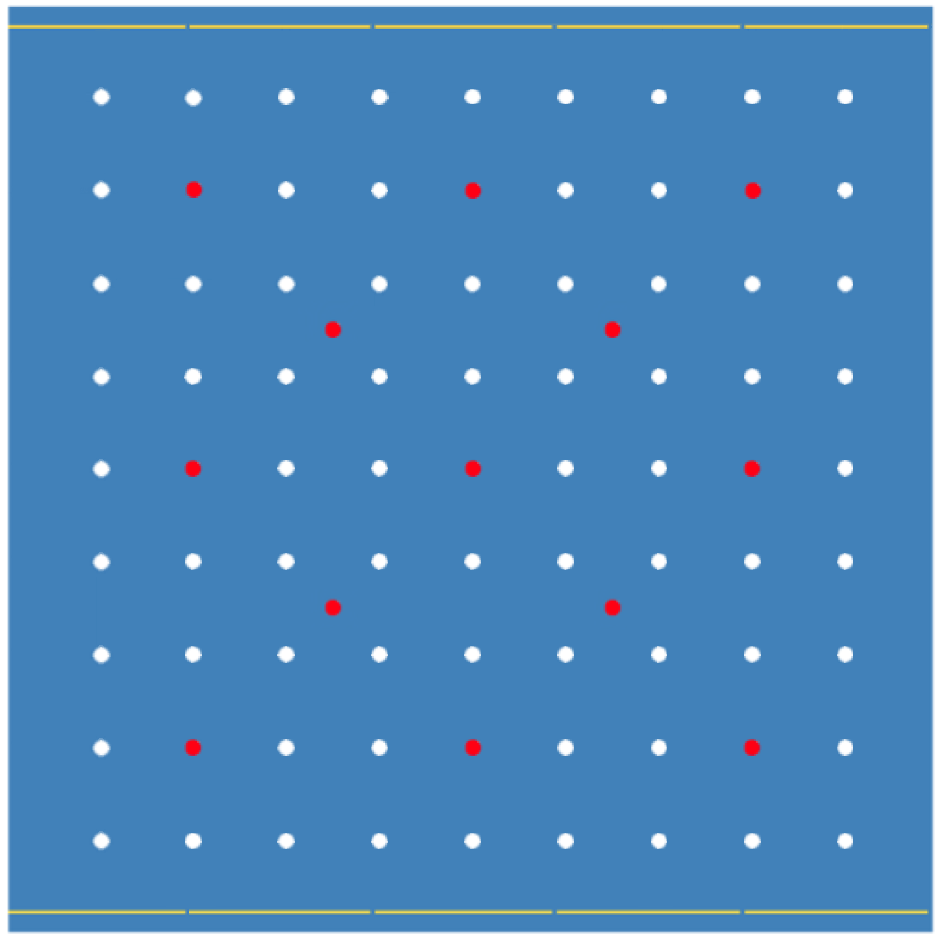}}
\caption{Position of gas gap spacers (red and white dots) and gas channelling (yellow segment) in a $1\unit{m^2}$ chamber. }\label{spacers}
\end{figure}

The glass plates are soda-lime float glass with a bulk resistivity of
$10^{12}\unit{\Omega cm}$. 
Assembly of each chamber begins with the cathode glass supported paint
side down on a flat support (aluminium honeycomb plate).  The gas gap
spacers are then glued onto the glass at 85 locations using Araldite$\circledR{}$ 
2011 epoxy (Figure~\ref{spacers}).  The majority of the spacers (white
dots) were the $1.2\unit{mm}$ ceramic balls mentioned above, but in 13
locations (red dots) the spacers were $7\unit{mm}$ diameter Vetronite$\circledR{}$ G11
cylinders precision machined to a height of $1.2\unit{mm}$.  These spacers,
which provide a greatly increased gluing surface compared to the
ceramic balls, were added to improve the robustness of the assembly and maintain the assembly integrity when the absence of High Voltage does not balance the overpressure needed for gas flowing.  
Our choice of using these two kinds of spacers was motivated by the attempt to reduce the dead zone percentage as much as possible and still have a robust structure. Ceramic balls, glued to one electrode, allow to keep the distance between the two electrodes constant while introducing negligible dead zone. The dead zone introduced by cylindrical spaces is sensibly higher but allow more robustness since they are glued to both electrodes. 

The gas channelling system consists of short lengths of PMMA fibers
(diam. $1.2\unit{mm}$) that are glued to the glass end-to-end, with small gaps
through which the gas is allowed to escape into the chamber (yellow
lines on each side of Figure~\ref{spacers}).

\begin{figure}
\centerline{\includegraphics[width=1.0\columnwidth]{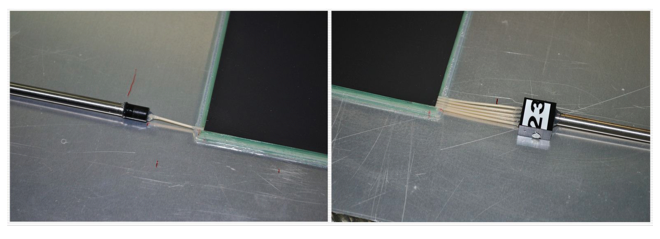}}
\caption{Inlet (left) and outlet (right) capillaries for gas transport.}\label{capillaries}
\end{figure}

The gas volume is closed by gluing G11 strips ($3\unit{mm} \times 1.2\unit{mm} \times 1000\unit{mm}$) around the perimeter of the glass.  On one side of the
chamber, gaps are left at two locations into which capillaries are
glued to allow direct injection / removal of the gas mixture into /
from the channelling system.  The capillaries have $1.2\unit{mm}$ external
diameter with a $0.2\unit{mm}$ wall thickness.  A single capillary is used for
the inlet whereas the outlet consists of 5 capillaries to minimize the
pressure drop (Figure~\ref{capillaries}), thus minimizing the pressure inside the chamber.  Outside of the chamber, the capillaries are
adapted to short lengths of $6\times 4\unit{mm}$ stainless steel pipe for easy
connection to standard gas fittings.  In the case of the outlet, where
5 capillaries must be grouped together, a custom adaptor was produced
in-house (numbered connector in Figure~\ref{capillaries}).

The assembled cathode glass is transferred to a pivoting table and
glue applied to the upper surfaces of the chamber frame, the
cylindrical spacers and the walls of the gas channels.  The table is
rotated to the vertical position and the anode glass is mated to the
gluing surfaces in this position (the glass plates may be handled
manually without supports when vertical).  A lip on the lower edge of
the pivoting table allows precise alignment of the anode and cathode
glasses.  The whole assembly is then returned to the horizontal
position and the glue is allowed to dry.  Weights are placed above the
gluing points during the curing period.

\begin{figure}
\centerline{\includegraphics[width=1.0\columnwidth]{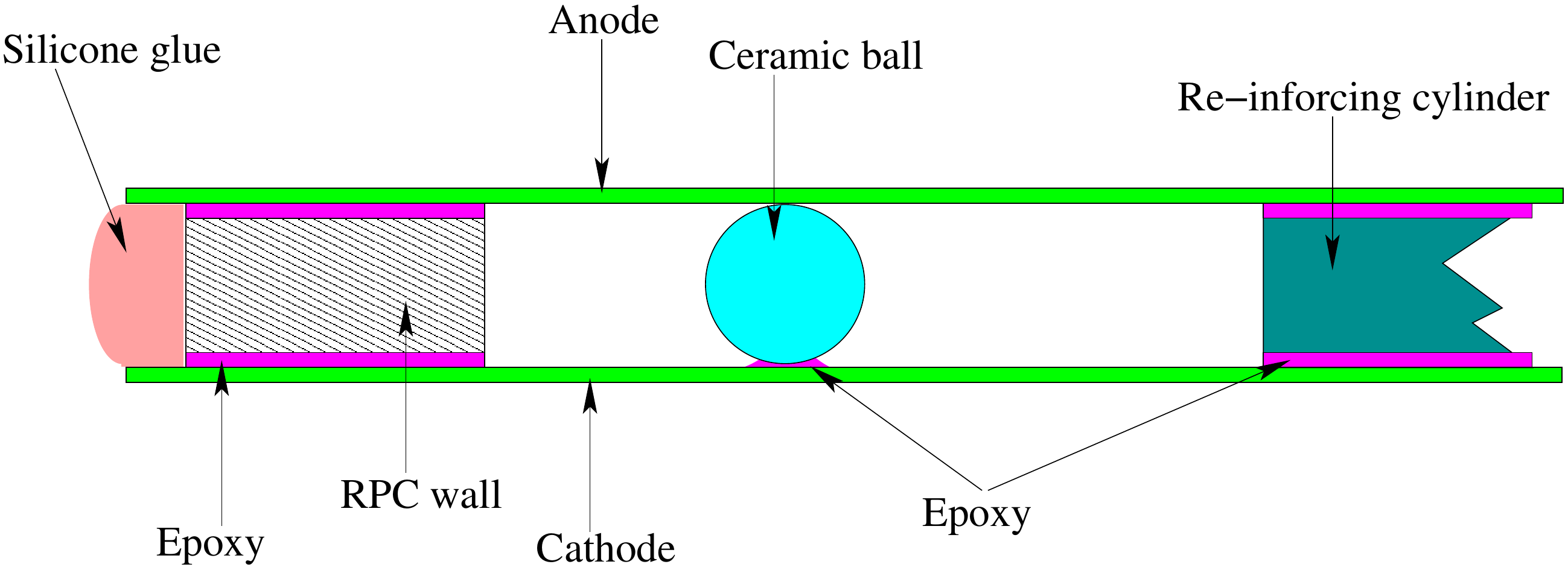}}
\caption{Summary of glue joints inside the RPCs.}\label{gluing_scheme}
\end{figure}

The G11 strips forming the chamber frame are recessed by $2\unit{mm}$ to allow
the later application of a bed of Dow Corning$\circledR{}$ 3140 silicone glue
around the perimeter of the chamber for additional gas tightness.  A
summary of the gluing scheme is given in Figure~\ref{gluing_scheme}.

\section{Electronics readout}
\subsection{ASIC}
\label{sec:ASIC}
\begin{figure}
\centerline{\includegraphics[width=.70\columnwidth]{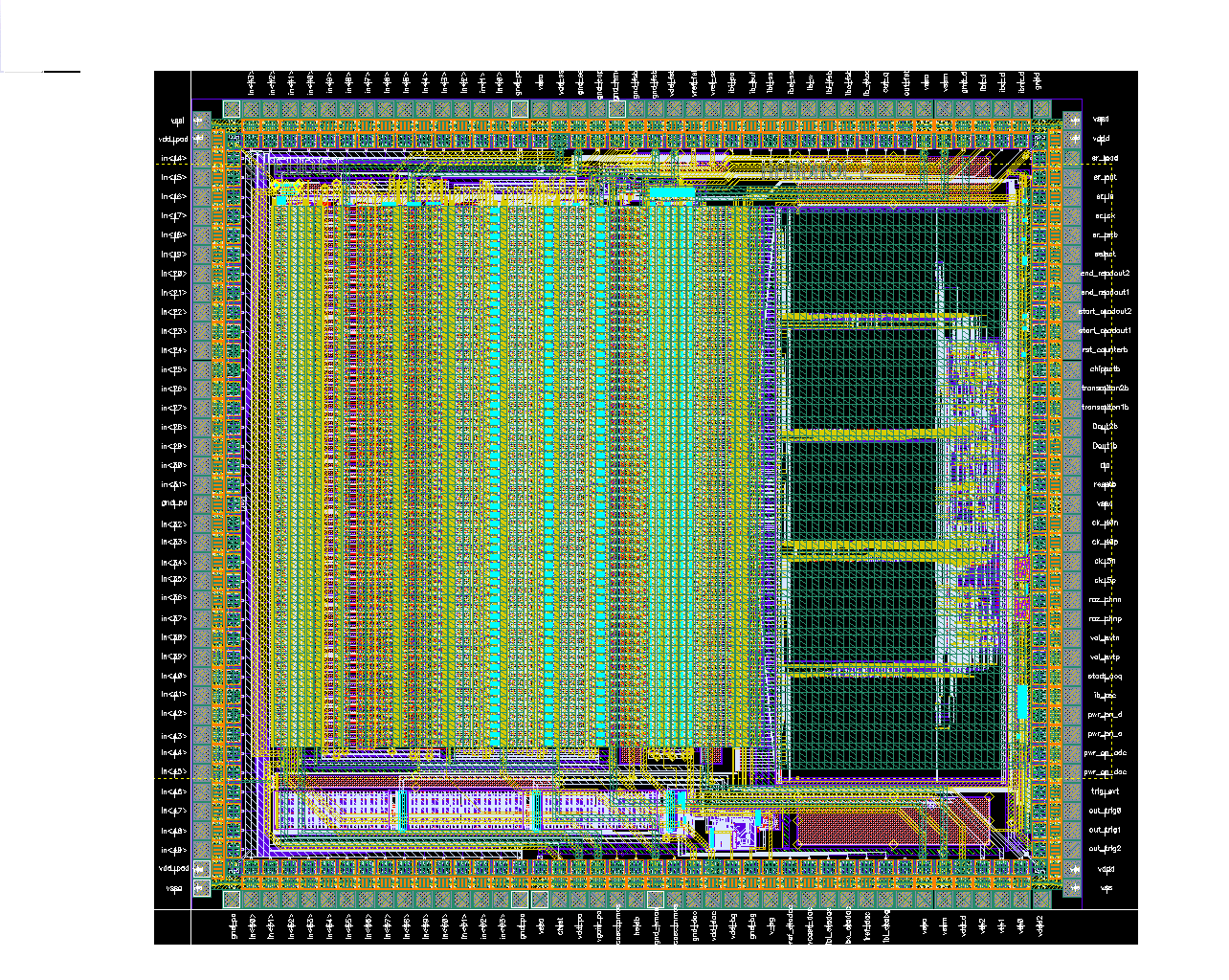}}
\caption{HARDROC layout.}\label{HR2}
\end{figure}

HARDROC (HAdronic Rpc Detector ReadOut Chip) is the very
front end chip (Figure~\ref{HR2}) that was designed for the readout of the RPC detectors foreseen for the
Semi-Digital HAdronic CALorimeter (SDHCAL) of the future International
Linear Collider.  It has been designed in SiGe $0.35\unit{\mu m}$ technology. There have been two  versions of this ASIC : (HARDROC1~\cite{HR1} and HARDROC2~\cite{HR2}). 
The main difference between these two versions is the package which is a 1.4 mm thick plastic Thin Quad Flat Package with 160 pins (TQFP160) for HARDROC2 instead of a 3.4 mm thick Ceramic Quad Flat Package with 240 pins for HARDROC1 (CQFP240). The thin package is more suitable to be embedded inside the detector. It is the HARDROC2 version that was used in the SDHCAL prototype construction. 

\begin{figure}
\centerline{\includegraphics[width=1.0\columnwidth]{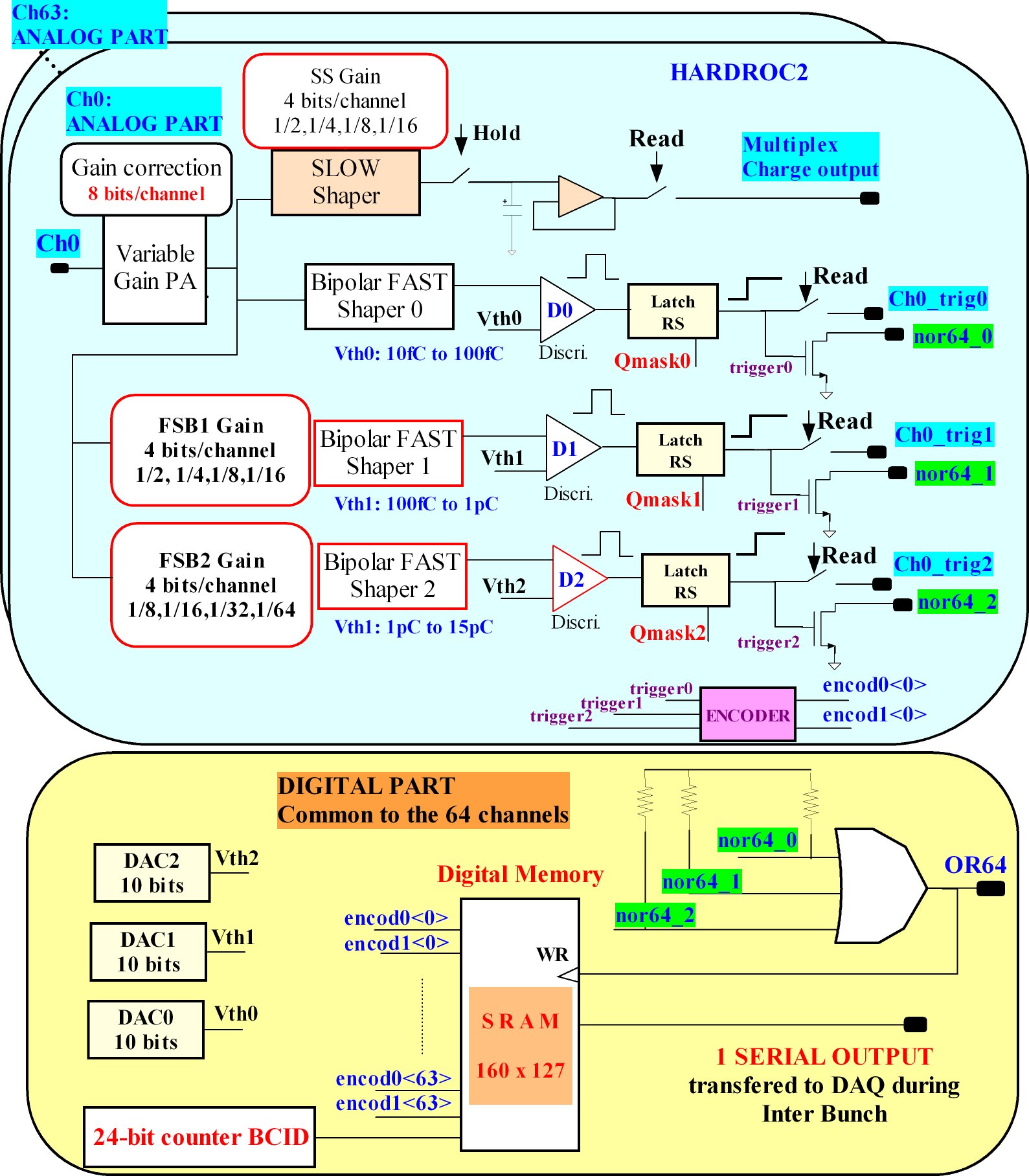}}
\caption{HARDROC2 simplified schematics.}\label{schematic}
\end{figure}

The HARDROC readout is a three-threshold (semi-digital) one
that  integrates on-chip data storage functionality.  Each of the 64 channels of
HARDROC (Figure~\ref{schematic}) is made of a fast low-input-impedance
current preamplifier with a  8-bit precision gain tunable  between 0 and  2, followed by 3 variable gain fast shapers (FSB) connected to
3 low-offset discriminators. The threshold of  one is set to a low value, that of the two others to
medium and  high values in order to auto-trigger down to $10\unit{fC}$ and up
to $15\unit{pC}$. This auto-triggering mode allows reducing the data
volume by recording only signals compatible with those produced by the passage of charged particles in the GRPC. The 3 discriminator outputs are encoded into 2 bits which are
stored in a 127 deep digital memory. Also recorded, the bunch crossing
identification which is coded using a 24 bits counter (ASIC BCID). This internal
digitization implies that only digital data are taken out. To minimize
the number of lines between ASICs, they have been designed to be daisy
chained and read out sequentially during the readout period. To reach a
power consumption of $10\unit{\mu W/ch}$ , the ASIC has been designed to be power
pulsed and a Power-On-Digital (POD) module for the 5MHz and $40\unit{MHz}$
clocks management during the readout phase has been integrated. 872
configuration registers with default values are integrated to
set the required configuration. These configuration parameters are
referred as Slow Control (SC) parameters here after.

\subsubsection {Trigger path}
The trigger path is made of the input preamplifier followed by 3 fast
shapers (with a peaking time tunable between 15 and 30 ns), each followed by three discriminators. The fast shapers are
designed around a band-pass architecture and are referred as FSB0, 1 and
2 for Bipolar Fast Shaper below.  FSB0 is dedicated for input charges
varying from $10\unit{fC}$ up to a few hundreds of fC, FSB1 for input charges
from $100\unit{fC}$ up to $1\unit{pC}$, and FSB2 for input charges from $1\unit{pC}$ up to $15\unit{pC}$.
The feedback network of each shaper can be changed independently
thanks to the SC parameters allowing peaking times between 15 and $30\unit{ns}$ and also a variable gain. The gain of FSB0 is typically about
$2\unit{mV/fC}$ using a feedback resistor of $100\unit{k\Omega}$ and a feedback capacitor of
$100\unit{fF}$. This gain can be varied by a factor of 4.  The gain of FSB1 and FSB2 can be varied using their
variable feedback network and also a 4-bit current mirror gain.  The
output of each shaper is connected to a low offset comparator. The
comparator reference levels (thresholds) are set using three
integrated 10-bit Digital Analog Convertors (DAC) allowing the
settings of the thresholds in the 0-1023 DAC units range. The slope is
$2.1\unit{mV/DAC}$ unit which corresponds typically to $1\unit{fC/DAC}$ unit for
FSB0. \\

\begin{figure}
\centerline{\includegraphics[width=1.0\columnwidth,trim=0 90mm 0 0]{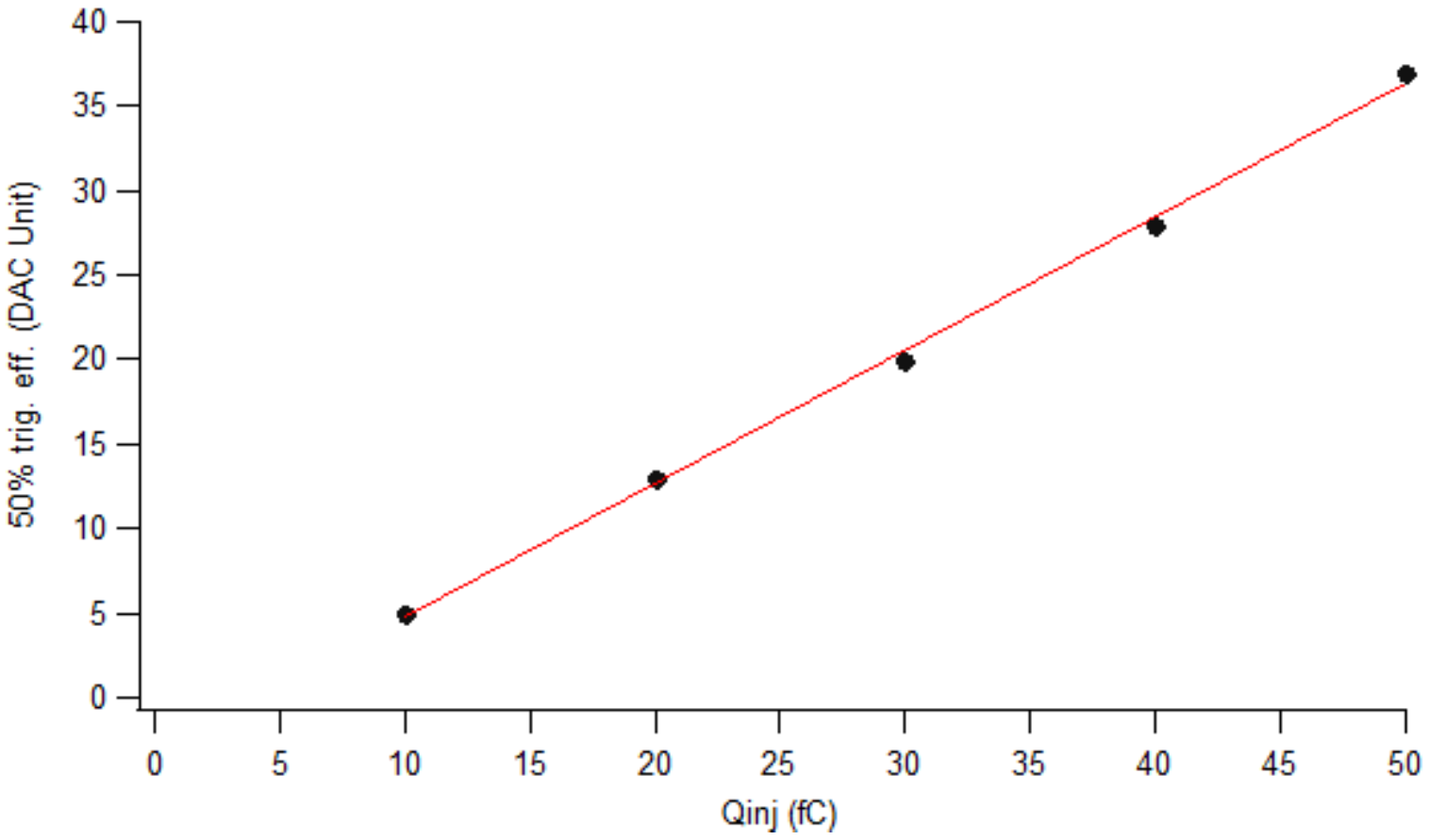}}
\caption{ Relation between the injected charge and the DAC value corresponding to the 50\% efficiency.}\label{conversion}
\end{figure}

\noindent
Finally, to estimate precisely the conversion factor between the injected charge and the DAC value of a given threshold  for the three FSB, a scan of charge  injection  was performed in the range for which the FSB response is linear. For each injected charge and for one of each of the thresholds  a scan of the threshold values is performed and the efficiency of each channel is measured by repeating the injection many times. In absence of electronic noise this curve should be a perfect step function but the efficiency curve obtained experimentally is not. It has a shape of inverse S and is commonly called the S-curve.  It represents indeed an {\it Erf} function and hence the threshold value in DAC units corresponding to the 50\% efficiency of the curve is used to determine the value of the injected charge  in the DAC units. 
Taking advantage of the linear behavior of the FSB in the studied range of charge, a linear fit of the various points obtained by the scan gives the  conversion factor between the injected charge and the DAC. This operation is repeated for the the three FSB. For instance, the linear fit of the response of FSB0 of one channel is shown in Figure~\ref{conversion}. In this case the conversion factor was found to be :   1 DAC Unit  = 0.8 * Q$_{in}$ (\unit{fC}).



\subsubsection {Internal digitization}

For each of the three comparators, a logic OR is made of the 64
channels. By Slow Control, the user can select which of the 3 OR should be used to trigger the digital part of the ASIC. 
The status of the selected OR is evaluated by the
common digital part. If one of the 64 comparators of the selected OR
is fired, the data are stored in the memory. Each comparator output
can be masked by Slow Control to avoid fake triggers due to a noisy
channel. This is known as the "auto-triggering" scheme. The ASIC can
store up to 127 frames in its internal memory. A frame stored in one
ASIC consists of a hit map of the 64 channels with 2 bits per channel
plus a time stamp of 24 bits and an 8-bit chip identifier. The
Gray-coded time stamp is derived by a 5MHz clock.

Each channel of the
ASIC integrates for each channel a test capacitor of $2\unit {pF}$ with a precision of $\pm 0.02\unit{pF}$ (given by the technology). This calibration capacitor is used to calibrate the response of each channel. 
The cross-talk between adjacent channels was measured by injecting an electric signal
of $5\unit{pC}$  through one channel. The signal observed in the other channels was found to be less than 2\% of the injected charge.  About 0.5\% of this crosstalk is given by the analog part of the electronics, the rest is given by the proximity of the other detector cells.  This 2\% crosstalk is low enough for this semi-digital readout as the lowest threshold of the three discriminators is always set to more than 100 fC. The effect of such cross-talk is thus limited to cases when large charge is deposed in one pad ($> 5\unit{pC}$) but this scenario takes place in the core of the hadronic and electromagnetic showers for which the adjacent pads are very often fired at the same time due the presence of many charged particles in the shower core.

The communication between the ASIC and acquisition system (DAQ) 
acts in two steps: an acquisition phase and a readout phase. The
acquisition phase is stopped at the end of the acquisition window or
when a full-memory state (RAMfull) is reached. Then, triggered by an external
signal provided by the DAQ, the second phase starts: after receiving
the StartReadout signal, the first ASIC enters its readout phase and issues
an EndReadout upon completion. The latter is transmitted to the next
ASIC which interprets it as a StartReadout command. To read out large
GRPC detectors with a $1\unit{cm^2}$ lateral resolution many ASICs need to be
assembled together. To limit the number of wires between the ASICs and
the DAQ system the digital readout signals are connected with an open
collector bus per board and daisy-chained in a ring topology, leading
to use only one wire for the link. Similarly, a daisy-chained ring bus
is used for the assignment of the SC parameters to all the ASICs of one board. The situation is
different for the acquisition phase where all the ASICs should start
at the same time. In this case, a StartAcquisition command is
broadcasted to all the ASICs to initiate this phase. Broadcasting signals are sent to all the ASICs in parallel. All the commands
sent to the ASICs by the DAQ are controlled by an FPGA device ( more details in section \ref{sec:DIF}).

To avoid inherent problems related to the daisy-chained system which could result in losing control over a whole branch of the readout system,  a redundancy system was implemented. The system uses two readout buses (one is designed as the default and the other as the spare bus) with the possibility to select one bus or the other through the configuration parameters of each ASIC. In case of an ASIC (number N in the chain) is found faulty on the default bus, the chain is then repaired as follows:
\begin{itemize}
\item ASIC number N-1 is configured to get its data on its default bus  and to output its data on its spare bus.
\item ASIC N is configured  to use its spare bus (both for input and output data).
\item ASIC N+1 is configured to use its spare bus for the input and the  default one for the output.
\end{itemize}

This works as long as two consecutive ASICs are not faulty and the spare bus of the ASIC number N is still operational.

It is worth mentioning here that a  more robust design is foreseen for the next generation of the HARDROC ASIC.  Two I2C (selectable by external programmable means for each ASIC) buses will be used for the configuration parameters transmission.  This is intended to eliminate the possible problems related to the use of daisy-chain technique to configure the ASICs.

\subsubsection {Power-pulsing mode}

Since the ILC duty cycle is expected to be one millisecond bunch
crossing every 200ms, the HARDROC was conceived to take advantage of
such a scenario using the power pulsing mode thanks to a special
module called Power-On-Digital (POD). This consists of switching off
the internal bias currents of the various ASIC channels during the
inter-bunches whereas the power supplies are kept on. There are 3
Power-On signals: Power-On-Analog (controlling the analog part),
Power-On-Digital (controlling the clock-gating) and Power-On-DAC
(controlling the setting of the discriminators thresholds). Each of
them can be forced through the parameters of the Slow Control
configurations.

\subsubsection{ASIC production and control}

\begin{figure}
\centerline{\includegraphics[width=1.0\columnwidth]{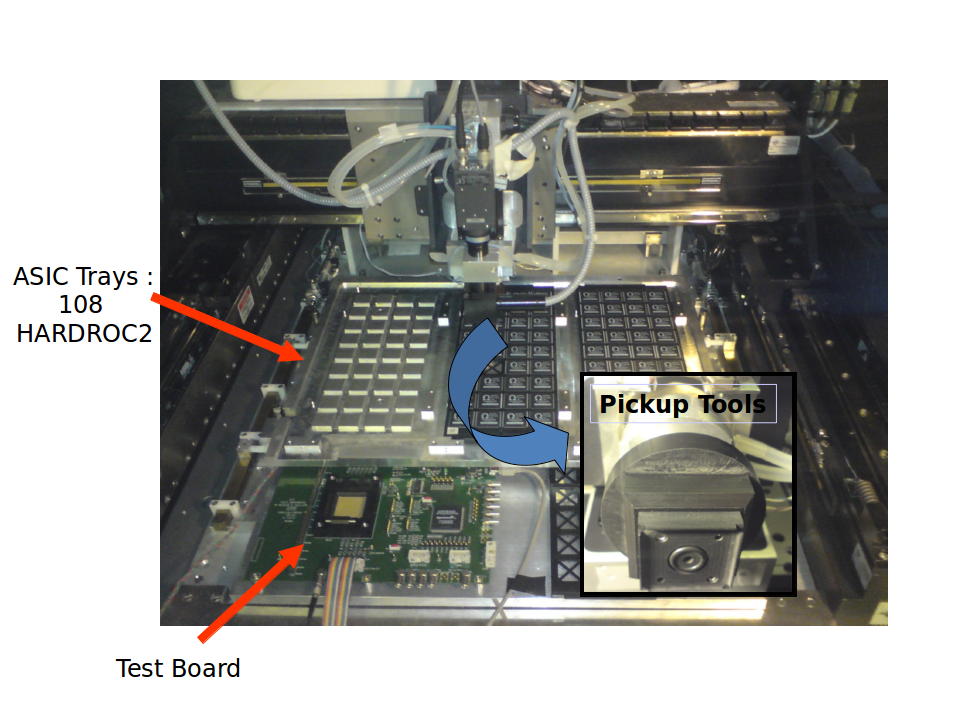}}
\caption{A picture showing the robot that was used to test and calibrate the ASICs as well as the tray with the tested ASICs.}\label{robot}
\end{figure}

\begin{figure}
\centerline{\includegraphics[width=1.0\columnwidth]{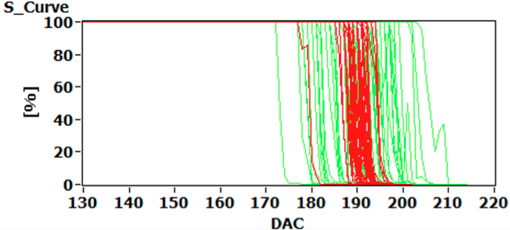}}
\caption{The S-curves of the 64 channels of one ASIC before (in green) and after (in red) gain correction. The S-curves display 
the comparator's triggering efficiency as a function of the comparator's threshold in DAC for a fixed value of the injected charge of $100\unit{fC}$. }\label{Scurve}
\end{figure}

To equip the 48 detectors 
of the SDHCAL, 6912 ASICs are needed. More
than 10500 ASICs were however produced. They were tested and calibrated
using a dedicated tool.  The tool consists of a robot allowing to pick up one by one the ASICs placed in a 
special tray containing up to 108 ASICs and to plug them in a test
board (Figure~\ref{robot}) conceived for this purpose.  A pattern recognition
system using a camera allows to guide the robot to achieve this
step. Once the ASIC is well placed, the proper test then starts using a
Labview$\circledR{}$ application that was elaborated with the purpose to realize all the tests and
the calibration operations automatically. The tests allow to check
that all the channels of one ASIC are operational.  This consists
of checking the response of the three comparators associated to the
three thresholds for a given injected charge. ASICs for which the 64
channels and their three comparators are found to be operational are
then calibrated and replaced in their original place in the tray.
The calibration operations are realized as follows: a scan of the first threshold
is performed by injecting the same charge of $100\unit{fC}$ in one given
channel after selecting the first threshold value in terms of DAC. 
$100\unit{fC}$ is a
typical threshold value used to read out GRPC chambers.
The operation is repeated for a large range of DAC to allow the
construction of the so-called S-curve (Figure~\ref{Scurve}).  The same operation is
performed for different electronics gain of the first comparator.
The 50\% trigger efficiency point of each channel  is extracted by fitting the S-curves with a sigmoid function. 
The dispersion of this 50\% trigger efficiency point gives the 
dispersion between channels for a  given gain.   As illustrated in 
Figure~\ref{Scurve}, this dispersion can be reduced to a few percent 
by tuning the gain of each pre-amplifier individually. In this way all the channels have their 50\% efficiency at almost the same point for a given threshold.The
gain corrections are then recorded in a dedicated data base 
for future use.
The whole batch of ASICs was tested and calibrated with a rate of 200
ASICs per day. A yield of 93\% was observed. It corresponds to ratio of ASICs
which are fully operational.

\subsection{Active Sensor Unit}

The Active Sensor Unit (ASU) is the electronic board that hosts the
different electronic readout components. In the SDHCAL prototype the
ASU was conceived to cope with the Daisy chain scheme. It contains all the connections allowing the transmission
of the different signals between the ASICs as well as those between the
ASICs and the acquisition board. The ASU hosts also the pads of $1\unit{cm^2}$ which are to be in contact with the GRPC detector.

\subsubsection {ASU structure}

\begin{figure}
\centerline{\includegraphics[width=1.0\columnwidth]{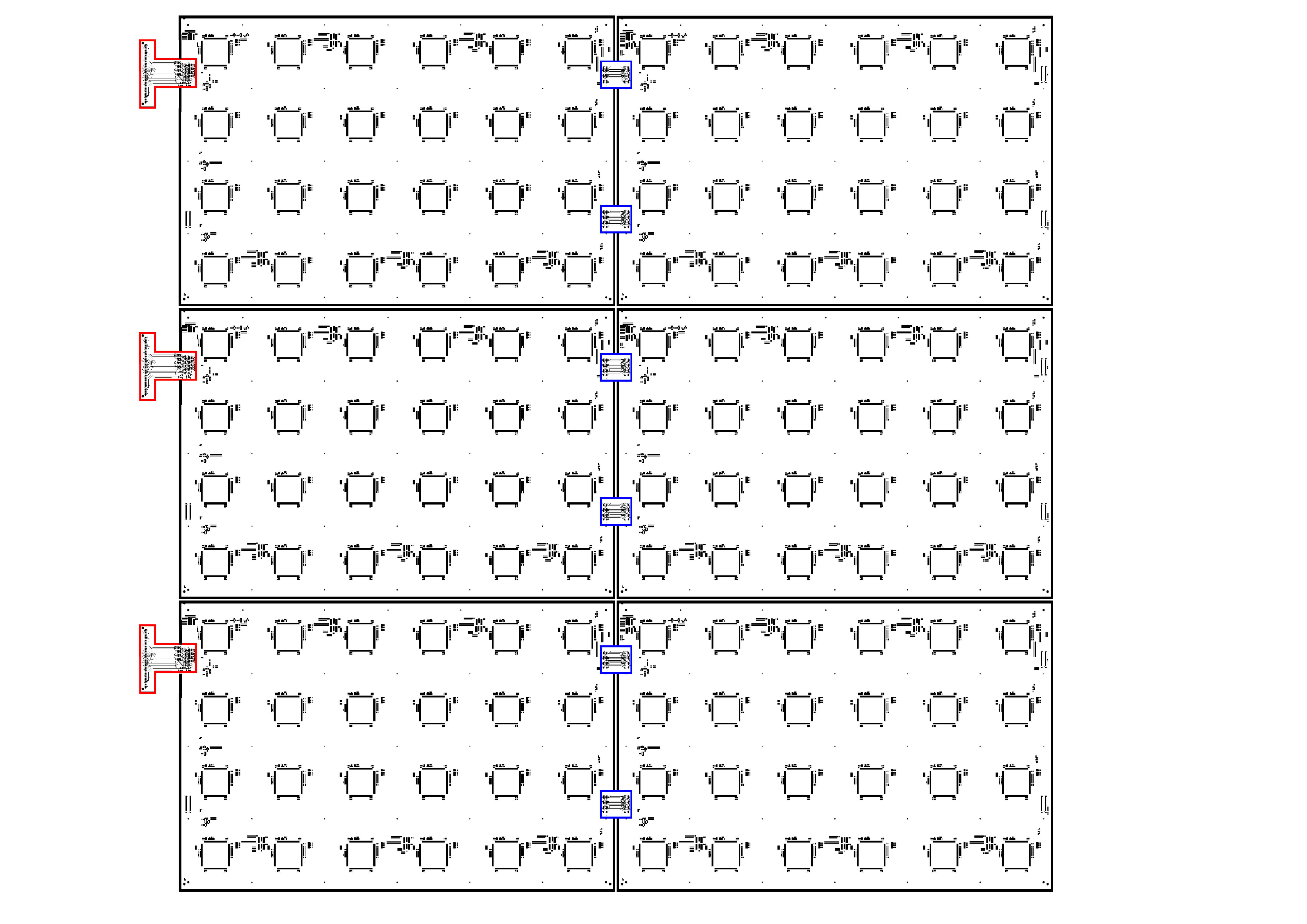}}
\caption{ A scheme of the ASU used to read out $1\unit{m^2}$ detector by assembling six small ASUs. Inside the blue and red shapes are 
boards ensuring connection between 2 ASUs (blue) and between ASU and DIF (red).  }\label{6ASU}
\end{figure}

To read out the GRPC detectors of the SDHCAL, ASUs of the same size
($1\unit{m^2}$) are needed.  Feasibility constraints make the tasks of
circuit production, components soldering, testing and handling of the
assemblies, exceedingly difficult in the case of a single PCB of one
square meter. The solution of dividing that circuit in 6 more
manageable ASU boards was adopted and is schematically shown in Figure
\ref{6ASU}.

\begin{figure}
\centerline{\includegraphics[width=1.0\columnwidth]{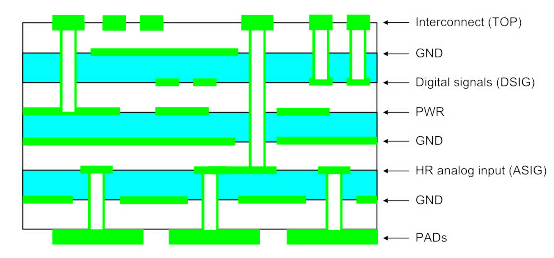}}
\caption{ Transverse structure of the 8-layer readout ASU }\label{ASU-structure}
\end{figure}

\begin{figure}
\centerline{\includegraphics[width=.8\columnwidth]{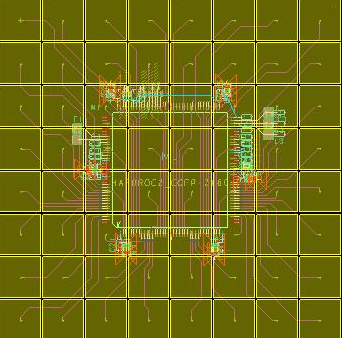}}
\caption{Semi transparent view of the ASU from the copper pads side. 
The image shows how a matrix of eight by eight $1\unit{cm^2}$ pads is connected to the ASIC's pins. 
Two neighboring pads are separated by $406\unit{\mu m}$. } \label{ASU-base}
\end{figure}

\noindent
Each of the six ASUs has been designed to host 24 ASICs. The pads
coupling the GRPC chamber with the front-end HARDROC must be laid
out on the bottom layer of a PCB without other components and
vias. This, with the added constraint of keeping the thickness of the electronics assembly to a minimum, poses the problem of routing
both the ASIC analog input signals and the digital clocks in order to
minimize crosstalk.  A solution based on an 8-layer integrated
circuit, with blind and buried vias, using the stacking diagram is shown
in Figure \ref{ASU-structure}. The designed base pattern of 64 square
pads arranged in a 8 x 8 matrix (Figure \ref{ASU-base}), has been
adopted.   To reduce cross-talk, adjacent pads are separated by a space of 406 microns. As described in ~\cite{smallRPC}, this space is not a dead one since the passage of a charged particle in this  could be detected thanks to the charge induction in the the adjacent pads.
The routing of each input signal from one pad up to the
 ASIC pin has been carefully optimized to reduce the crosstalk:
all input signals are laid out in the same analog signal layer (ASIG
in Figure \ref{ASU-structure}) which is sandwiched between two GND
layers. Great care has been taken to keep the routing of digital
signals well separated from the vias connecting signals in ASIG to the
TOP layer.  The HARDROC base pattern is then replicated 24 times in
the $33.3 \times 50\unit{cm^2}$ boards following a $4 \times 6$ form
factor.  Furthermore, the form factor chosen for the ASU board makes
it possible to use a single Detector InterFace (DIF) board to control a set of 48 HARDROCs:
the pair of two ASU boards (called a slab) is then powered through the
same DIF board and a complete square meter of front-end electronics
requires 3 DIFs.

One of the main challenges in the design of the ASU board
relies on the routing of the LVDS and single-ended digital control
signals while keeping their distribution tree well balanced along the
slab. The LVDS control signals issued by the DIF board are buffered in
a separate interconnect board (DIF-ASU) for each slab of two ASU
boards. Those as well as other signals, together with the power supply,
need to be passed to the second ASU board in the slab.  The
interconnect solution adopted for the board is shown in Figure
\ref{6ASU}:  2 ASU-ASU boards (outlined in blue) interconnect the
signals of the same slab (there are 6 per 1m$^2$), while one DIF-ASU (outlined in red) for each
slab connects the slab to the DIF board (there are 3 SIF-ASUs per 1m$^2$).  The main
constraint in the design of such interconnects relies in keeping the
total thickness of the assembly within $1.4\unit{mm}$ so it does not exceed that of the HR2 package: this has been achieved
using a $0.4\unit{mm}$ pitch board-to-board connector with $0.8\unit{mm}$ stacking
height. The ASU board uses four receptacle connectors on its two short
sides in order to allow the interconnection with the DIF and with a
second ASU board.

\begin{figure}
\centerline{\includegraphics[width=1.0\columnwidth]{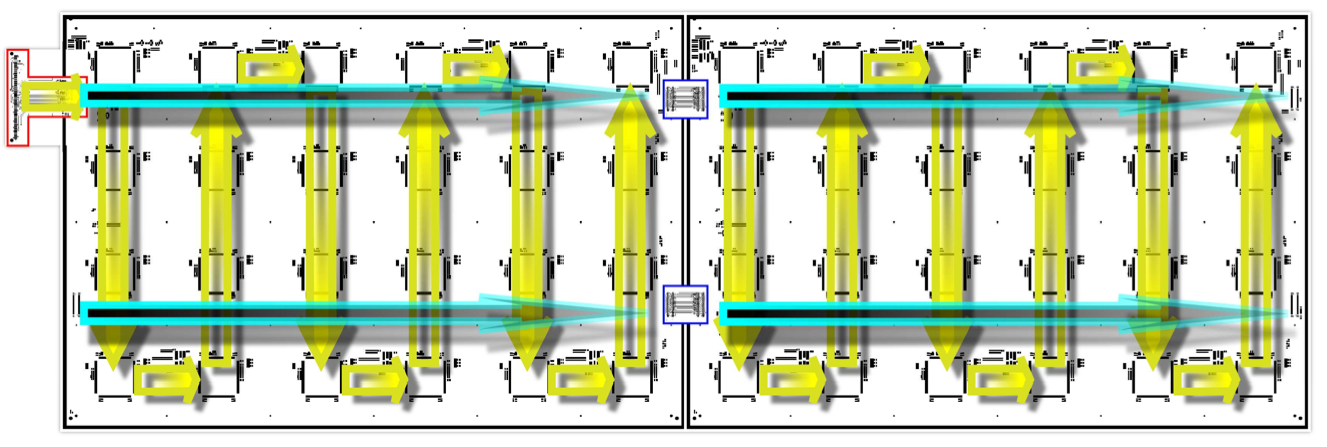}}
\caption{ A scheme representing a slab built by connecting two small ASUs. 
The yellow arrows show the path of the SC line. 
The blue arrows show two of the four LVDS lines used for 
clock and DAQ command distribution.}\label{slab}
\end{figure}
\noindent
Figure\ref{slab} shows the routing distribution flow of the HARDROC
control signals all along the slab: the slow-control signals proceed
connecting the ASIC column wise, with the possibility foreseen on the
board of being buffered at each column of 4 ASICs. 
Actually, two buffers are placed on each ASU : 
one at the beginning of the SC line and the other at its end.  The LVDS signals
(the 40 MHz readout clock amongst them) are routed instead along
differential lines interconnecting 6 ASICs each, which are terminated
on the second board of the slab.

\subsubsection{ASU production and quality control tests}

A pre-production run of 20 PCB has been required to the manufacturer
in order to primarily assess the required planarity under production
conditions. Three of those PCB has then been fully populated with 24
HARDROCs, connectors and components and the assemblies tested in a
validation test-bench. Again, production-like conditions were asked to
the assembly manufacturer, in order to assess the critical capability
of maintaining the correct alignment tolerances between the connectors
throughout the full production run. A positioning tool has been built
by the manufacturer in order to prevent any misalignment of the four
board-to-board connectors during the solder reflow process while an
assembly tool testing the mechanical tolerances of the populated board
assembly has been used for validating the pre-production.  The
validation test-bench has been developed and refined to assess the
production process and has been used throughout the mass production to
check out each assembled ASU. It consists of a GRPC detector  in a
steel cassette and its HV supply with the necessary setup for reading
out the data produced by one ASU and then by a full slab of 48 HARDROCs. An acquisition
DIF board connected to a PC with an adapted version of the DAQ
software collects for each ASU the data generated by the 24 HARDROCs,
triggered only by the noise, until an adequate statistics is
collected. The analysis software rejects any board showing more than
20 total channels dead (i.e. collecting no data) or more than 5 dead
channels on the same HARDROC. All the ASUs produced by the manufacturer
are in a test-only configuration allowing the readout of the 24 ASICs:
after a board has been validated for the integration into the SDHCAL,
it is reconfigured as one of the two possible configurations used
within a slab of two boards. The assembly consisting of two ASU boards
and two interconnecting boards (ASU-ASU) is then tested again checking
the connectivity between the ASUs: the four boards are labeled
together to be integrated into a square meter circuit as part of the
same slab.  A square meter assembly consists of three slabs soldered
together: in order to respect the dimensional tolerances of the
assembled board, a framing support has been manufactured allowing the
correct relative positioning of the 6 ASU composing the 3 slabs
before their soldering.  Only ASU boards having been checked out with
the production test-bench are integrated in a square meter assembly:
the boards failing this preliminary test are labeled as faulty and
enter a debug phase. 

The faulty ASUs are due to problems related to the  ASIC or to the ASU assembly. Completely failing ASICs and  pin-soldering faults preventing the correct functional of some ASICs were the most common in the first category.  
The faults falling into the second category concern shorts and opens found on the assembly, mainly at
board-to-board connectors pins.  


Less than 1\% of the 306 manufactured ASU boards failed to pass the
test-bench and needed some kind of debug and repair: the largest
percentage of those faults was brought by ASICs failing after having been
assembled on the boards. This large number of failing ASICs (53/7344), taking into account that all the
HARDROCs used for the integration were previously tested in a
dedicated chip-level test-bench, require an explanation. Indeed, no Burn-In strategy has been adopted
during the manufacturing. In the first part of the board assembly
production the packaged ASICs were sent to the manufacturer after a
prolonged period of non-dry storage and then they were submitted to
the reflow soldering without a previous baking for moisture
removal. The introduction of a baking routine has cured almost completely the problem of failing ASICs. The adoption of a
lead-based soldering has completely solved the problem with the pin-soldering problem.

\section{The SDHCAL data acquisition hardware}
\begin{figure}
\centerline{\includegraphics[width=1.0\columnwidth]{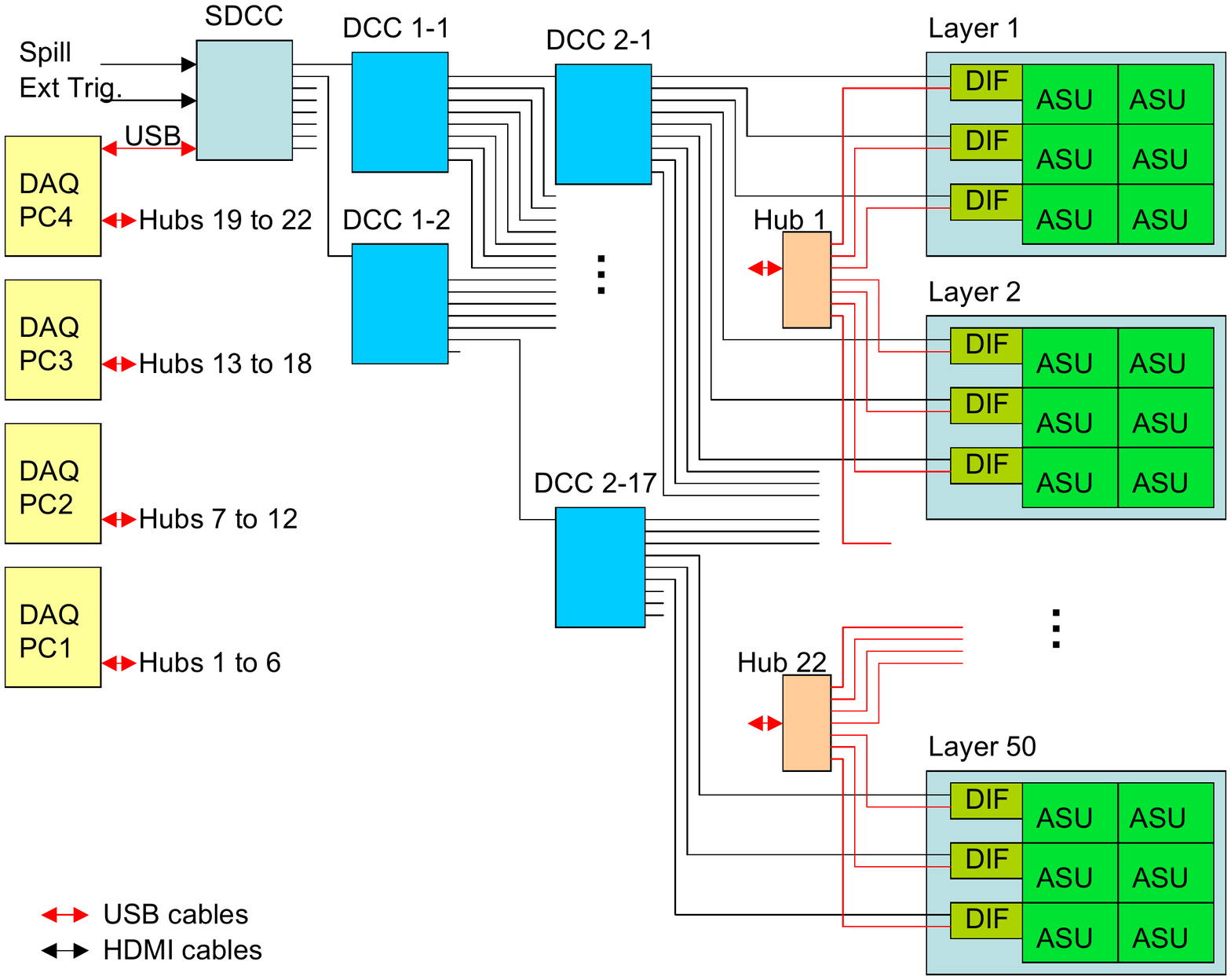}}
\caption{SDHCAL DAQ Architecture.}\label{DAQHW}
\end{figure}

The general architecture of the acquisition system hardware is shown
in Figure \ref{DAQHW}. The DAQ is connected to the computer network in
two ways: the first one allows to manage the system synchronization
using HMDI transmission protocol and the second one, using a USB
transmission protocol, is responsible for SC configurations and the data transmission.  The Synchronous Data Concentrator Card
(SDCC) manages the synchronization of the system. It receives commands
from the computer network and sends them synchronously using private protocol on HDMI cables, to the different Detector InterFace boards (DIFs),
which are connected to the detectors (see next subsection).  The limited number of the SDCC  HDMI
outputs (9 outputs) made it necessary to use  additional boards called Data Concentrator Cards (DCCs). 
They are used as FAN IN/OUT devices to send the commands to a large number 
of DIFs  but with one level of DCC (9 of them) with each having 9 HDMI outputs also, only 81 DIFs could
be connected to the DAQ.  Since each active layer of the the prototype  
needs 3 DIF to be read out,  only 27 cassettes  could thus  be connected to the 
DAQ with one level of DCC. In order to operate the totality  of the 
prototype units, a second level of DCC was  added. This allows to read out  
up to 729 DIF  corresponding to to a number of 243  SDHCAL layers exceeding largely the needed number for the prototype's layers.\\
\noindent
To operate the DAQ system, only one PC is needed to send the commands to the SDCC 
while several PC are used to receive the data. 
\noindent
In the following we give  a detailed description of the different boards used in the acquisition 
system and  their functionalities. 

\subsection{Detector InterFace(DIF) card}
\label{sec:DIF}

\subsubsection{DIF architecture}

\begin{figure}
\centerline{\includegraphics[width=1.0\columnwidth]{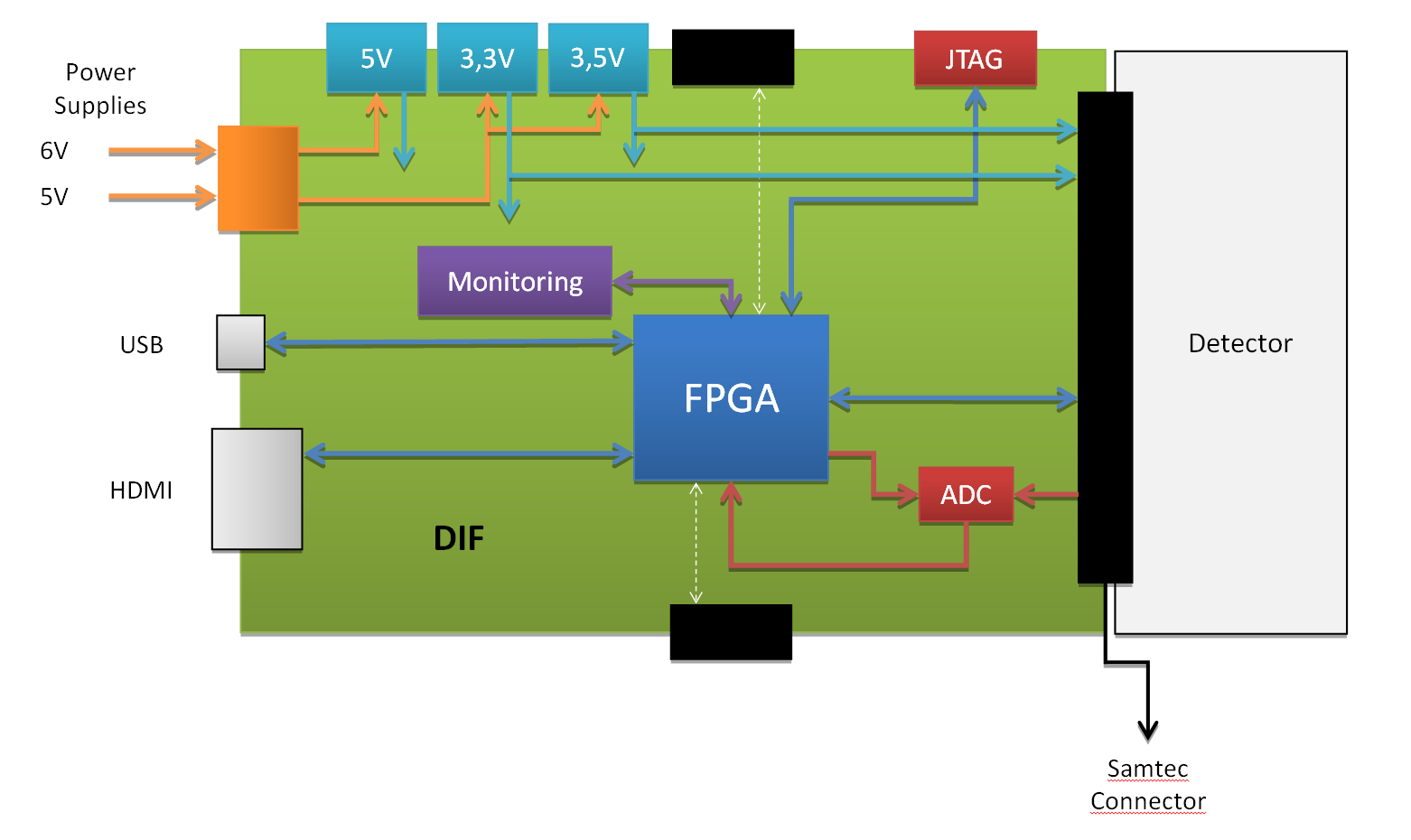}}
\caption{The DIF architecture.}\label{DIF_architecture}
\end{figure}

The general architecture of the DIF is shown in Figure
\ref{DIF_architecture}. The DIF intelligence is based on an Altera$\circledR{}$
FPGA Cyclone 3. The FPGA is connected through USB and HDMI connectors
to the DAQ and through a Samtec$\circledR{}$ connector to the detector's ASU. The data
transmission from the active layer is digital and goes directly to the
FPGA.  
The DIF is also equipped to monitor the current consumption and the DIF temperature. The DIF needs 2
different power supplies, one of $6\unit{V}$ to create, with a regulator, $5\unit{V}$
for the USB devices.  Another power supply of $5\unit{V}$ is also needed to create, with regulators,
3.5, 3.3, 2.5 and $1.2\unit{V}$ for the other devices on the DIF. This concerns in particular
the FPGA (3.3, 2.5, $1.2\unit{V}$). It is also responsible for powering the active layer
electronics (3.5, $3.3\unit{V}$). Each DIF has an identification number (ID),
it can be read directly from the ID tag on the DIF or from the network
computer accessing the EEPROM of the USB device using a USB FTDI245 processor.

\subsubsection{DIF interfaces} 
 The DIF plays a central role in the DAQ system. It links the active layer to the external DAQ devices. This is done through the following interfaces:
 \begin{itemize}
\item  {\bf Detector interface}\\
\noindent
A connector made by the Samtec$\circledR{}$ Company is used to establish the connection
between the DIF and the ASICs of the ASU. It  has 80 pins allowing to transmit and receive the different signals.  It allows also to power the ASU.


\noindent
\item {\bf  Data Acquisition (DAQ) interfaces}\\
 \noindent
 The two interfaces linking the DIF to the DAQ are the following :\\
 \noindent
 {\bf 1-USB Protocol.} The USB transmission system is used for data
 transmission from the DIF to the computer network and for receiving
 non synchronous commands (like the ASIC configuration one) from the
 computer network. It is used also for register control to read and write actions.
 To handle the USB communication protocol an FTDI chip is used. \\
\noindent {\bf 2-HDMI Link.} The HDMI transmission system is used
for sending the clock and synchronous commands to the ASIC.  It allows in Test Beam configuration to provide the Trigger and Spill\footnote{Spill signal is used to drive the ASU power pulsing.} signals. 
It is also used to concentrate Busy and RAMfull signals from the DIF
to the SDCC. Those 2 signals are mandatory for the synchronization of
the whole system (see Section \ref{sec:SDCCfeatures}). 
The protocol used for the HDMI commands is a homemade protocol which includes a header, the command
number and a checksum. Almost all  the HDMI signals are real LVDS signals\footnote{Except pins 15 and 16 that can be used as LVDS signal but there are not twisted pairs.}.
\end{itemize}
\subsubsection{DIF features}
\label{sec:DIFfeatures}
The FPGA on the DIF is the key element  of the board. 
It controls everything on the board and performs all the features of the DIF. These features are summarized here after: 
\begin{itemize}
\item {\bf ASIC configurations}\\
As mentioned before the SC parameters are used to configure the ASICs of the detector's ASUs.  
Because of some spread in pedestal, gain and thresholds, each ASIC has
a specific configuration. The SC parameters allow to homogenize all
the ASIC responses.  The command to transmit these parameters  does not require synchronization; it is
done by the USB transmission data system.  When the command is
started, the computer network sends the needed amount of data to
configure all the ASICs.  872 bits are necessary to configure one
HARDROC. There are 144 ASICs on each active layer unit. Therefore, more than 6~Mbits are sent to the prototype's ASICs from the
computer network.  For each DIF, a 16 bit Cyclic Redundancy Check
(CRC) is sent at the end of the SC parameters transmission to verify that it was
correctly done.  After receiving all the configuration parameters, the DIF
forwards them to the ASIC through the Samtec connector.  Since the
configurations transmission is  daisy chained, it is possible to
return the configuration parameters from the ASICs to the DIF if the detector Printed
Circuit Board (PCB) is routed for this purpose. In this case the DIF
sends twice the configuration parameters to the ASICs so that the first
configuration can be checked against any modification during the transmission process inferring that  
the 2nd configuration transmission is correct.\\
\item {\bf Digital Readout}\\
\noindent
When the ASICs are configured, the data acquisition can be started and 
run according to one of the 2 following modes:\\
\noindent
{\bf 1-Trigger mode}: In this mode the ASICs are active after the StartAcquisition signal. Upon the arrival of an external signal (for instance a coincidence signal 
from scintillator PhotoMultipliers system tagging the passage of a particle) a StopAcquisition signal is received by all the  ASICs  from the SDCC through the DIFs and followed by a StartReadout allowing to transmit the collected data to the DIFs. Since the ASICs are self-triggered,  it may  happen that the memory of one of them gets full before the arrival of the Trigger signal. In this case a RAMfull signal is produced by the ASIC and transmitted to the DIF which forwards it to the SDCC.  A global signal is then sent from the SDCC to all the ASICs through their own DIF  and their memories are reset before starting the acquisition process again.\\
Trigger mode is used generally when the time information of beam particles going through the detector cannot be extracted from the data only and one needs to discriminate the collected data due to the particles from those due to the noise. This is typically the case when dealing with a small number of chambers. This mode is also useful when particle identification signals (from e.g. Cherenkov detectors or a muon veto) enter the coincidence to enrich the purity of the event sample. \\
\vglue .2 cm
\noindent
{\bf 2-Triggerless  mode}: In this mode the ASICs are put into acquisition state  at the start of a
Spill signal (for instance the accelerator clock). When any of the ASICs has its memory full, it sends a  RAMfull signal and, as for the Trigger mode, this signal is sent to the SDCC through the  DIF and a global signal is sent back from the former to stop the acquisition process  in all the ASICs, followed by a StartReadout signal to initiate the  readout process .\\

It is worthwhile to mention that in both modes when the acquisition process is stopped and the readout is started, a timestamp called Bunch Crossing IDentifier (BCID) , whose counter is synchronized with the ASIC's one, is recorded by the DIF and included in the data stream. This allows an easier reconstruction and analysis
of the data.\\
\vglue .2 cm
\noindent
\item {\bf Power-pulsing (PP) mode}\\
 In the case of the future ILC, the active time corresponding to the beams Bunch Crossing (BC) is expected to last $1\unit{ms}$ every
 $200\unit{ms}$. During the relatively long inactive period, the ASICs can be
 switched off in order to reduce the power consumption of the
 electronics  and thus the heat dissipation in the calorimeters. Therefore  a mode which consists in 
 powering  ON the ASICs  during the BC and the
 subsequent readout and powering them OFF the rest of the time is extremely helpful. The same mode  could be used during the
 test beam operations. In this configuration  the power pulsing is synchronized with the  accelerator clock by using the particle spill signal delivered in the
 control room.   Nevertheless, in both the ILC and Test Beam conditions,  when  the chip
 is turned on, a certain programmable delay is applied before any
 detector signal can be recorded to the memory. This delay accounts
 for the stabilization of the various voltages and currents inside the
 chip.  When the  delay is too short  the detector occupancy is dominated by noise. Measurements with oscilloscopes show that the
 typical delay to have the voltage stabilized is around $25\unit{\mu s}$
 (Figure~\ref{fig:oscilloPP}). To keep a safety margin, 
 the delay applied before detector signal can be recorded was set
 to $100\unit{\mu s}$. The PP mode could be used with either the Trigger  or the Triggerless mode.

\begin{figure}
\centerline{\includegraphics[scale=1]{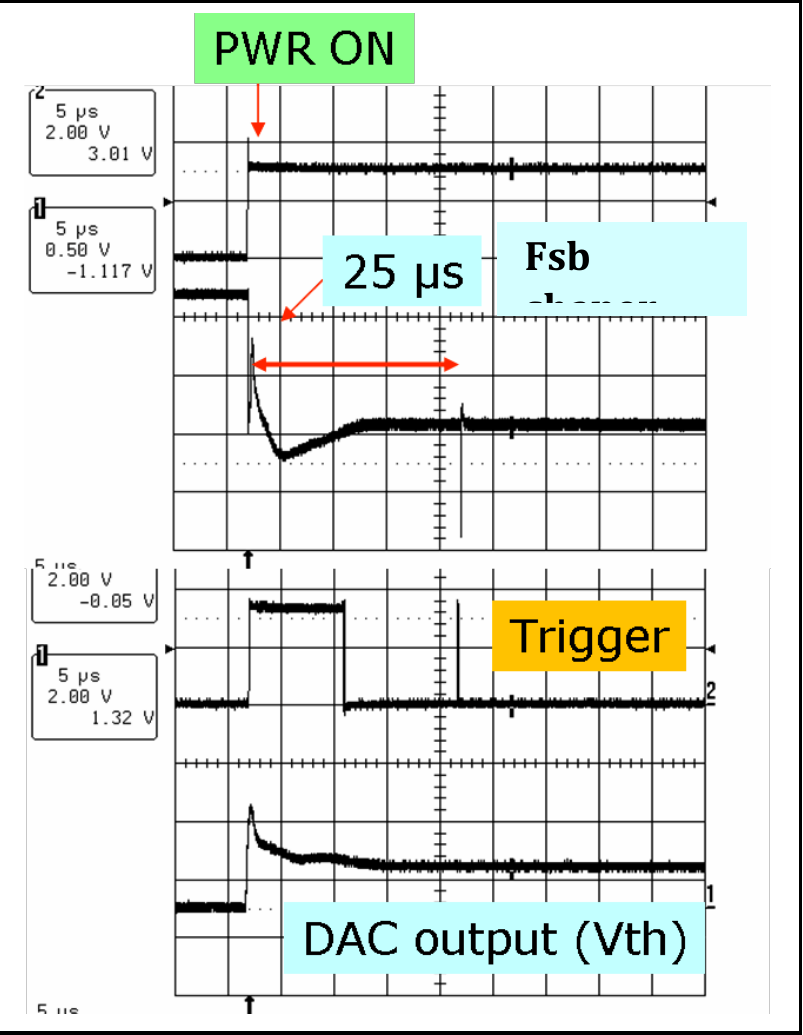}}
\caption{Voltage measured on various HARDROC lines showing the stabilization of a HARDROC threshold and FSB line voltages at a Power-On-Analog signal.}\label{fig:oscilloPP}
\end{figure}

\noindent
An important point to mention here is that for the various modes, the readout of the ASICs  is automatically performed at the end of the
acquisition controlled by the SDCC. The DIF FPGA performs the readout
of all ASICs under its control. The data are stored in the FPGA memory
and sent with other information specific to the data format to the computer network through the USB link.
While the FPGA performs the readout, it also generates a Busy signal
which reaches the SDCC in order to avoid starting a new acquisition process, ending when all data are read out. Indeed, as soon as the last ASIC has sent his
data, the busy signal is released and the DIF can accept again a new StartAcquisition command from the SDCC.\\
\noindent
\item {\bf Data format}\\
In order to facilitate the reconstruction and analysis of the data,
additional information are added to the data stream according to a
specified format.  For every readout operation, the ASIC data are
encapsulated by the DIF's FPGA with a header, a trailer and the
following information before they are sent to the computer:
\begin{enumerate}
\item DIF Trigger Counter (DTC): Coded in 32~bits, it counts the number of readouts. It is
reset at the first acquisition of each run.\\
\item Information Counter (IC) :  Coded in 32~bits. Bits 23 to 0 are used to count the dead
time. This occurs when ASICs are not acquiring. It is reset every acquisition. Bits 31 to 24 are used for  BCID counter overflow. It is reset at the first acquisition of each run.\\
\item Global Trigger Counter (GTC) : Coded in 32~bits. It counts the number of triggers received by the DIF when the Trigger mode is run and  counts the number of readouts in the Triggerless mode. It is reset at the first acquisition of each run.\\
\item Absolute BCID : Coded in 48~bits. It is incremented with the $5\unit{MHz}$ clock received 
from the SDCC. It is reset at the first acquisition of each run.\\
\item BCID DIF : Coded in 24~bits. This is incremented  with the $5\unit{MHz}$ clock coming from the SDCC and is synchronized with the ASIC BCID mentioned in  section \ref{sec:ASIC}.\\
\end {enumerate}
\noindent
\end {itemize}

\subsection{Data Concentrator Card (DCC) and Synchronous DCC (SDCC)}

\subsubsection{(S)DCC architecture}
The two boards have the same hardware. 
A (S)DCC board has a VME format. Each board has 9~HDMI connectors on one side and 1~HDMI, 1~USB and 2~lemo connectors on the other side. 
All the connectors are connected to an FPGA so that the FPGA can send the signal form one side to the other.
The DCC differs from the SDCC by the FPGA firmware. 

\subsubsection{DCC interfaces}
The two interfaces the DCC has are the following: 
\noindent
\begin{itemize}
\item  {\bf Synchronous Data Concentrator Card (SDCC)  interface}\\
The DCCs and the SDCC send each other information through the HDMI
transmission system. The SDCC sends clock and commands to the DCCs and 
the DCC FAN IN sends the Busy and RAMfull signals from the DIFs to the
SDCC.\\
\item{\bf DIF interface}\\
The DCC can be connected up to 9~DIFs. The DCC sends to all the DIFs the clock and the commands received from the SDCC in a synchronous
way. Each DIF sends its Busy and RAMfull signals to the DCC.
\end{itemize}

\subsubsection{SDCC interfaces}
The SDCC has two kinds of interfaces. One is related to the DIF directly or through the DCC. The other one manages the link with the computer network. 
\begin{itemize}
\item {\bf Data Acquisition (DAQ) interface}\\
\noindent The SDCC is connected either directly to up to 9~DIFs or
to up to 9~DCCs to increase the number of DIFs connected to the DAQ.  \\
\item{\bf Computer network interface} The SDCC is connected to
the computer network to allow the user to control the DAQ. 
\end{itemize}
\noindent The commands which need synchronization are sent from the PC
to the SDCC through a USB transmission system, and then the SDCC sends
the command to all the DIFs  through the HDMI transmission system.  
These commands are sent with the $5\unit{MHz}$ clock, but they are serialized
(SDCC) and deserialized (DIF) with a $50\unit{MHz}$ clock. The commands contain
one header, the command number and a checksum.
Some
commands are only meant for the SDCC and are not forwarded to the
DIF.

\subsubsection{(S)DCC features}
\label{sec:SDCCfeatures}
The DCC acts like a FAN OUT with the information coming from the SDCC
and like a FAN IN with the information coming from the DIFs.

The SDCC is connecting the DAQ PC through its USB interface with the DIFs through 9 output HDMI connections. 
The main commands and signals controlled by the SDCC FPGA are summarized here after.
\begin{itemize}
 \item{\bf Trigger signal}: In Trigger mode, the DAQ needs the
trigger signal to stop the acquisition and start the readout of the
detectors. A lemo connection is foreseen for this purpose on the SDCC
board. 
\item{\bf Spill signal}: When the ASIC  are power pulsed, the DAQ needs 
the duty cycle or the Spill signal from the accelerator.  The signal should be active
during the BC or the spill but also slightly a bit before ($100\unit{\mu s}$).  This permits
the DAQ to power ON the system when the spill signal is active and to
power OFF the system when the spill is inactive. Another lemo
connection is foreseen on that purpose on the SDCC board.
\item{\bf Busy and RAMfull signals}: The Busy and RAMfull signals are used to maintain the
synchronization and the automation of the different processes of the DAQ system. 
Both signals are sent from the DIFs and then concatenated in the DCCs up to the SDCC.
Depending on the running mode, Trigger or Triggerless, the RAMfull signal can have different consequences. In the former it is followed by a reset of the ASIC memory while in the latter it initiates the readout phase.  Concerning the Busy signal, it 
is active when the ASICs are being read out. While the
Busy signal is active, a new start acquisition
can not be sent. But as soon as the last Busy is unset 
from the DCCs connected to the SDCC, the SDCC sends a 
new StartAcquisition command to the different DIFs to launch a new 
acquisition. This is how the automation is done in the DAQ system. 
\end{itemize}

\section{The SDHCAL data acquisition software}

The data acquisition software is split in three parts: the low
level hardware access that hides  effective hardware implementations,
the configuration data base software handling devices description
and SC parameters and finally the data collection and monitoring. All
packages are written in C++ with interactive scripting in python.
Low level and database C++ libraries are all parsed to python object
with the Swig tool. It allows an interactive instantiation and debug
of single code pieces.

\subsection{The low level hardware access}

\subsubsection{USB readout}

DIFs and SDCC FPGAs are interfaced with the same USB chip. As mentioned before this is  an
FTDI chip. The DIF is thus uniquely identified by its FTDI device identifier (id) stored in an EEPROM and access to
a specific device id is done using either the proprietary library
FTD2XX or the free version library libFTDI. A device driver library (UsbDeviceDriver) was develop
ed to implement a set of low level access (read and write of registers) of the two boards.

\subsubsection{DIF and SDCC readout}

An upper layer software is dedicated to DIF configuration and
DIF readout. The class called BasicUsbDataHandler aggregates a pointer to
a UsbDeviceDriver and to a configuration buffer handling all DIF and
SC parameters. Specific methods are used to implement the DIF configurations,
the ASIC configurations and a single event readout to an internal
buffer. Similarly, an SDCCDataHandler implements commands associated to the SDCC.

\subsubsection{DIF Manager}
\label{sec:DIFManager}

\begin{figure}
\includegraphics[width=0.9\textwidth]{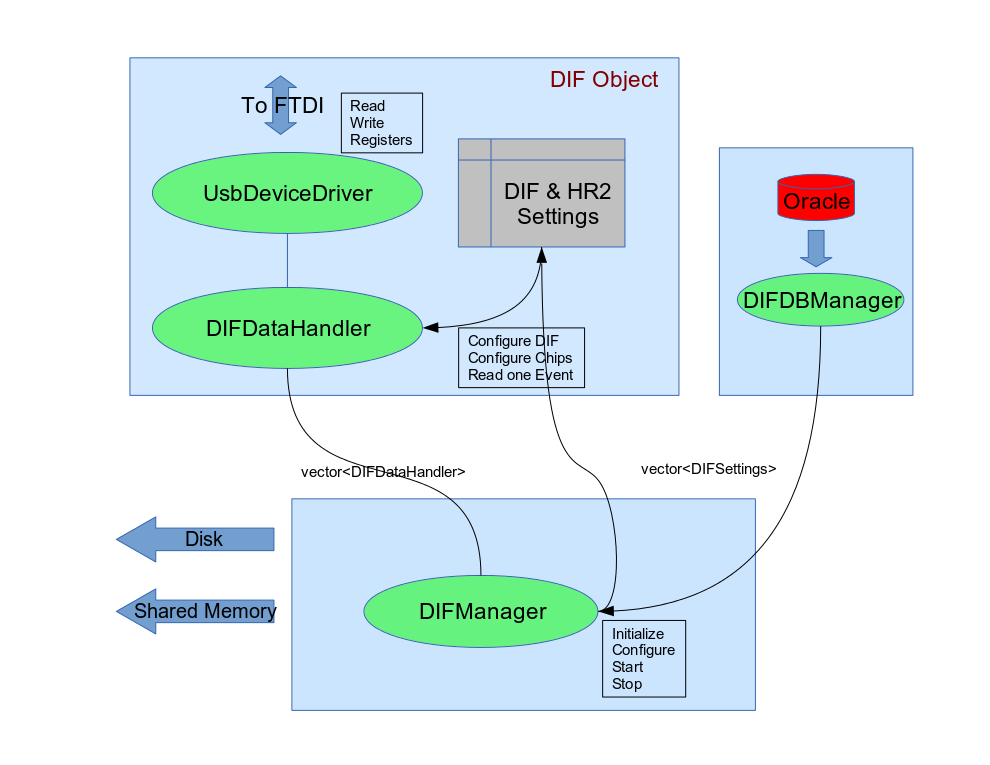}

\caption{Low level software architecture and configuration database. The black arrows show the path of configuration data.}
\label{software_archi}

\end{figure}

Eventually one DIF manager class per PC is responsible of the data
taking of the DIFs through the USB connected to it. It handles the DIF and ASIC
configuration parameters via an interface called  DIFManager. This scans and
detects all connected DIF, instantiates one DIFDataHandler class per detected
device and distributes configuration parameters. 
When data taking starts, it starts a polling thread continuously
reading events on all connected DIFs. Events can be directly dumped
to disk in LCIO\cite{Aplin2012} format or transferred to the central data acquisition.
Figure~\ref{software_archi} summarizes this architecture.

\subsection{The configuration database}

The configuration database gives the possibility to store and retrieve
all parameters needed by the DAQ system. The database itself is hosted
on an Oracle server at CC IN2P3 (Villeurbanne, France). To interface
this SQL database with the DAQ software and to allow users to insert
and query data without knowledge of SQL, a C++ library has been written.
Part of the DAQ system being written in Python, we used Swig to generate
a Python version of the C++ library. The system has been designed
to easily allow the addition or modification of existing object parameters.
It is built on 2 levels: the database itself and the C++ library.

\subsubsection{Database model}

The database model is conceived to deal with  extendable number of ASICs. It can also handle different kinds of ASICs using different  settings of parameters to take into account addition of other sub-detectors. 
 In this model, each ASIC has a unique entry in the ASIC table, containing its type. The actual configuration parameters
are contained in 2 tables: the first is  ASIC\_CONFIG for the parameters common
to all ASIC types and the second   <ASIC\_TYPE>\_CONFIG for the parameters specific
to a given type. For instance the configuration parameters for a HARDROC
ASIC will be stored in the ASIC\_CONFIG table for the common part
and in the HR2\_CONFIG table for the HARDROC specific parameters\footnote{Here label HR2 is to indicate that we are using the version 2 of the HARDROC ASIC.}.  As a
consequence, supporting a new type of ASIC requires only to create
a new table containing the specific configuration parameters. When
downloading the parameters, the system will automatically choose the
accurate tables associated to the type of the concerned ASIC. 

\subsubsection{C++ Library}

All accesses to the database are done through the C++ library. There
are no mention of the actual configuration parameters in the library: 
all names of parameters are retrieved from the database itself at
runtime. The methods used to set or to get parameters value in the
application (API) use the parameters name as an argument (ex: setInt(\textquotedbl{}B0\textquotedbl{},5)
and getInt(\textquotedbl{}B0\textquotedbl{}) ) to respectively set and
retrieve the integer value of the B0 parameter). As a consequence,
there is no need to modify the C++ library when modifying parameters
or adding new types of objects. This will be done dynamically from the
database itself. Each stored configuration is associated with a version
number having the form "MajorID.MinorID". A configuration with version
number 1.0 will contain all the parameters for all the DAQ objects,
whereas a configuration with version number 1.$<$X$>$ will only contain
the modifications from version 1.$<$X-1$>$. This allows a reduction in the
amount of data to transfer and to store.

\subsubsection{Tools}

\begin{figure}
\centerline{\includegraphics[width=0.9\textwidth,trim=0 100mm 0 0]{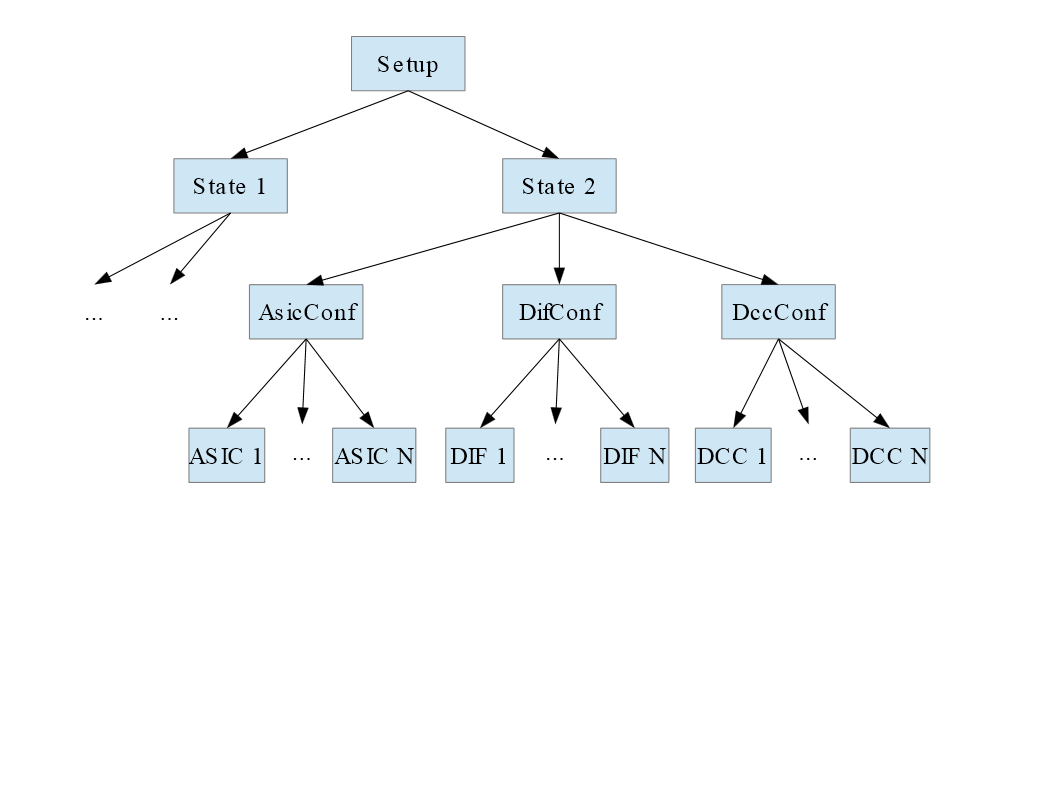}}
\caption{Structure of the data from the setup to the basic objects parameters.}
\label{DBSetupStructure}
\end{figure}

The C++ library gives a high level access to the database, providing
the possibility to manipulate the concepts of ASIC, DIF or Configuration
without any SQL knowledge. It allows to create/modify/upload or download
a complete setup defined as follows:

\begin{enumerate}
\item  Setup: a set of \textquotedbl{}states\textquotedbl{} (one per sub-detector).
\item State: a coherent set of \textquotedbl{}configurations\textquotedbl{}.
\item Configuration: a list of basic objects. 
\item Basic object: a unique object
with all its parameters (ASIC, DIF, DCC, \ldots{}) . 
\end{enumerate}
Figure~\ref{DBSetupStructure}  illustrates the structure of such a setup.

\begin{figure}[h]
\centerline{\includegraphics[width=0.9\columnwidth]{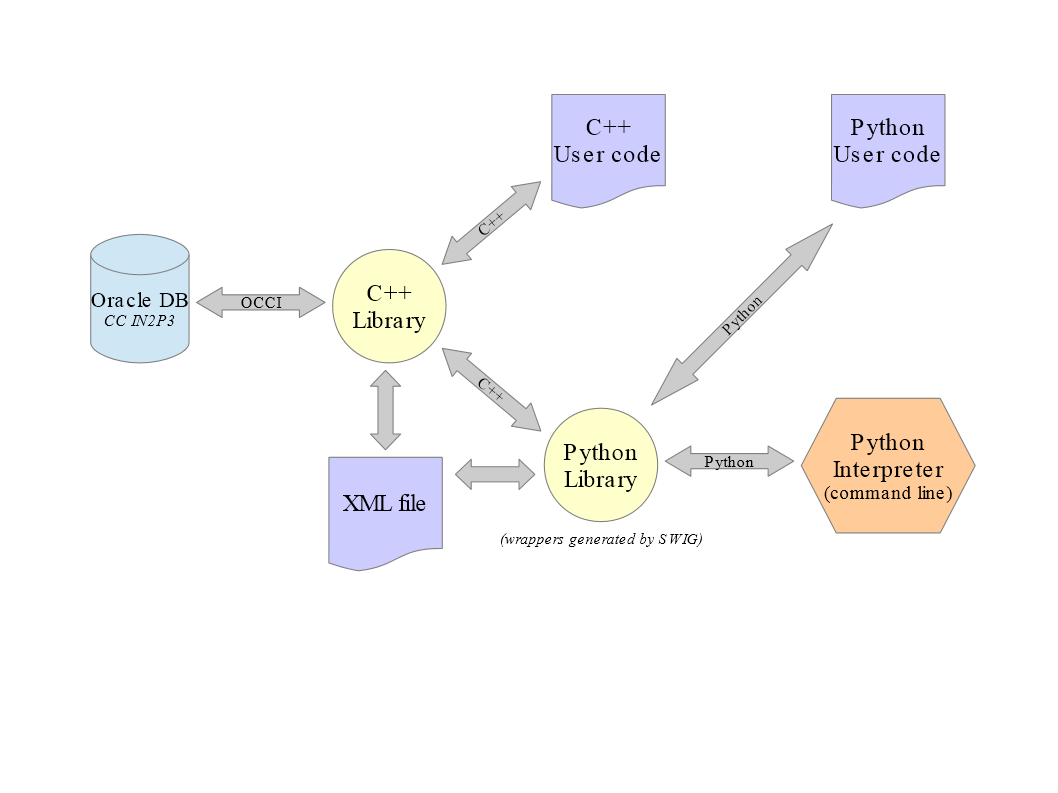}}
\caption{Scheme of the database software.}
\label{DBSoftWare}
\end{figure}

In the case of the  SDHCAL prototype with 50 active layers, a typical setup contains a single state with 150 DIFs,
each having 48 ASICs connected. This corresponds to roughly 550 000
parameters. The download of such a setup takes around 5-7 seconds.
In addition, the library allows to store information about a run
(time, date, type, XDAQ configuration\cite{Xdaq}, setup used, ...). Using Swig,
a Python version of the C++ library is provided. It is in fact a collection
of python wrappers to the C++ functions. It gives the possibility
to access the database from Python scripts and allows the usage of
the python interpreter as a command line tool to manage configurations.
Finally, configuration setups can be imported/exported to/from the
database using XML files. These files can be used to run an acquisition
without access to the database which could be useful when access to database is interrupted for some reasons.
Figure~\ref{DBSoftWare} summarizes the structure of  database access using the different tools.

\subsubsection{Data acquisition access}

In order to minimize database access during data acquisition, a class called
the DIFDBManager (Figure \ref{software_archi}) is responsible of the download of all the DIF and ASIC configuration parameters
of a given setup and to cache it for fast access. Data are stored in
indexed maps that provide an instantaneous access to the data needed
by one specific DIF. Each separate DIFManager instantiates one instance
of DIFDBManager and uses it as a configuration cache.

\subsection{The data collection}
\label{sec:data_collection}

\subsubsection{Local data acquisition and coherence}

 In order to  keep the coherence of collected data, DIF  readout outputs are synchronized using the absolute Bunch Crossing ID identifier defined in section \ref{sec:DIFfeatures}. This is a 48 bits  5 MHz counter that is reset when the first acquisition starts after the DIFs are powered on.
Data from different DIFs may be read out at  different times but will have the same absolute BCID for a given Trigger (or RAMull) signal.
The logical way to keep synchronicity is to store in a BCID indexed
map the buffers of all read DIFs but this requires to manage memory
allocation, access and cleaning. Recent Linux kernels offer the possibility 
to use shared memory based file, ie, /dev/shm. This special file system
is directly mapped in the system memory and data can be written, listed
and read with extremely fast access. Each DIF data block is written
as a single file named Event\_BCID\_DIFID in /dev/shm and an empty file
/dev/shm/closed/Event\_BCID\_DIFID is created once the event file
is closed by the DIFManager. Standard C functions are used to list events available,
to read, write and delete data. This method allows to separate the
process reading data from those making data collection, writing,
debugging and monitoring in a single computer without special protocol
or API to be used.

\subsubsection{Global data acquisition with XDAQ}
\label{sec:DAQxithXDAQ}

\begin{figure}
\centerline{\includegraphics[clip,width=0.85\paperwidth,height=0.4\textheight,keepaspectratio]{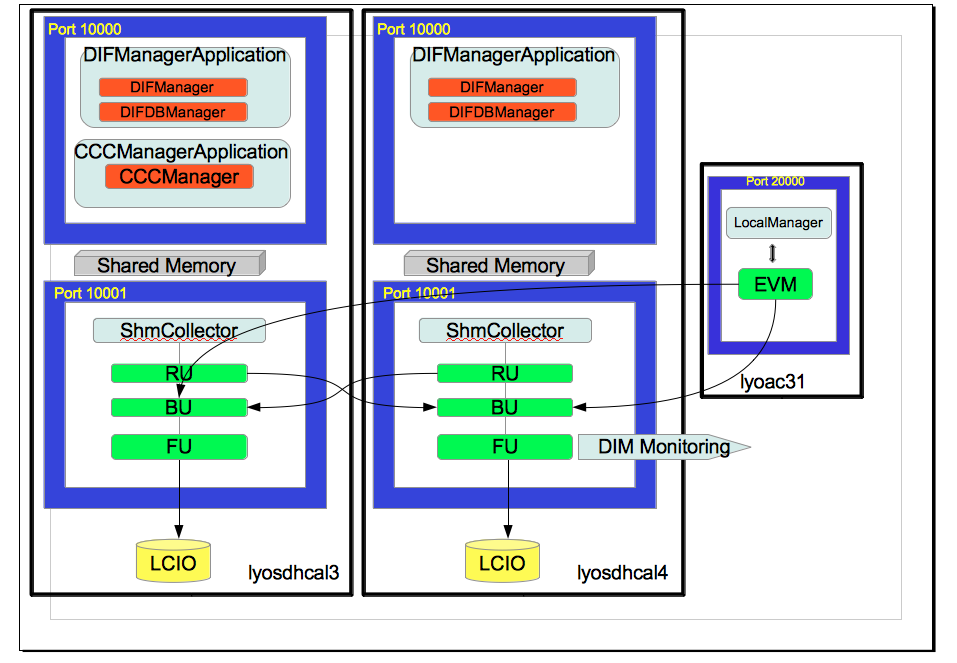}}
\caption{SDHCAL DAQ structure}
\label{SDHCALDAQStructure}
\end{figure}

Whenever several computers are involved in the data taking, a communication
framework is needed. We choose to use the CMS data acquisition XDAQ
framework~\cite{Xdaq}. It provides:
\begin{itemize}
\item A communication with both binary and XML messages.
\item An XML description of the computer and software architecture.
\item A web-server implementation of all data acquisition applications.
\item A scalable Event builder.
\end{itemize}
Each PC handling DIFs hold a DIF manager XDAQ application obeying
to a message driven state machine responsible for initialization (
USB scan (section \ref{sec:DIFManager}) and DB download), configuration (DIF and ASIC settings)
and running (DIF readout and /dev/shm storage). A second application
(ShmCollector) scans the shared memory and pushes completed events
to the local event builder application (Readout Unit). The main advantage
of the CMS event builder is the scalability. It is possible to add
any number of collecting applications (Builder Unit) that will merge
data from all Readout Units for a given trigger. Those Builder Units
distribute merged events to any analysis and data writing application
(Filter Unit) declared to them. In the SDHCAL case, the computing capability
is handled by the PC reading the DIFs, so one Readout Unit, one Builder
Unit and one Filter Unit application are created on each PC as described
on Figure~\ref{SDHCALDAQStructure} and each Filter Unit writes the associated data in a separate stream. 

The whole system is controlled by the Event Manager (EVM). It collects the list of free slots on all Builder Units and generates an associated number of tokens
sent to the LocalManager. This last application is the trigger master of the system. As long as it has free tokens, it sends software trigger to the Event Manager that 
 distributes them to one of the free Builder Units. Once an event is built, a new token is sent to the LocalManager. Any trigger strategy (pause for calibration, software back pressure...) has to
be implemented in the LocalManager.   

\subsection{Run Control}
The run control developed in CMS could have been used here but its deployment for small 
Beam Test experiments is heavy. Alternatively a set of python script was developed 
to create XML description of application content of each computer 
(DIFManager, CCCManager (controls SDCC), Database). 
Then a second set of python scripts uses the web capabilities of the XDAQ 
application and sends SOAP or XMLRPC messages to each server to trigger the 
final state machine of the applications. 
The global state machine (Initialise, Configure, Enable, Stop, Halt, Destroy) 
is coded in these python scripts. Eventually a Qt based graphical user interface 
is provided, using PyQt \cite{PyQt} framework to bind python methods 
to graphic widgets. 

\subsection{Online Monitoring}

The online monitoring is a two steps process:
\begin{itemize}
\item A XDAQ application LCIOAnalyzer runs in a standalone mode, waiting for events to 
be sent by each of the Filter Unit in a sampling mode ($\simeq$ 1\%). Events 
collected are stored and a separated thread analyzes the events using the ILC 
analysis program MARLIN \cite{Gaede:2006pj} allowing any developer to deploy 
its own analysis code. The ROOT\cite{ROOTsys} histograms are booked in memory by a 
singleton class that keeps a map of all histograms booked by any of 
the analysis modules.

\item The LCIOanalyzer answers SOAP commands to
return the list of histograms names and an XML encoding of the requested histograms. 
Clients, using XMLRPC or SOAP request in python, can 
access the XML answers remotely and using PyROOT \cite{PyROOT} can recover 
requested ROOT histograms and display them. Once again a PyQt graphical interface is 
provided to ease the user access. 
\end{itemize}
This distribution of ROOT histograms is useful since users are able 
to manipulate the histograms (zoom, fit...) and not only their image.  This distribution  
is limited however to a relatively small number of clients ( $\simeq$ 10).

\subsection{Acquisition system performance}

The readout mode we used in Test Beam configuration is the ILC mode. The ASIC acquisition is 
started via a command from the SDCC, each chip acquires a 20 bytes frame 
synchronized with the 5 MHz clock whenever one of the 64 channels fired a 
threshold.
Once an ASIC reaches 127 frames buffered, it triggers a RAMfull signal to the SDCC 
which sends back a StopAcquisition to all ASICs and a StartReadout command to 
empty their buffer and send data on the USB bus. Once all ASICs are read a new 
StartAcquisition is triggered for the next event. In a pure ILC mode (1 ms data 
taking , 200 ms inter crossing) the StopAcquisition is forced at the end of the 
crossing.

With an optimal noise of 1 Hz/pad, each ASIC is able to acquire frame during 2~s. 
The average size of data read by a DIF is then 
$ 48 (ASIC) \times 127 (frame) \times 20 (frame size) \simeq 120 $ kbytes. 
Those 60 kbytes/s are read using a serial line at 5 MHz on the ASU and 
a 10 Mbit/s USB1 bus. So we expect in this best case a dead time of 180 ms 
due to the ASU readout and 90 ms due to the USB one. The slow control commands 
handling by the DIF and SDCC is limited to less than 20~$\mu$s and can be 
neglected. That leads to an average availability of 88~\%, that can be increased 
to 92~\% if an USB2 readout is used.

Unfortunately this readout mode is driven by the most noisy ASIC in the SDHCAL. 
Local variation of the gap, of the painting resistivity or of the temperature may 
lead to hot spots where the noise can reach hundreds of Hz.  A few ASICs
($ \simeq $ 2 \%) located on the edge of very few  chambers reach up to 4 kHz. The reason behind such behavior was found to be  a slightly reduced frame height leading to a locally increased electric field.
Still, with such high noisy spot the readout rate could reach 30 Hz  and the availability decreases to 8 \%. Masking the most noisy ones and using USB2 protocol  a data taking availability of 20 \% was achieved during Beam Tests.

One should notice that in pure ILC mode (1ms crossing, 200 ms off), the tiny 
acquisition time guaranties full availability of the detector despite those local 
high-noise ASICs.

\subsection{Recent developments}

One critical point in the previous architecture is the number of USB devices that 
are to be read simultaneously. Moreover, since USB bus was originally foreseen 
as a debug channel the chip only support USB1 (10 Mbit/s) protocol. 
Reading  150 USB devices on a single PC is then extremely slow and we 
split it on 4 different PCs with up to 7 buses (42 devices) read in parallel.

With new pocket size, ARM based PC, availability like the rapsberry Pi 
\cite{raspberry}, we were able to bring low cost (30 \$) and low consumption 
(< 3 W) computers and buses near the detector. We developed a homemade USB hub 
plugged on the raspberry Pi with 12 USB channels available on one  bus and connect 
it to 4 chambers. Thirteen raspberrys are used to read the SDHCAL. Since XDAQ 
software is not yet available on ARM architecture, the central DAQ is rewritten 
using the lighter DIM \cite{Gaspar:1993ji} acquisition middleware.

In parallel the USB1 chips on the DIF was replaced with a USB2 equivalent chip.  
New buses speed and lower number of devices on each bus improves performance by 50 \%.

\section{The SDHCAL prototype mechanical structure} 
\label{mechanicalstructure} 
\subsection{The Design}
\begin{figure}[h]
\centerline{\includegraphics[width=0.41\textwidth]{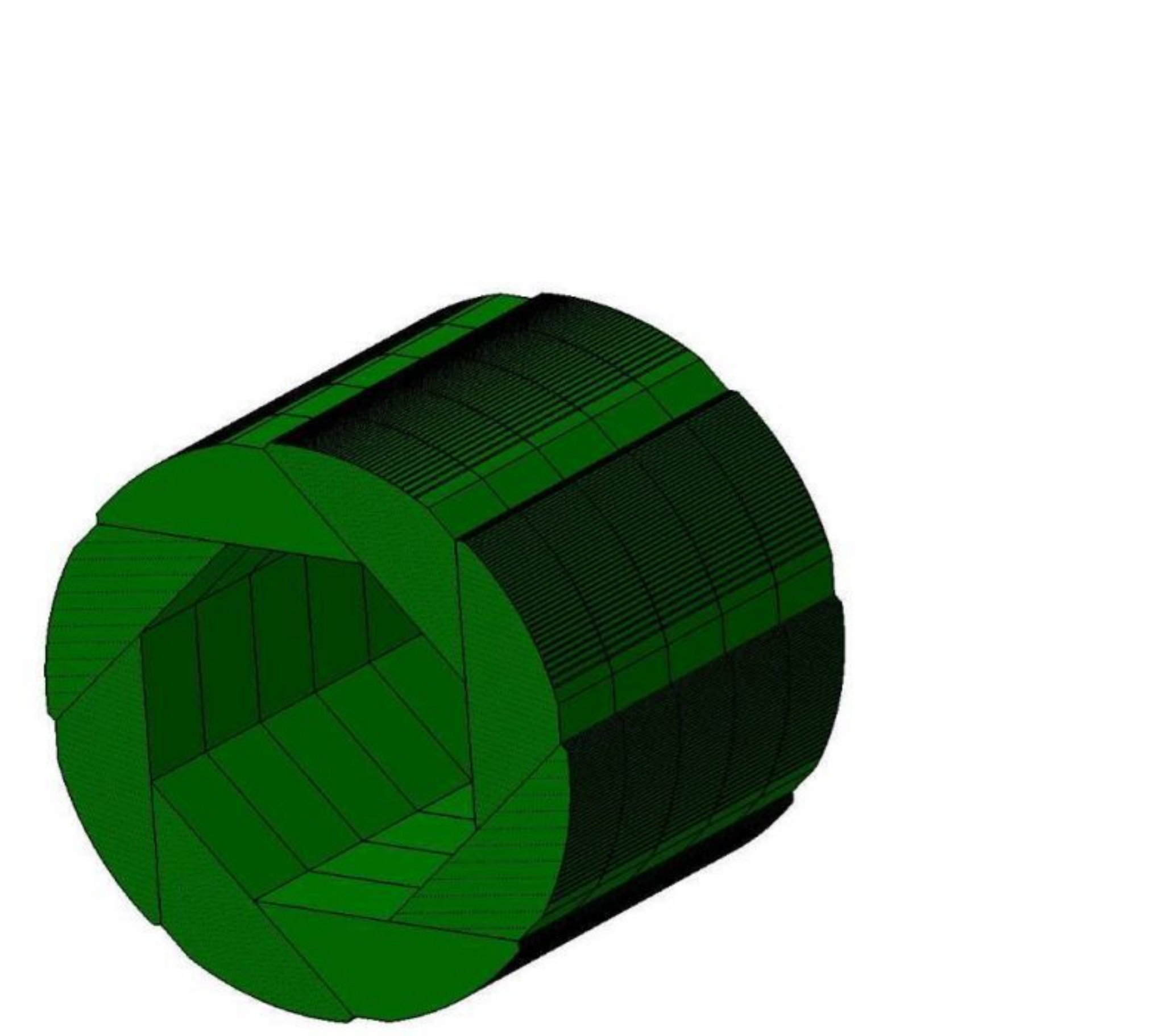}\hfill
\includegraphics[width=0.50\textwidth]{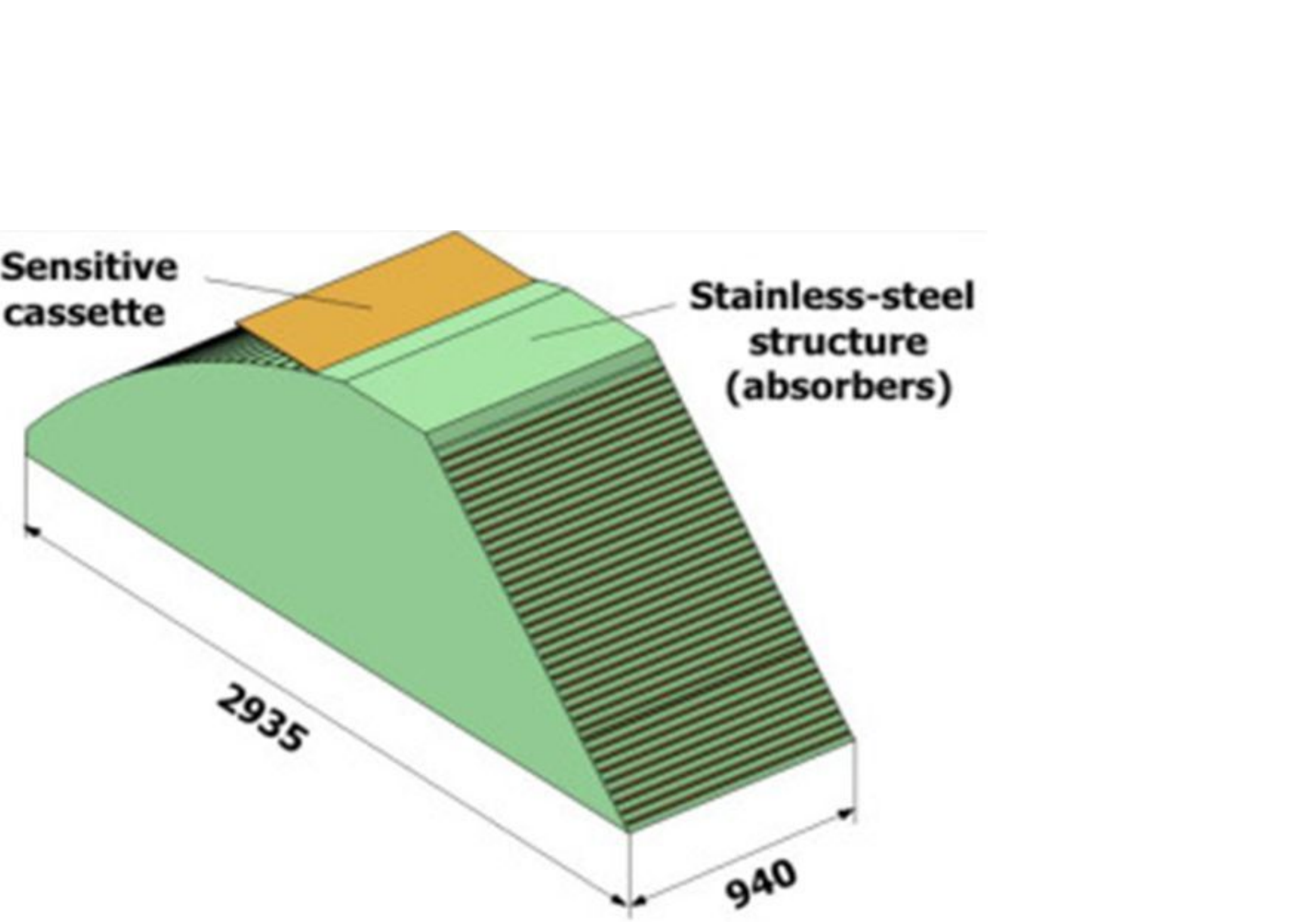} }
\caption{Design layout of the ILD Barrel HCAL (left) and one single module (right). Dimensions are in mm.}   
\label{fig:ILDDesign} 
\end{figure} 
 The mechanical structure of the SDHCAL prototype was conceived as a demonstrator of the one we propose for the International Large Detector (ILD)~\cite{ILDDBD}.
Figure~\ref{fig:ILDDesign} left shows the design layout for the barrel part of the ILD SDHCAL. The design is self supporting and has been 
optimized to reduce cracks. The barrel consists of 5 wheels each made of 8 identical 
modules. Each module is made of 48 stainless steel absorber plates interleaved with 
detector cassettes of different sizes as shown in Figure  \ref{fig:ILDDesign} right. 
A simpler geometry has been adopted for the 
prototype: a cubic design with all plates and detectors having 
the same dimensions of $\sim 1\times 1\unit{m^2}$. 
\begin{figure}[h]
\includegraphics[width=0.42\textwidth]{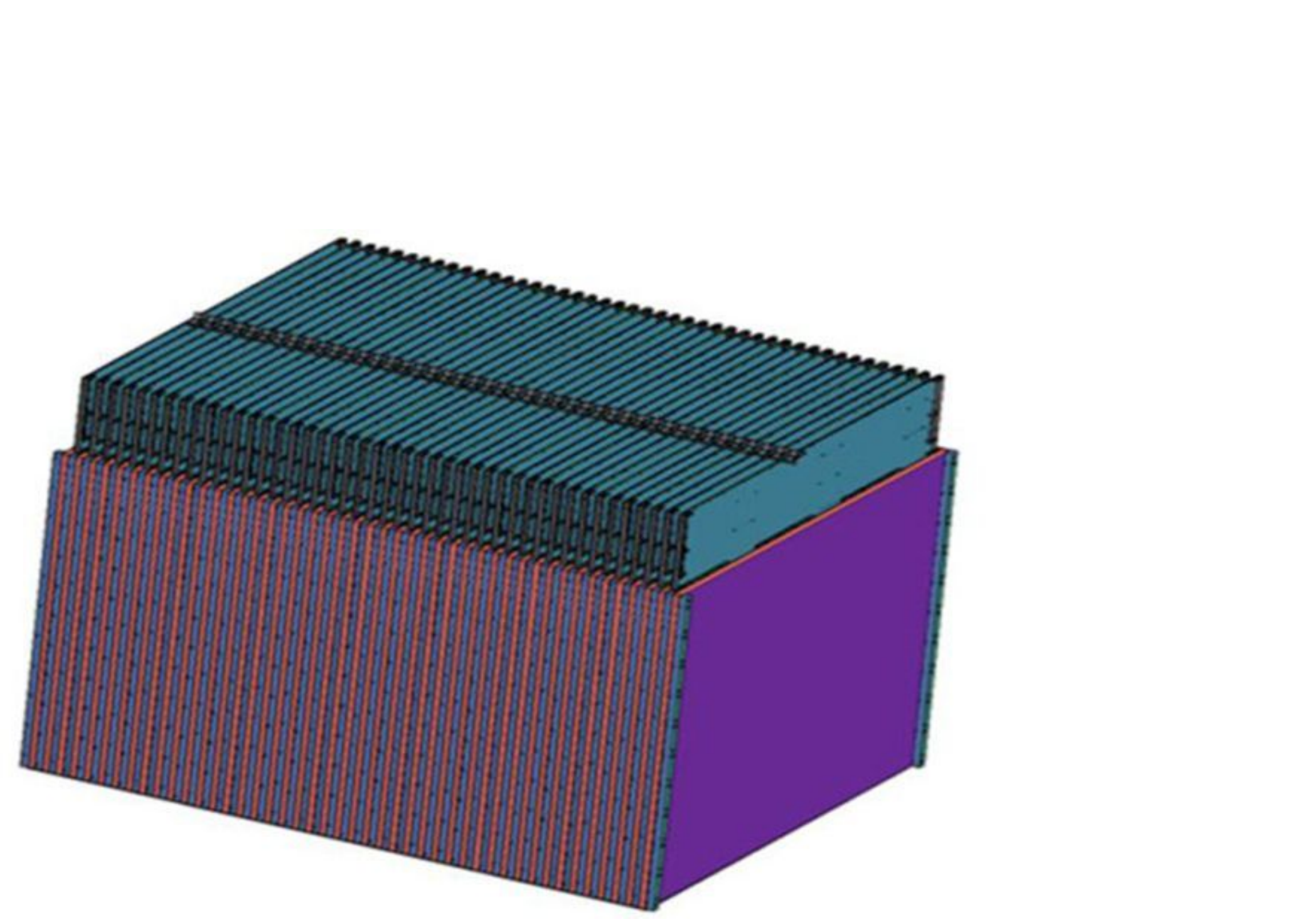} 
\includegraphics[width=0.54\textwidth]{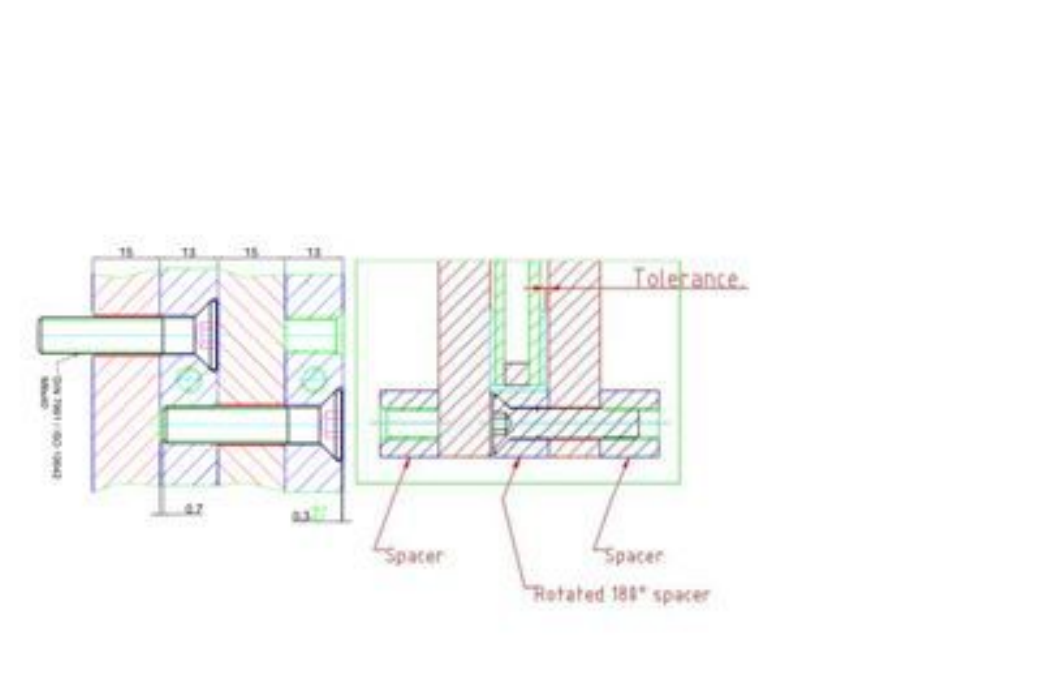}
\caption{$1\unit{m^3}$ SDHCAL prototype design (left).
Details of lateral (center) and top (right) corners of the design. It shows 
some assembly details including the plates, spacers and bolts.}
\label{fig:1m3DesignAndPlanDetails} 
\end{figure}

Figure~\ref{fig:1m3DesignAndPlanDetails} (left) shows the prototype design. It consists 
in a mechanical structure, made of the absorber plates, that hosts the detector 
cassettes. The design allows an easy insertion and further extraction of the cassettes. 

The mechanical structure is made of 51 stainless steel plates assembled together 
using lateral spacers fixed to the absorbers through staggered bolts as can be 
seen in Figure~\ref{fig:1m3DesignAndPlanDetails} (center and right). The dead spaces 
have been minimized as much as possible taking into account the mechanical tolerances 
(lateral dimensions and planarity) of absorbers and cassettes, to ensure a safe 
insertion/extraction of the cassette. 
The plate dimensions are $1010\times1054 \times 15\unit{mm^3}$. The thickness tolerance 
is $0.05\unit{mm}$ and a surface planarity below $\sim$500 microns was required. The 
spacers are $13\unit{mm}$ thick with $0.05\unit{mm}$ accuracy. The excellent accuracy of plate 
planarity and spacer thickness allowed reducing the tolerances needed for the safe 
insertion of the detectors. This is important to minimize the dead spaces and reduce 
the longitudinal size in view of a future real detector.  

\begin{figure}[h]
\includegraphics[width=0.40\textwidth]{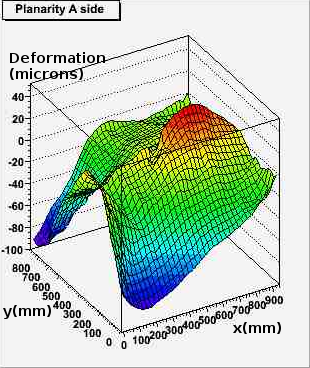}\hfill
\includegraphics[width=0.55\textwidth]{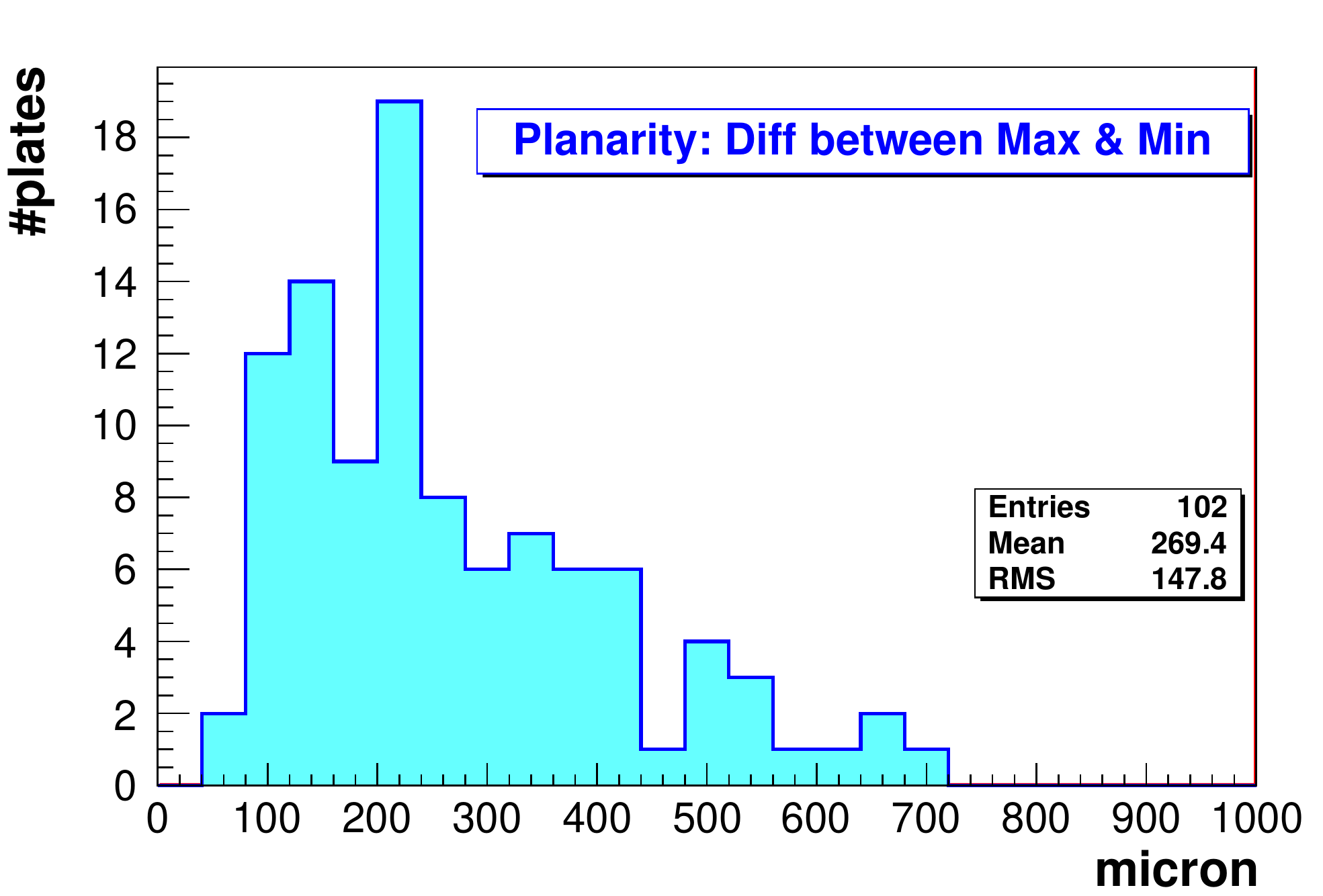} 
\caption{ Left: Planarity distribution for one side of one of the plates. 
Right: Maximum planarity differences for all (51) plates (measurements from 
both faces).}
\label{fig:Planarity} 
\end{figure}

The thickness and flatness of the plates used for the prototype have been verified 
using a laser interferometer system in order to certify they were inside the 
tolerances. Figure~\ref{fig:Planarity} (left) shows, as an example, the planarity 
distribution for one face of one of the plates. For this particular plate the 
maximum deviation from planarity is lower than $150\unit{\mu m}$. For most plates the 
maximum does not exceed the required $500\unit{\mu m}$. Figure  \ref{fig:Planarity} (right) 
shows the maximum deviations from planarity for all plates.

\begin{figure}[h]
\includegraphics[width=0.44\textwidth]{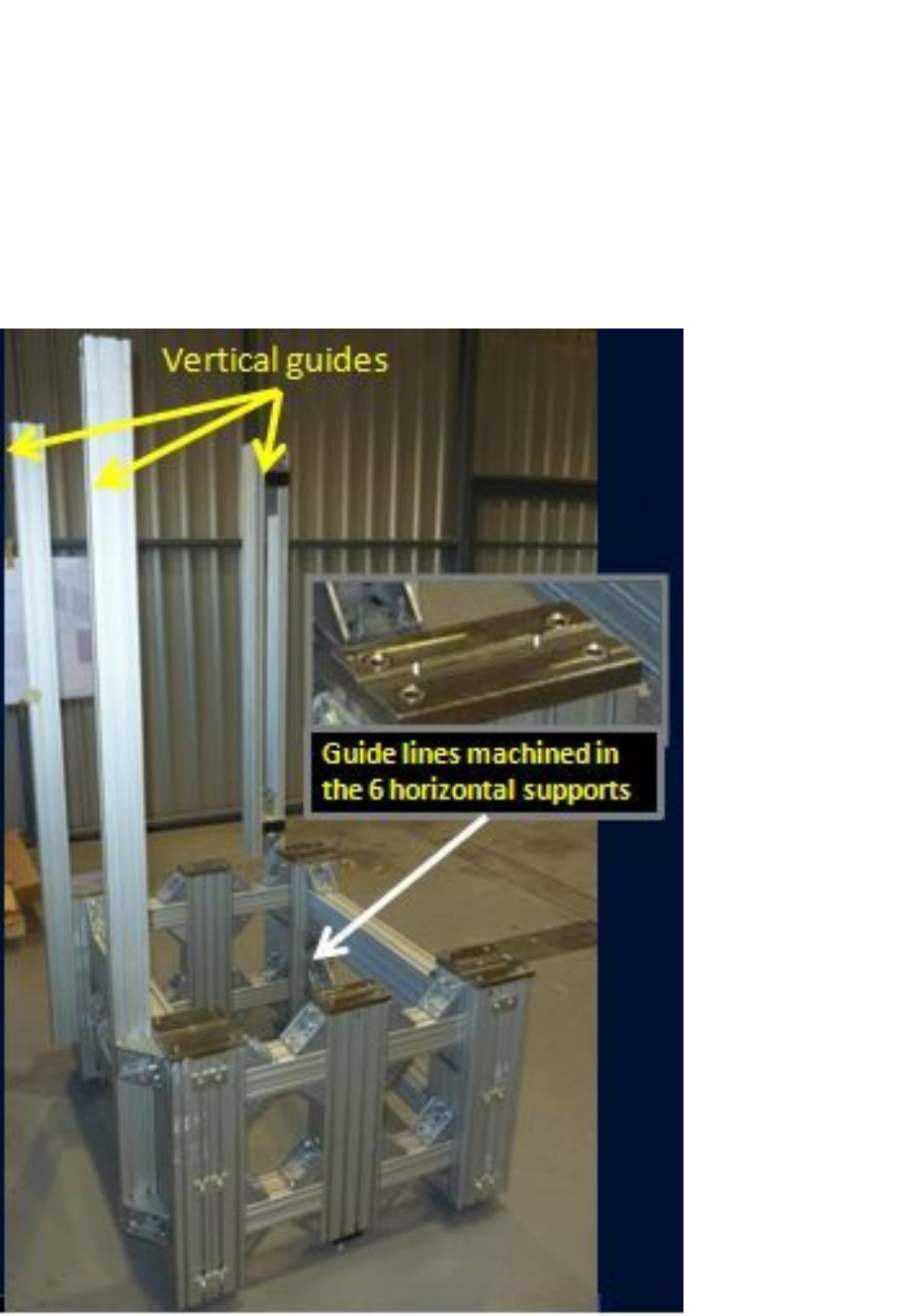}\hfill
\includegraphics[width=0.52\textwidth]{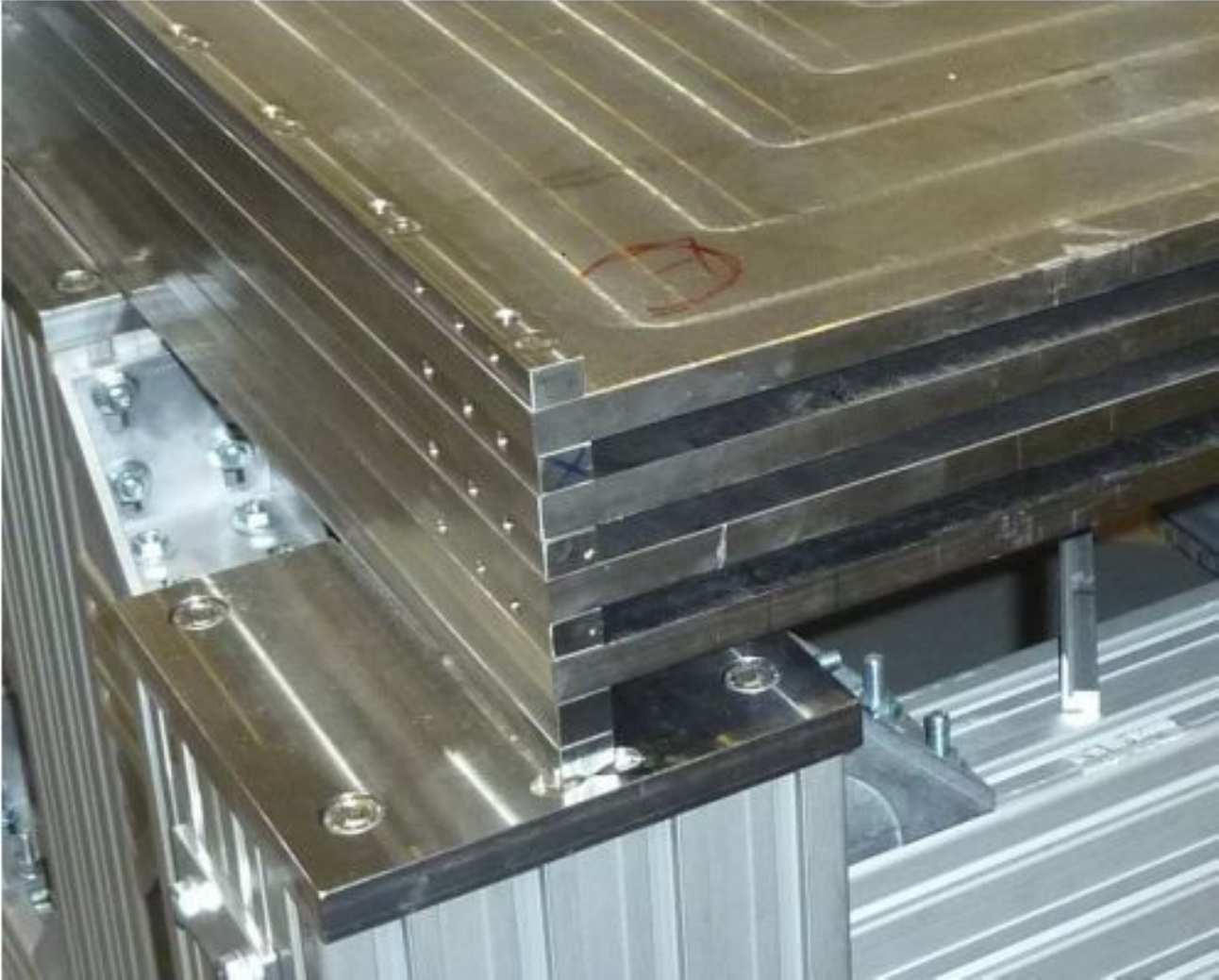} 
\caption{Left: Table for the mechanical structure assembly. It contains horizontal 
and vertical guide lines.
Right: Detail of one corner during the assembly.}
\label{fig:TableAndAssemblyConner}
\end{figure} 

\subsection{Assembly of the mechanical structure}\label{sec:meca}
For the assembly of the mechanical structure a special table has been designed and 
built (Figure~\ref{fig:TableAndAssemblyConner} left). This table must 
support a weight of about 6 tons.  The table has vertical guides attached to the table
and horizontal guide lines machined in 6 supports for the positioning of the first 
spacers. Plates and spacers are piled up and screwed together. 
Figure~\ref{fig:TableAndAssemblyConner} right  shows a detail of one corner of the 
assembly of the first plates.

\begin{figure}[h]
\includegraphics[width=0.27\textwidth]{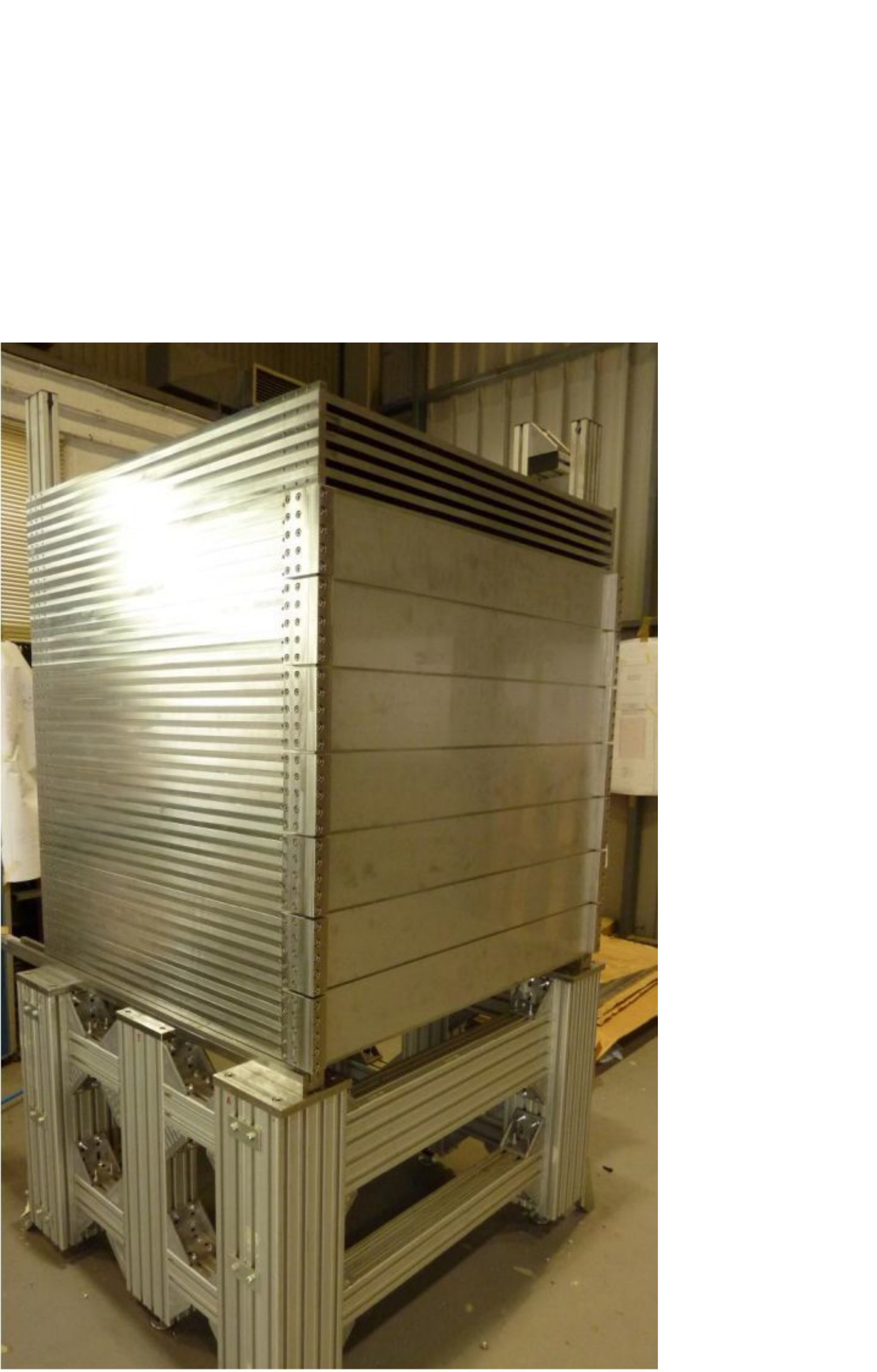} \hfill
\includegraphics[width=0.315\textwidth]{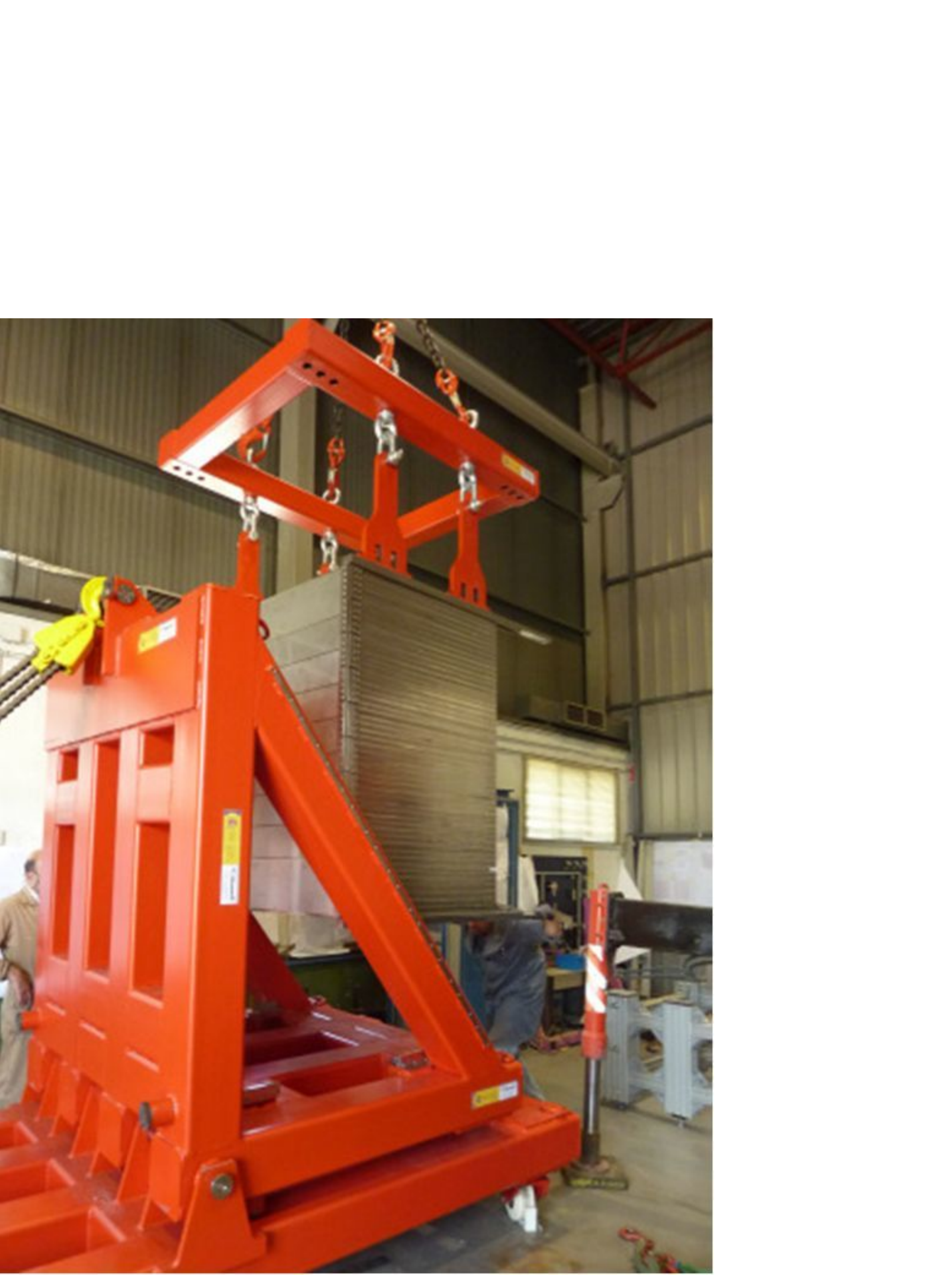} \hfill
\includegraphics[width=0.39\textwidth]{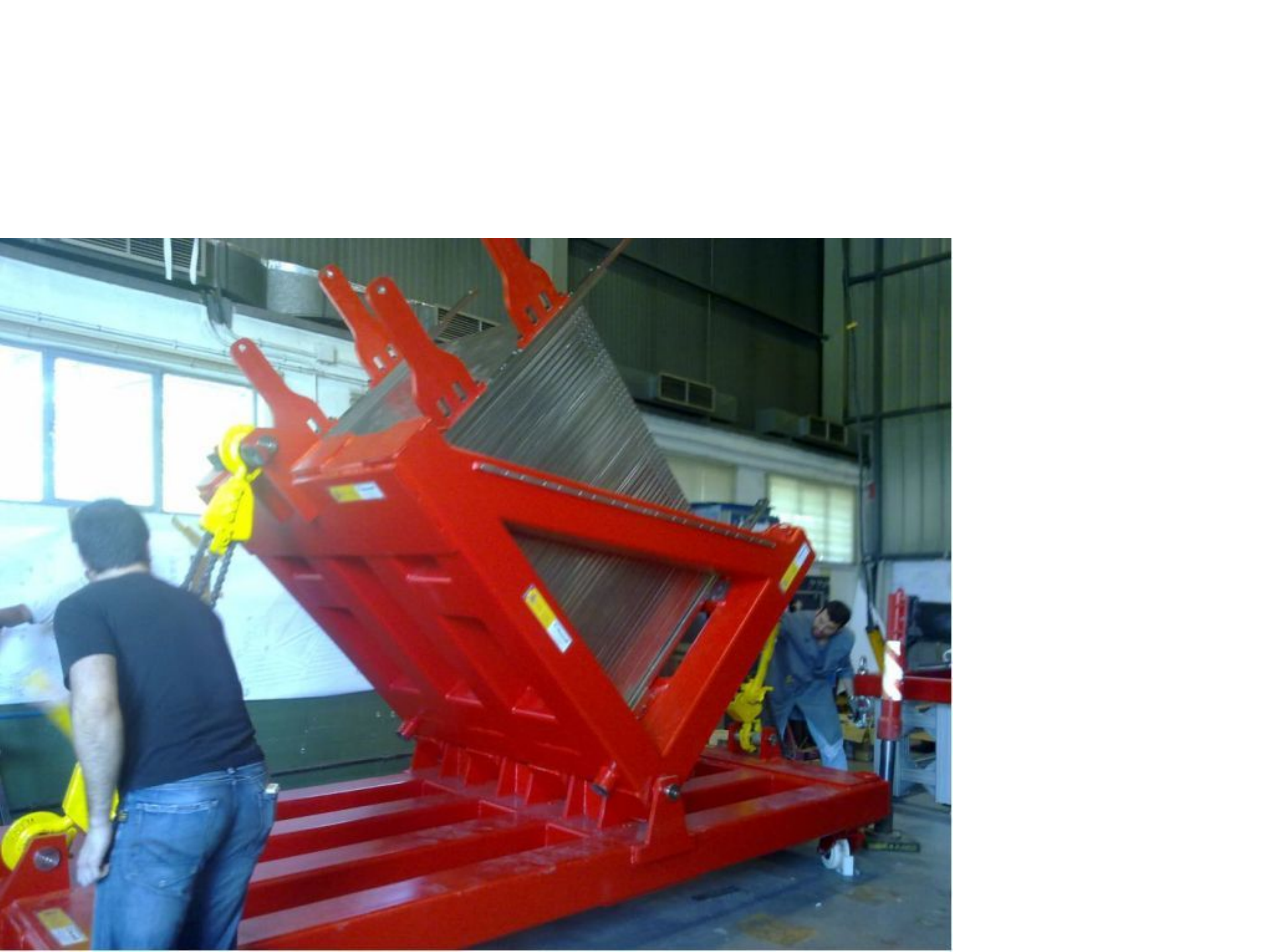} 
\caption{Left: Mechanical structure almost finished.
Center: Mechanical structure being placed on the rotation tool.
Right: Mechanical structure during the rotation from horizontal to vertical position.
}
\label{fig:MechanicalStandRotation}
\end{figure} 

Figure~\ref{fig:MechanicalStandRotation} left shows a picture of the mechanical 
structure almost finished. 
The bottom of the structure is closed with $1090\times 135 \unit{mm}$ stainless steel plates of $3\unit{mm}$ thickness, fixed with bolts to the structure. 
Once the structure was completed it was placed 
(Figure~\ref{fig:MechanicalStandRotation} center) in a rotation tool specially 
designed and it was rotated (Figure~\ref{fig:MechanicalStandRotation} right) to 
vertical position. This rotation tool serves not only to rotate the mechanical 
structure but also the full prototype once it is equipped with the detector 
electronics and cables. This could be useful to change the orientation to test 
it in beam tests (vertical) and cosmic rays (horizontal).

The structure deformation has been checked during the assembly and after rotation 
using a laser interferometer and a 3D articulated arm.   
The total length of the prototype was $+2.2\unit{mm}$ larger than the 
nominal, and the parallelism between the first and the last 
absorber plate was $\sim0.2\unit{mm}$. In fact some extra length was 
expected due to the cumulation of the extra thickness in the 
spacers and plates, because, in order to guarantee the space 
needed between the absorber plates for the GRPC insertion, the 
thickness tolerance of the spacers was $[0\unit{mm},+0.05\unit{mm}]$, 
and, in addition, the thickness of the absorber plates has the 
same tolerance.
The vertical misalignment between plates was below $\sim5\unit{mm}$. 
This misalignment was due to an unexpected error during the 
manipulation of the structure that introduced some deformation, 
but, nevertheless, it does not interfere with the GRPC 
insertion.
During the rotation the induced deformation of the structure was 
below $0.05\unit{mm}$, this guarantees the possibility to rotate the 
structure with the GRPC installed inside without introducing 
forces that could damage the detectors. 

\subsection{The Detector mechanical structure: the cassette}

\begin{figure}[h]
\includegraphics[width=0.45\textwidth]{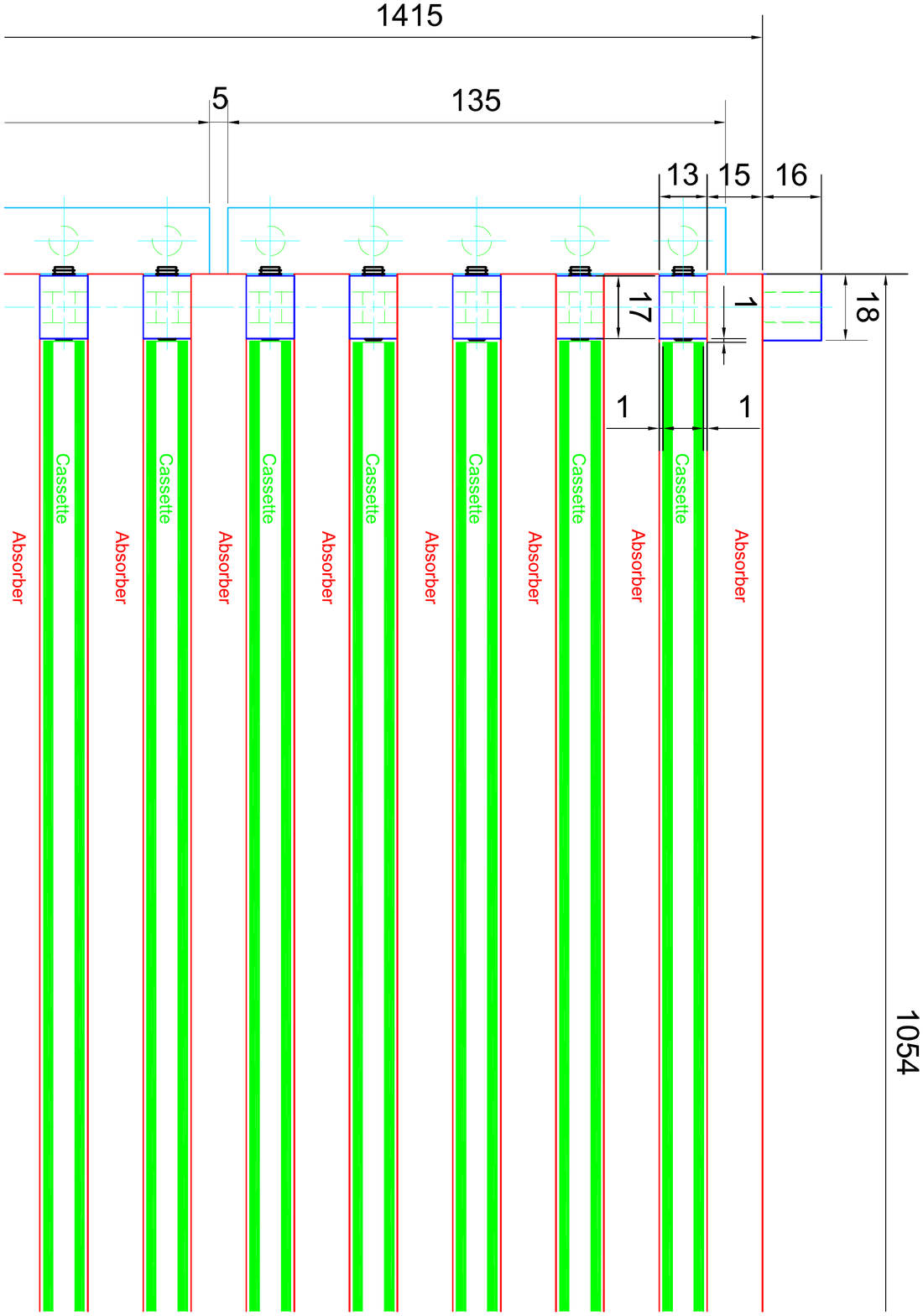}
\includegraphics[width=0.45\textwidth]{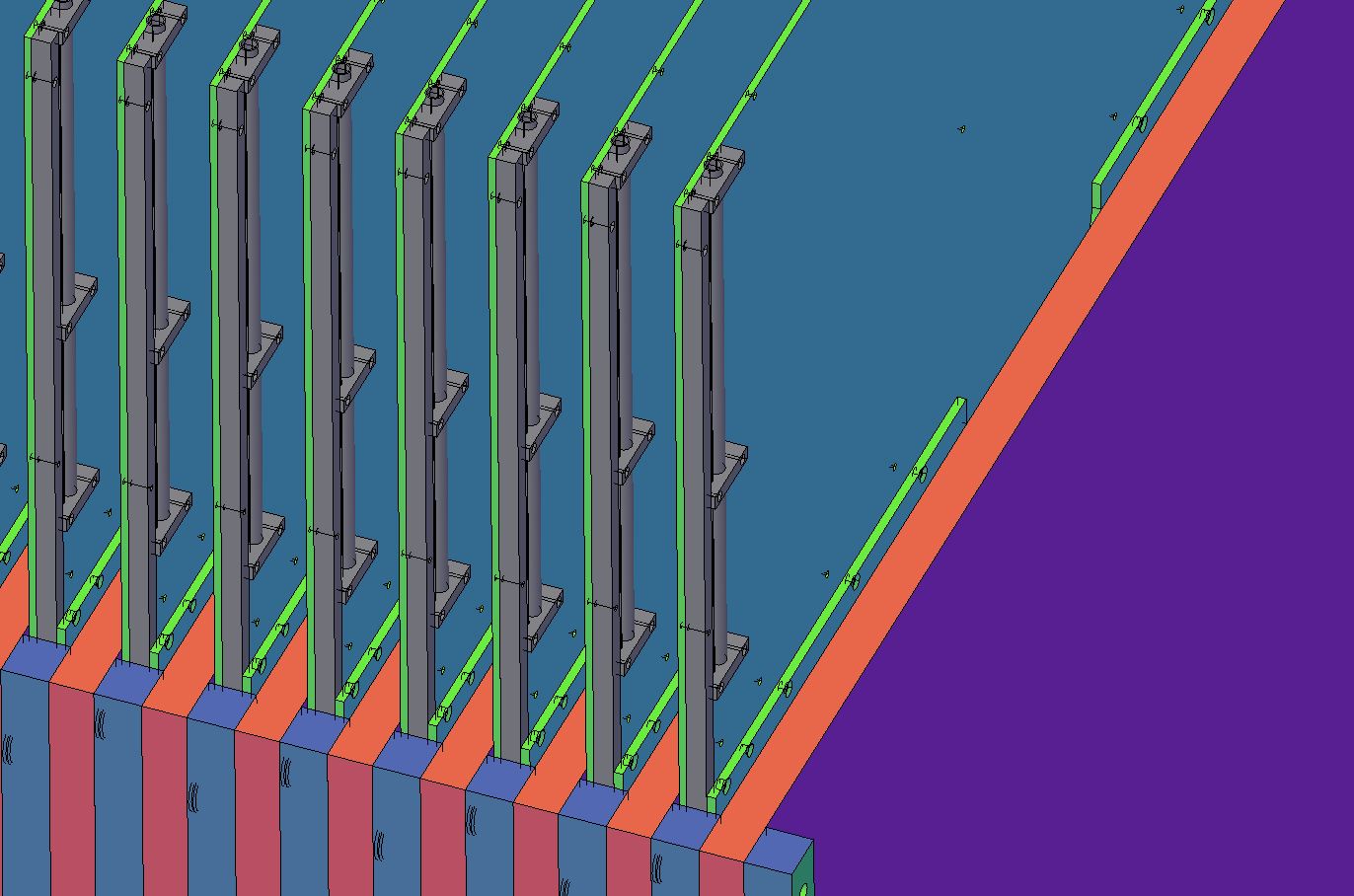}
\caption{Left: A schematic showing the position of the cassettes inside the mechanical structure. Right: An artist view of the cassettes within the mechanical structure.} 
\label{fig:scheme23} 
\end{figure}

\begin{figure}[h]
\includegraphics[width=0.99\textwidth]{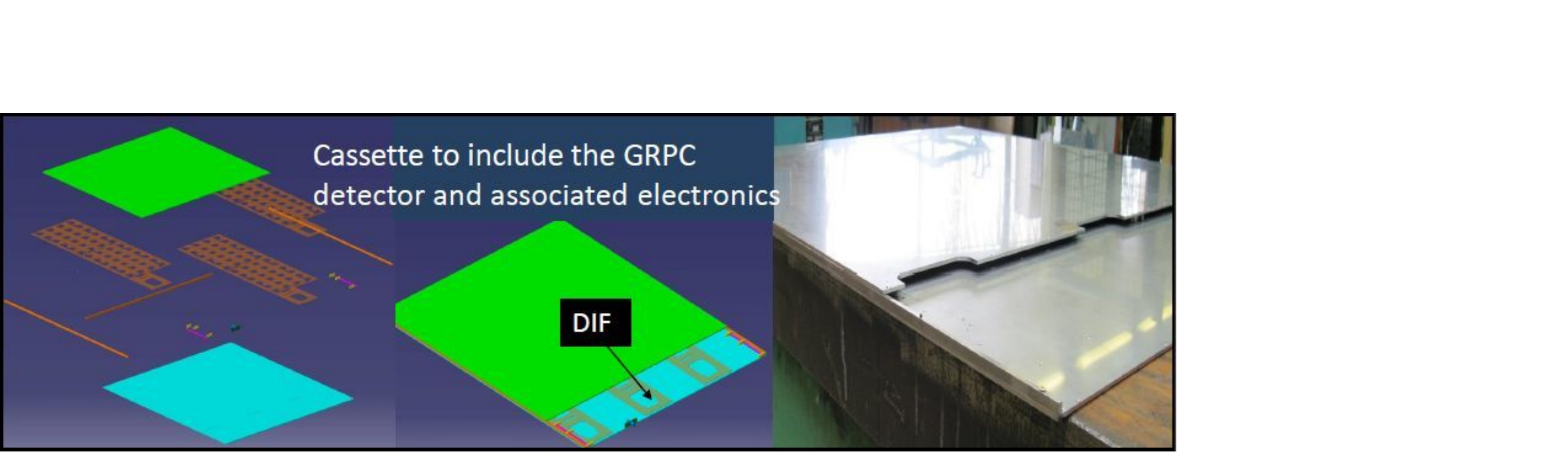}
\caption{Left : exploded view of the different cassette components without the GRPC. Center: cassette external view drawing. 
Right: picture of an empty GRPC cassette.} 
\label{fig:cassette} 
\end{figure}

The GRPC detector together with its associated electronics is hosted into a special 
cassette which protects the chamber, ensures a good contact of the readout board 
with the anode and simplifies the handling of the detector.  The cassette 
(Figure~\ref{fig:cassette}) is a box made of 2 stainless steel plates $2.5\unit{mm}$ thick and 
$6\times 6\unit{mm^2}$ stainless steel spacers machined with high precision closing 
the structure. One of the two plates is $20\unit{cm}$ larger than the other. It allows 
fixing the three DIFs and the detector cables and connectors (HV, LV, signal 
cables). A polycarbonate spacer, cut with a water jet, is used as support of the 
electronics; it fills the gap between the HARDROC ASIC improving the rigidity of 
the detector. A Mylar foil ($175\unit{\mu m}$ thick) isolates the detector from the box. 
The total width is $11\unit{mm}$, 6 of them correspond to the GRPC and electronics, and 
the rest is absorber. Figure~\ref{fig:scheme23} shows a schematic of the cassettes position inside the mechanical structure as well as an artist view of it. 

\subsection{Integration of the GRPC in the structure}

\begin{figure}[h]
\includegraphics[width=0.50\textwidth]{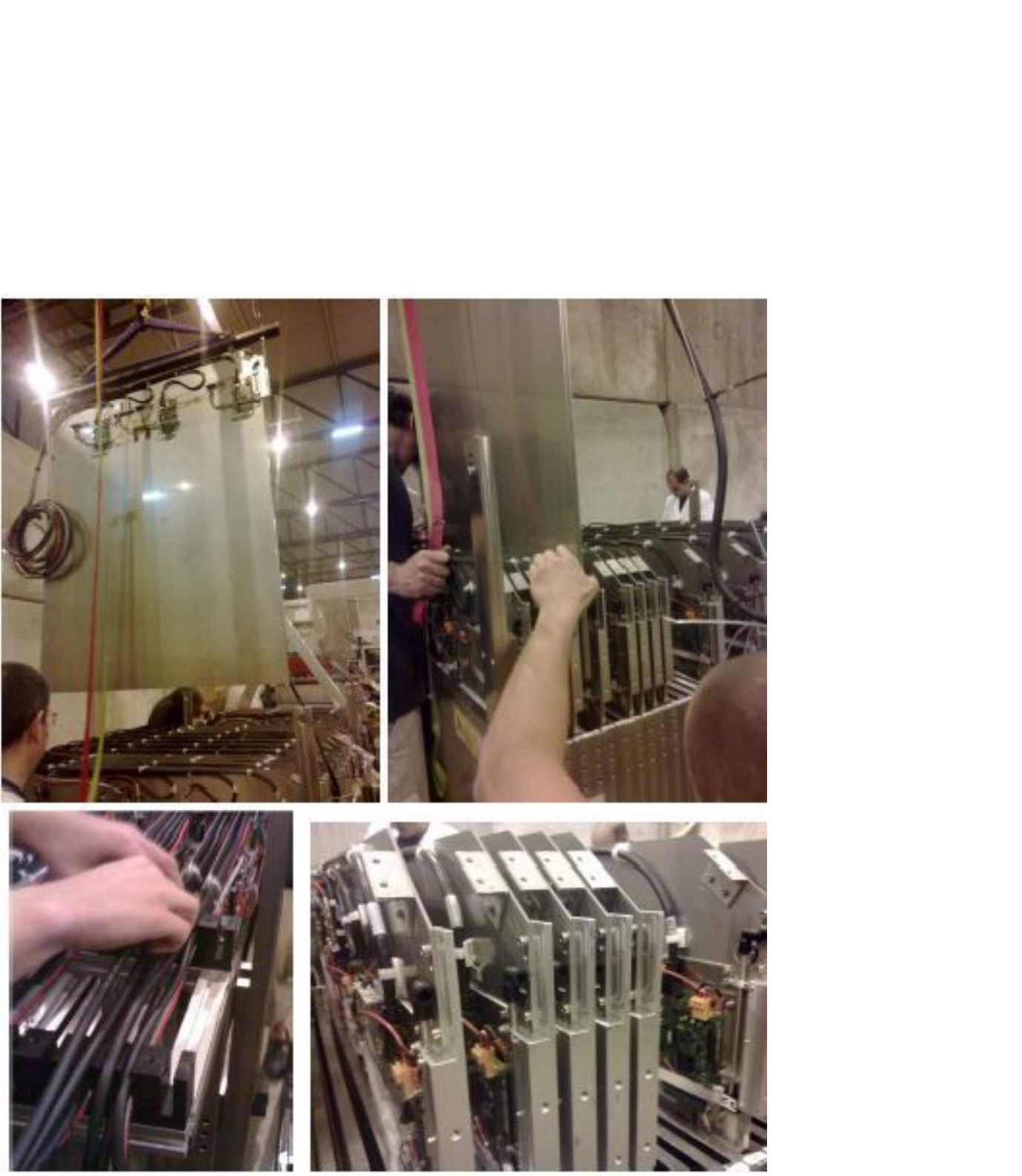}\hfill 
\raisebox{1.5cm}{\includegraphics[width=0.48\textwidth]{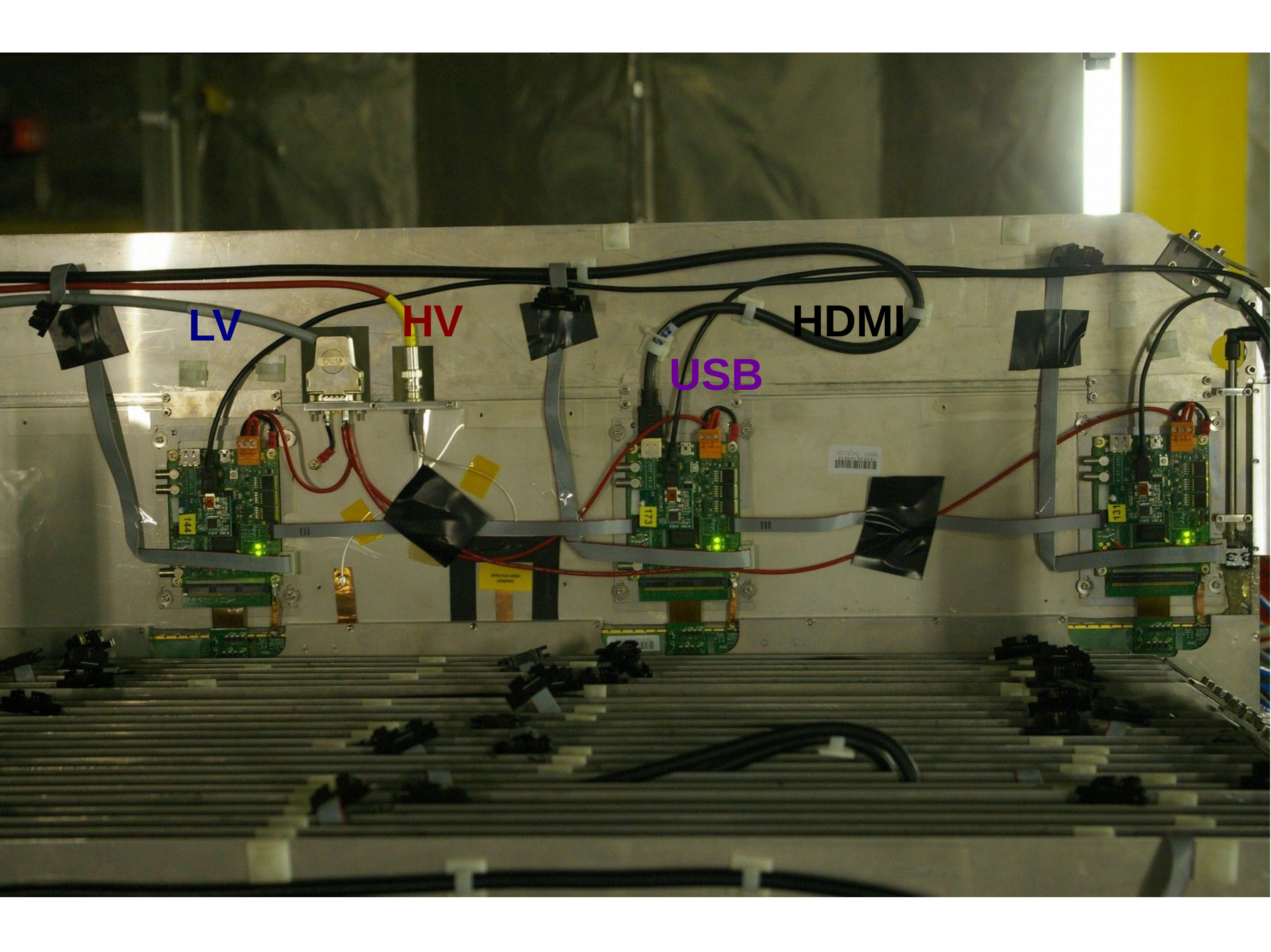}} 
\caption{Left: Several pictures showing the insertion of GRPC inside the mechanical structure and final cabling.
Right: Detail of the cassette cabling.}
\label{fig:FinalAssemblyAndCableServices}
\end{figure}

\begin{figure}[h]
\begin{center}
\includegraphics[width=1.0\textwidth]{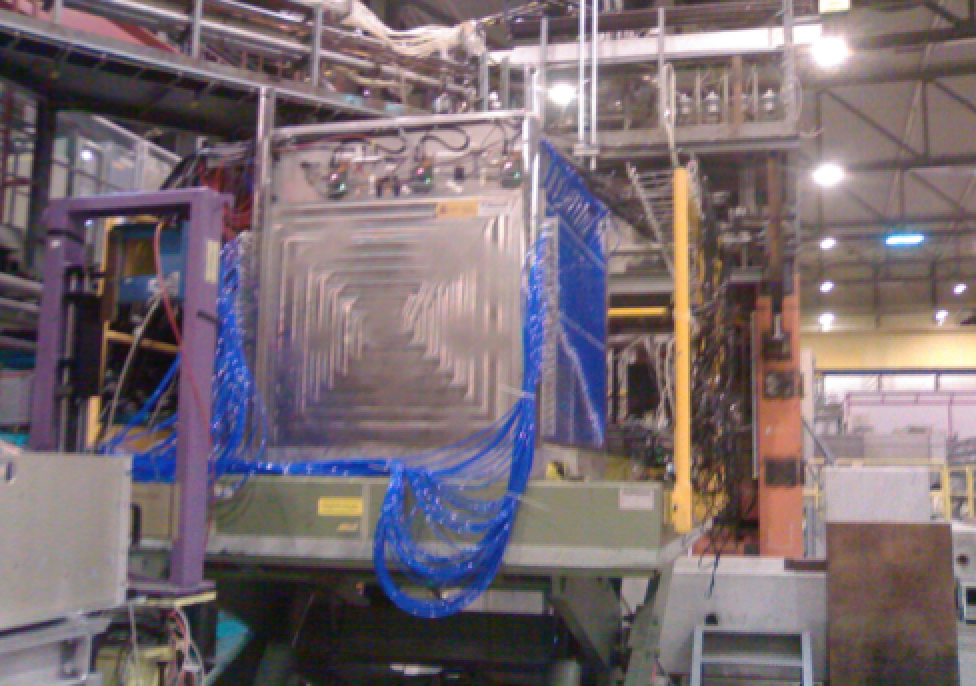} 
\end{center}
\caption{Final prototype at the SPS test beam area.}
\label{fig:PrototypeAtSPS}
\end{figure}

A total of 48 cassettes equipped with detectors and their embedded electronics  have been built and inserted in the mechanical 
structure. The insertion has been made from the top with the help of a small crane 
as it is illustrated in Figure~\ref{fig:FinalAssemblyAndCableServices}. Vertical 
insertion minimizes the deformation of the cassettes making easier the procedure. 
Each cassette is connected to 8 cables.  Three of them correspond to the USB 
readout connections, three others correspond to the HDMI connections, and the others carry the high and low voltage. The gas is 
distributed individually to each GRPC. Figure~\ref{fig:FinalAssemblyAndCableServices}
shows a detail of the external cabling distribution in the cassettes. Figure~\ref{fig:PrototypeAtSPS} shows the final prototype during one of the beam tests.

After the final assembly several cassettes have been extracted for reparation of some electronic connectors, the operation was done always smoothly without problems.

\section{Event building}\label{sec:trivent}
When running in the Triggerless mode  in beam tests (See Section~\ref{sec:DIFfeatures}), 
each ASIC auto-triggers  and stores the  information according to the scheme described in section \ref{sec:ASIC}. In this scenario, the acquisition system reads the full detectors when the memory of at least one ASIC is full. The collected data 
thus include not only the information about the incoming 
particles (pions, muons or cosmic rays ...) but also the intrinsic 
noise of the detector. The average duration of one acquisition 
window was found to be typically of $\sim 30\unit{ms}$. The time of each fired channel (called hit here after) with reference to the start of the acquisition is recorded by the BCID counter (See Section~\ref{sec:DIFfeatures})
increasing by a step of $200 \unit{ns}$.  

\subsection{Preliminary data format}
The raw data are then stored in LCIO format\cite{Aplin2012} by 
the acquisition system. Each acquisition window is stored in a 
preliminary LCIO file as an {LCEvent}. The 
{LCEvent} contains a collection of {LCGenericObject} 
containing the raw DIF data blocks (see Section~\ref{sec:data_collection}). The first data processing
is to convert these data blocks into a collection  of hits including Bunch Crossing ID ({BCID}), the {ChannelID}, {AsicID} and {DifID}. 

\subsection{The algorithm}
The physical event candidates are built from hits collection using a time clustering method. 
First, the time slot (i.e. BCID) with at least $M_{hit}$ hits are selected. 
For each of these selected time slots, hits belonging to the 
adjacent time slots in a window of $\pm~t_{win}$ are combined 
to build a physical event. Care was taken to ensure that no hit 
belongs to two different events. Indeed if after the time clustering procedure, two events are found to have a common time slot, the hits of the common slot are assigned to the first event\footnote{This scenario was never observed during several beam tests at CERN. This is due to the reduced beam intensity used during beam tests given the limited GRPC rate capability.}.
The information related to the 
coordinates of the hits, determined from the location of the 
fired pad and the related active plate are then saved together 
with the threshold reached (either 1, 2 or 3).  

\begin{figure}[!h]
  \centering
  \includegraphics[height=0.5\textwidth]{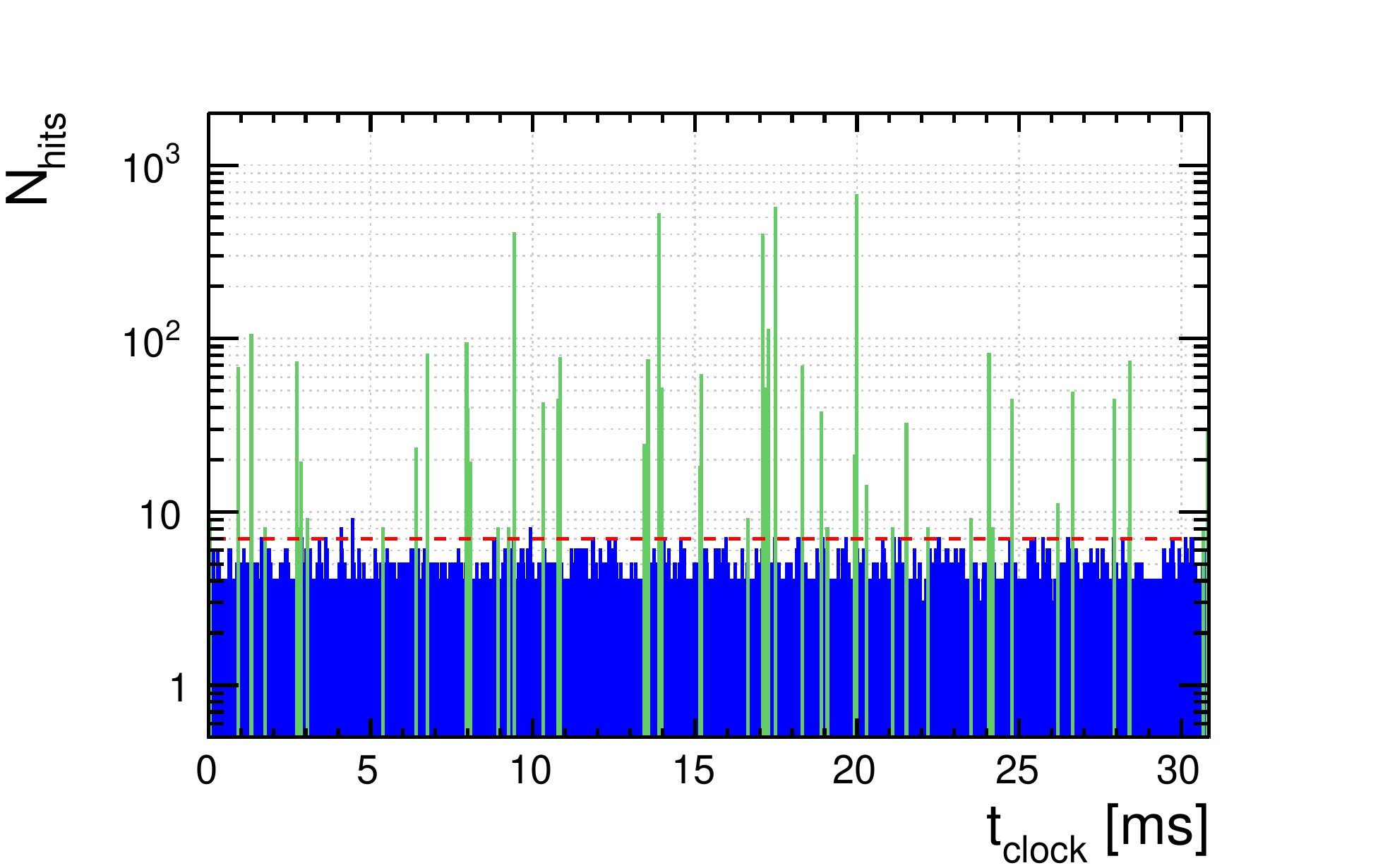}
  \caption{Time spectrum of an acquisition window with a granularity of $200\unit{ns}$. The physical event candidates are highlighted (green) over the background noise (in blue). The red line represents the threshold over which the events are considered.}
  \label{fig:time-spectrum}
\end{figure}

For the SDHCAL prototype the value $M_{hit}=7$ is chosen as minimum of hits required to build an event (Figure \ref{fig:time-spectrum}). This allows the rejection of intrinsic noise while eliminating a negligible fraction of hadronic showers produced by pions of energy larger than $5\unit{GeV}$.
\begin{figure}[!h]
  \centering
  \includegraphics[height=0.5\textwidth]{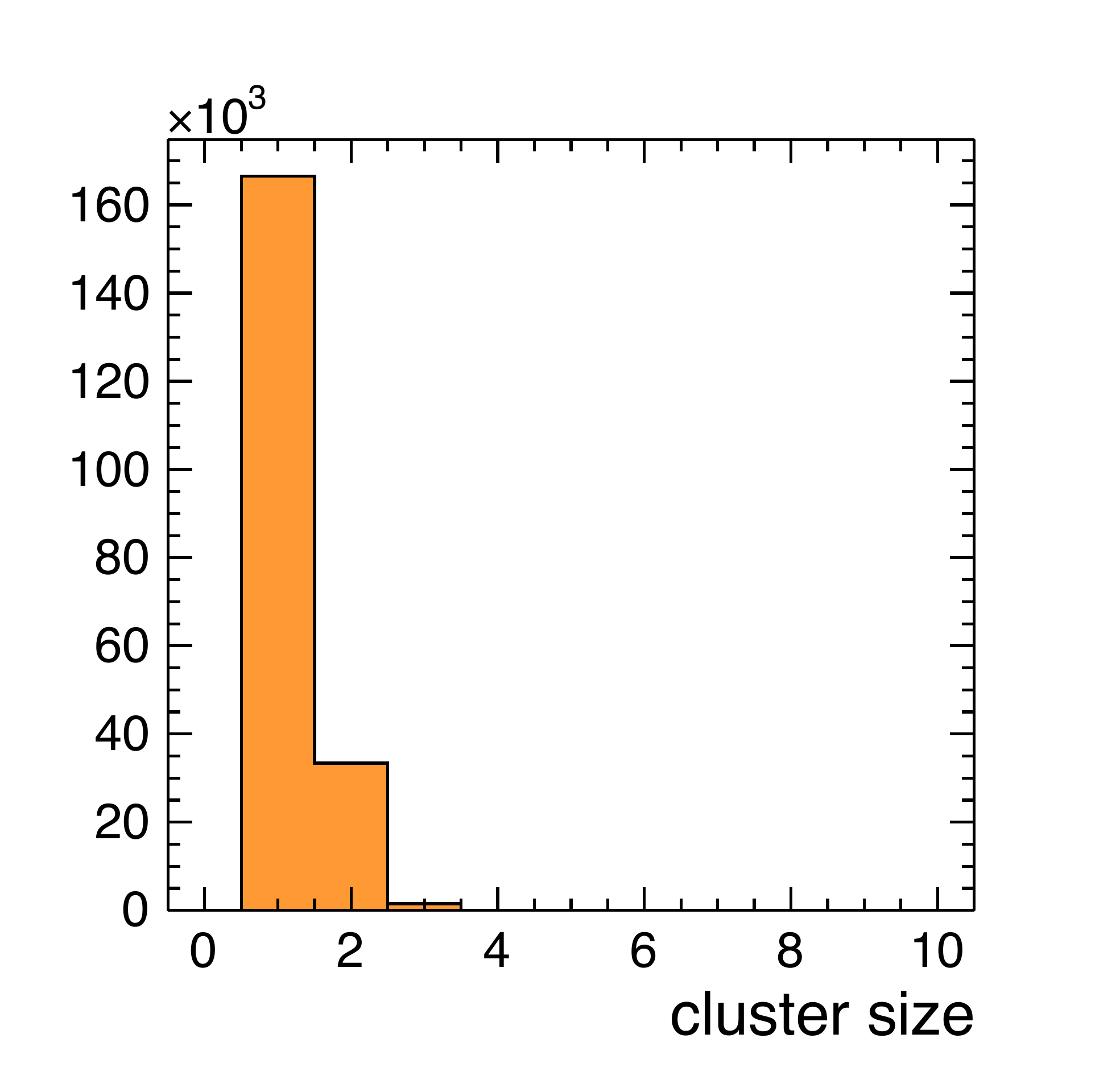}
  \caption{Time cluster size of hadron interaction events measured in number of time slots of $200\unit{ns}$.}
  \label{fig:timepeak}
\end{figure}
The width of the time window can be determined from the time cluster size built around
$t_{\rm peak}$ (time slot with $N_{hit}> M_{hit}$) and including adjacent time slots in which at least one hit is recorded. 
 As shown in Figure \ref{fig:timepeak}, hadron interaction events are contained essentially in two time  slots of $200\unit{ns}$.  In our event building a width of $3\times200\unit{ns}$ was chosen. Extending this width to 4 or 5 time slots was found to be of no consequence on our study.\footnote{The increase of number of hits associated to such events in these cases, was found to be compatible with an increase due to the noise which is less than one hit on average.}

{


\subsection{Geometry building}
The information provided by the acquisition chain about the hit 
contains identification numbers 
({DifID}, {AsicID}, {ChannelID}) and the 
position of the {DifID} in the layer and the detector. 
The relative position of pads in the ASIC, and the position of 
ASICs in the {DIF} are known from the hardware 
configuration. These information combined with the relative 
positions of the {DIF}  and chambers in the calorimeter 
allow the reconstruction of the hit coordinates. 

\subsection{Coherent noise effects and mitigation}

Some of the collected events in the prototype are not related to particles interaction. They  are characterized by the 
recurrence of many hits belonging to the same electronic slab or to the whole electronic layer (made of three slabs), as shown in 
the \fref{fig:coherent-noise}.  The reason behind the occurrence of such events  was identified to be the prototype's grounding quality. The frequency of this coherent noise was indeed reduced  by improving the grounding of the whole detector. This was achieved by connecting the cassettes to each other using metallic connectors adapted to the cassette geometry. In addition, for each active layer of the prototype  the grounding of the electronic board was connected to that of the acquisition board and both of them were then connected to the cassette grounding.
The grounding of the HV module was also connected to that of the LV grounding at the power supplies level. Applying this scheme improved the grounding significantly but did not eliminate all the related problems and events related to coherent noise, although very rare,  are still present. It is clear that a global grounding of the detector is to be necessarily taken into account at  early stages of the design to minimize loops and ensure an effective and reliable grounding of all the system.

\subsection{The \cpp package}
The introduced event-builder was implemented in a C++ framework 
called {Trivent}. 
The core algorithm is implemented as a {Marlin}\cite{Gaede:2006pj} processor. 
It takes as input an LCIO file with a collection of 
{LCGenericObejcts} containing all the information saved 
by the detector, where a {LCEvent} corresponds to a RAMfull. 
A configuration file in xml format containing the relative 
position of {DIF} in the detector is required. 

\begin{figure}[!h]
  \centering
  \includegraphics[height=0.5\textwidth]{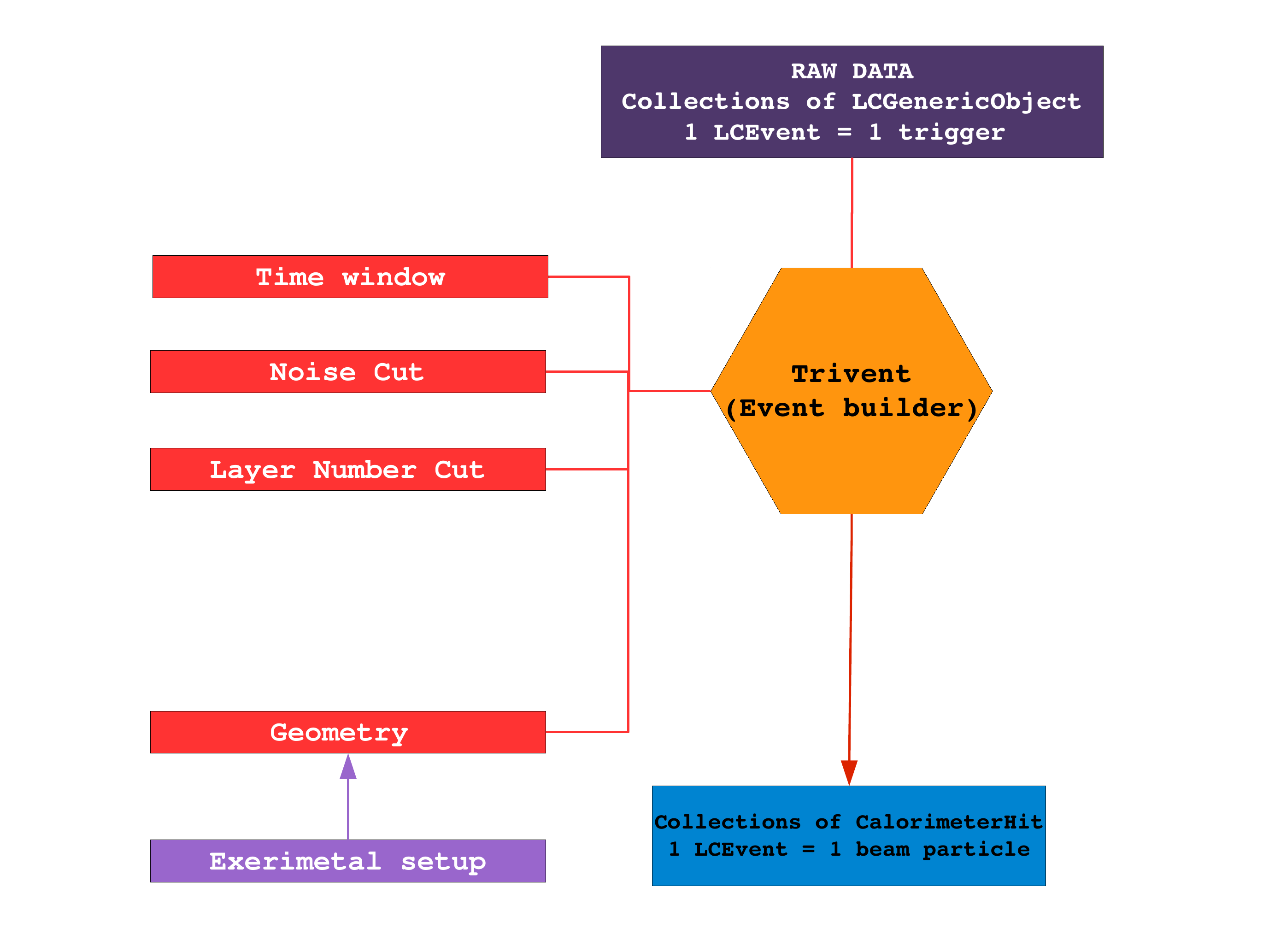}
  \caption{Trivent event builder workflow.}
  \label{fig:trivent}
\end{figure}

The running of the program is done by supplying a steering file 
containing the processor parameters (cut-list, file paths, etc). 
{Trivent} provides, after running, a LCIO file 
containing {CalorimeterHit} collections, in which each 
{LCEvent} corresponds to a physical event candidate. 
The \fref{fig:trivent} summarizes the {Trivent} working flow. 

%
\subsection{Data quality control}
During the data taking of the beam test periods, several 
controls have been done to check the validity of the saved data. 

\subsubsection{Online monitoring}
First, an online monitoring reads the data stream from the 
{DAQ} and makes a fast analysis in order to estimate the 
chip occupancy (chip noise) and the detection efficiency of each 
chamber. The online monitoring constitutes an important step in 
the data taking, since noisy channels can be isolated and masked 
at very-frond-end electronic level during data taking.

\subsubsection{Offline monitoring}
Using the event-builder previously introduced, an offline 
monitoring is performed. It is focused on the measurement of 
noise, and the performance of chambers and their stability over 
the time.  

\subsubsection{Noise estimation}
\begin{figure}[!h]
  \centering
  \includegraphics[height=0.5\textwidth]{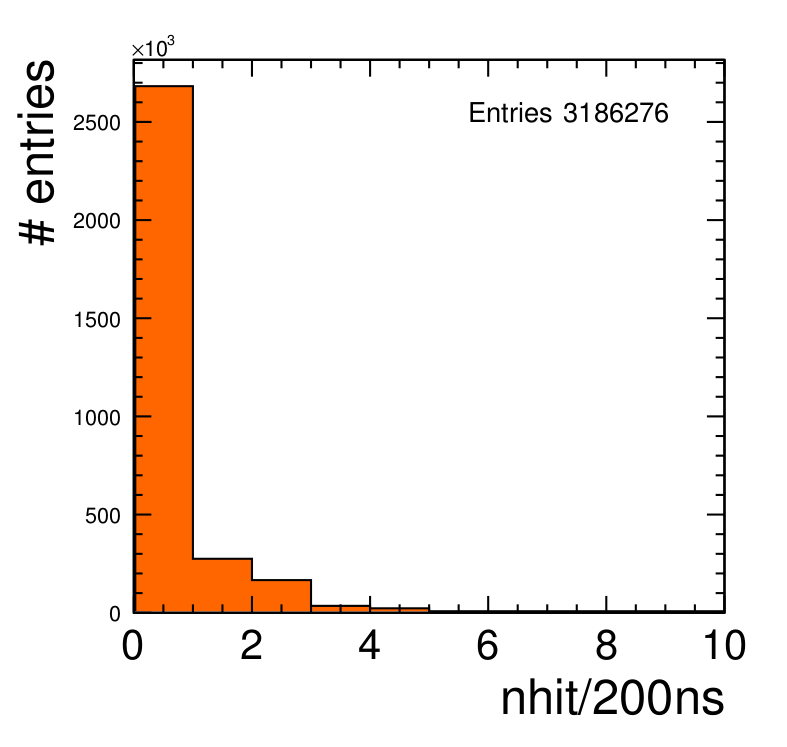}
  \caption{Distribution of number of hits per clock tick due to intrinsic noise of the GRPC sensors.}
  \label{fig:intrinsic-noise}
\end{figure}
Two kinds of noise can be distinguished: intrinsic and coherent 
noises. Intrinsic noise is made of hits not related to a particle-interaction event 
(\fref{fig:time-spectrum}).  It is essentially due to the gain fluctuation in 
GRPCs in some zones like the ones around the spacers.  Its intensity is a function of the temperature and the 
polarisation high voltage values.  Pads whose intrinsic noise frequency  was found to exceed 100 Hz were identified. 
Their number was found to be around few per mille.  

The \fref{fig:intrinsic-noise} shows a typical distribution of this noise in units of the DAQ clock tick with 
an average of $\sim 0.35 \rm hit/200\unit{ns}$. This measurement 
gives an estimation of the contamination of noisy hits along 
physics events and shows that the GRPCs are almost noise-free.

\begin{figure}[!h]
  \centering
  \includegraphics[width=0.4\textwidth]{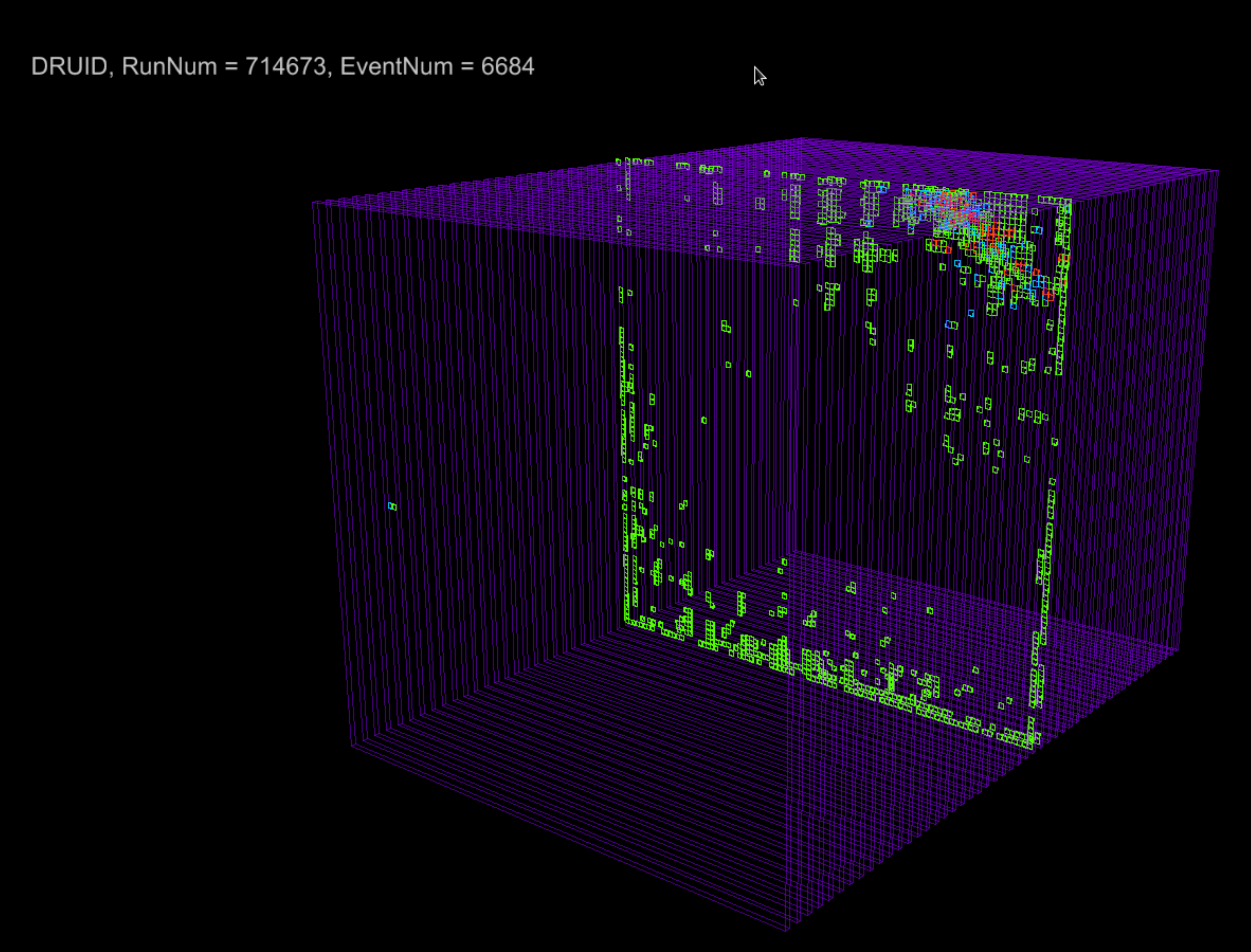}
  \includegraphics[width=0.405\textwidth]{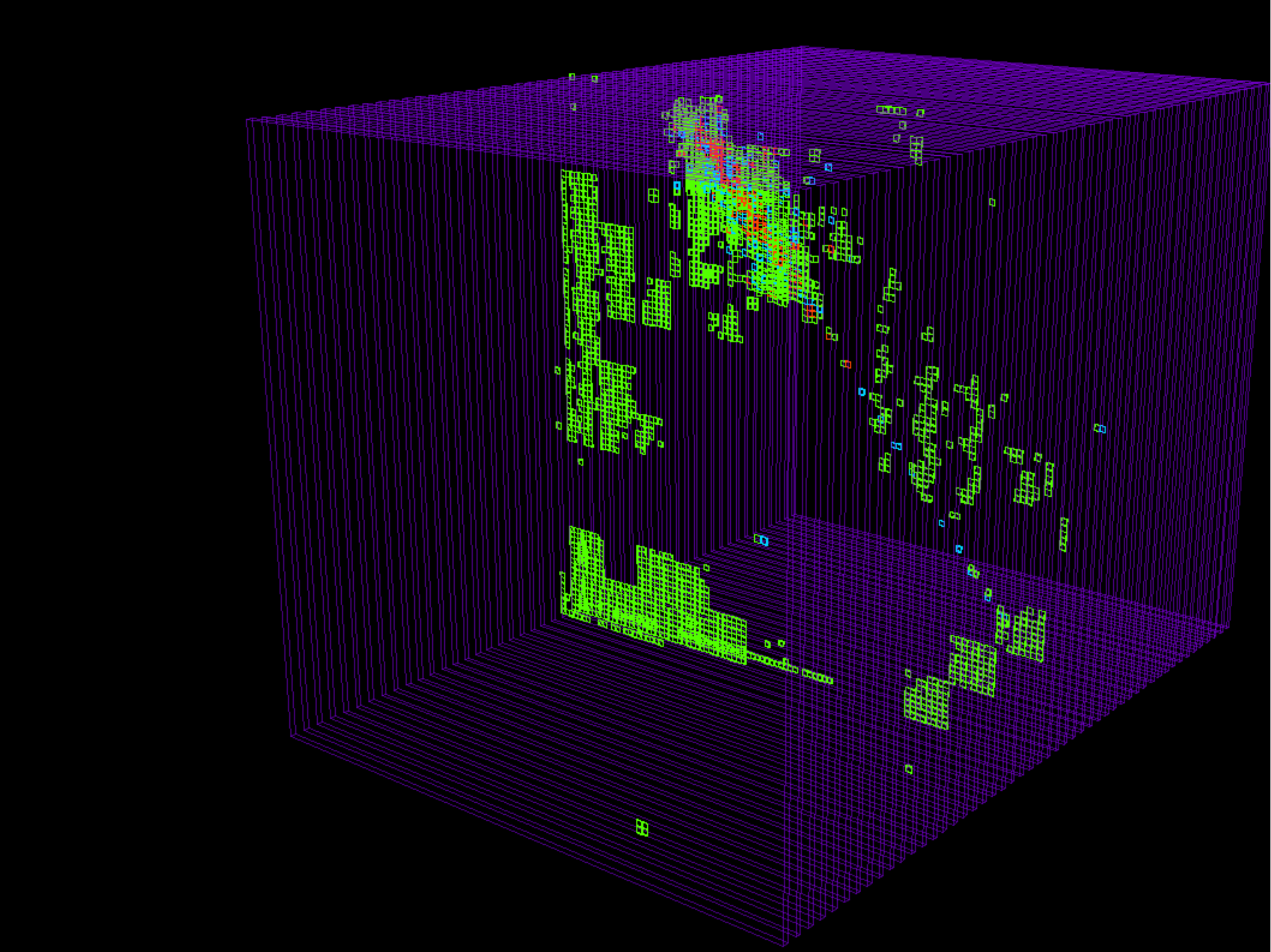}
  \caption{Example of coherent noise events in the SDHCAL prototype.}
  \label{fig:coherent-noise}
\end{figure}

The coherent noise events are related to grounding problems as previously mentioned.  They are easily identified and removed since the hits of such events  are concentrated in one or two layers.  The ratio of such common-mode noise events was found to be $2/10^{6}$ of the total events. 
Further selections, based on more sophisticated variables, can  be used to avoid the events piled-up with coherent noise.

\subsection{SDHCAL data quality}\label{sec:GRPCMuon}

The high granular calorimeters equipped with digital 
(semi-digital) readout can be characterized by the measurement 
of the detection efficiency and pad multiplicity. These two 
quantities condensate the intrinsic properties of the used 
sensor. They are needed for the monitoring and the calibration 
of the detector and are indispensable for a proper modeling in 
the simulation of the prototype (and any large detector). The 
reconstruction of the tracks left by the muons in the 
calorimeters, thanks to its tracking capability, allows the 
measurement of these quantities.

\subsubsection{Track reconstruction}
Only events with a total number of hits less than $200$ are 
considered for the further study\footnote{This number is chosen 
from the assumption that in the extreme case, a muons track can 
induce 4 fired pads in each of the $48$ plates, which corresponds 
to $\sim 200$ hits in total.}. Events with more hits are typical of interacting particle showers.  

\paragraph{Neighbour clustering} 
All hits in a given layer are clustered using a nearest-neighbour 
clustering algorithm. It consists in merging in each 
GRPC plate the hits sharing a common edge 
(\fref{fig:hit_cluster}). The position of the cluster is 
determined as the unweighted average in the two directions 
$x_{c}$ and $y_c$ of the position of the centers of the fired pads. 
The errors on the position $\sigma_{x_c}$ and 
$\sigma_{y_c}$ are calculated as $X$ and $Y$ spread divided by 
$\sqrt{12}$ (assuming a uniform distribution ranges in $x\in [0,1]$). 

\begin{figure}[!h]
  \centering
  \includegraphics[height=0.2\textwidth]{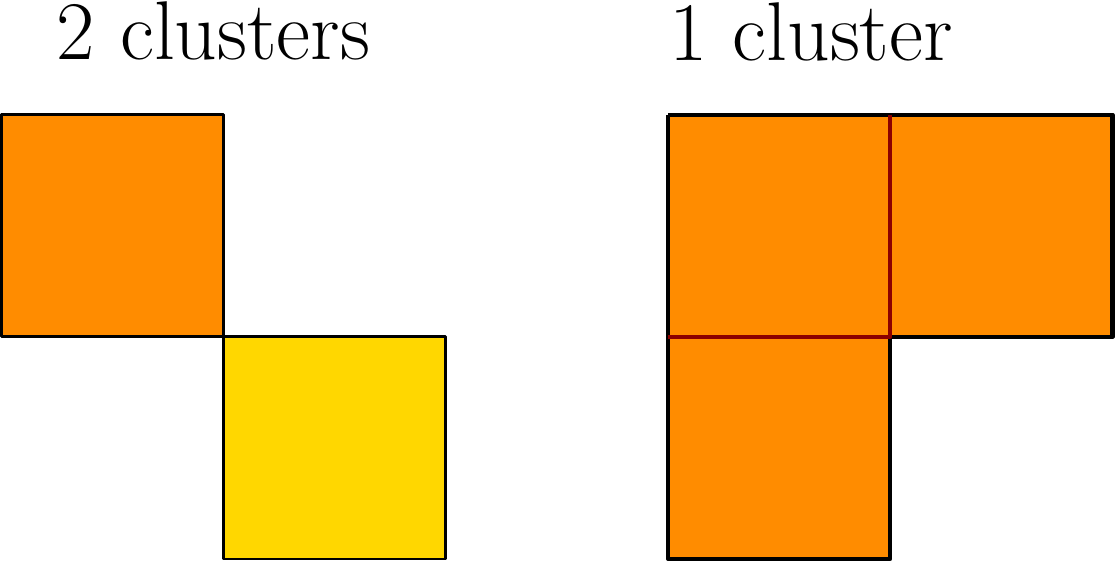}
  \caption{Example of two cluster configurations. In the left the hits are joined by a vertex, yielding two clusters. In the right configuration, all hits are sharing edges two by two, forming an unique cluster.}
  \label{fig:hit_cluster}
\end{figure}

\paragraph{Clusters cleaning} 
A removal of the farther clusters using a distance between a 	cluster ($c$) and the event cluster set ($\mathbb{E}$) is applied to the 	cluster set. This distance is defined by 
\begin{equation}
  \delta(c,\mathbb{E}) = \min \{\forall ~c' \in \mathbb{E}-\{c\}~|~ d(c,c') \}
  \label{eq:20}
\end{equation}
were $d(c,c')$ is the euclidean distance between two clusters in 
the GRPC plane. The clusters farther than $12\unit{cm}$ are then 
dropped. Event having at least one remaining cluster with 
$N_{hit}>5$  are skipped, to  exclude any possible hard muon 
interaction in the calorimeter.

\paragraph{Track reconstruction}
Tracks are reconstructed by performing $\chi^2$ minimisation. 
The reduced $\chi^2/\rm{ndf}$ of the fits is calculated 
\footnote{z-axis is perpendicular to the plates, x-axis and y-
axis are respectively horizontal and vertical.} as
\begin{equation} 
  \chi^2 / {\rm ndf} = \sum_{i}^{N_{plate}} \left( \frac{x(z_{i}) -x_{c,i}}{\sigma_{x_c,i}} \right)^{2} + \left( \frac{y(z_{i}) -y_{c,i}}{\sigma_{y_c,i}} \right)^{2}
\end{equation}
where  the sums run over all tracking clusters and 
\begin{equation} \left\{
    \begin{array}{cc} 
      x(z_i) = & p_0 + p_1 z_i \\ 
      y(z_i) = & p_2 + p_3 z_i
    \end{array} \right.
  \label{eq:line_eq}
\end{equation}
define a parametric equation of the straight line in the space 
with four parameters ($p_{i\in{0,1,2,3}}$). The errors 
$\sigma_{x_c,i}$ and $\sigma_{y_c,i}$ refer to the standard 
spread of each cluster defined previously. 

The minimisation is performed using the MINUIT package \cite{James2004} 
implemented in ROOT framework \cite{ROOTsys}. 
One example of muon event is shown in Figure \ref{fig:evt_cosmic} representing a cosmic ray muon. 

\begin{figure}[!h]
  \centering
  \includegraphics[trim ={0 0 0 9mm}, clip,height=0.8\textwidth]{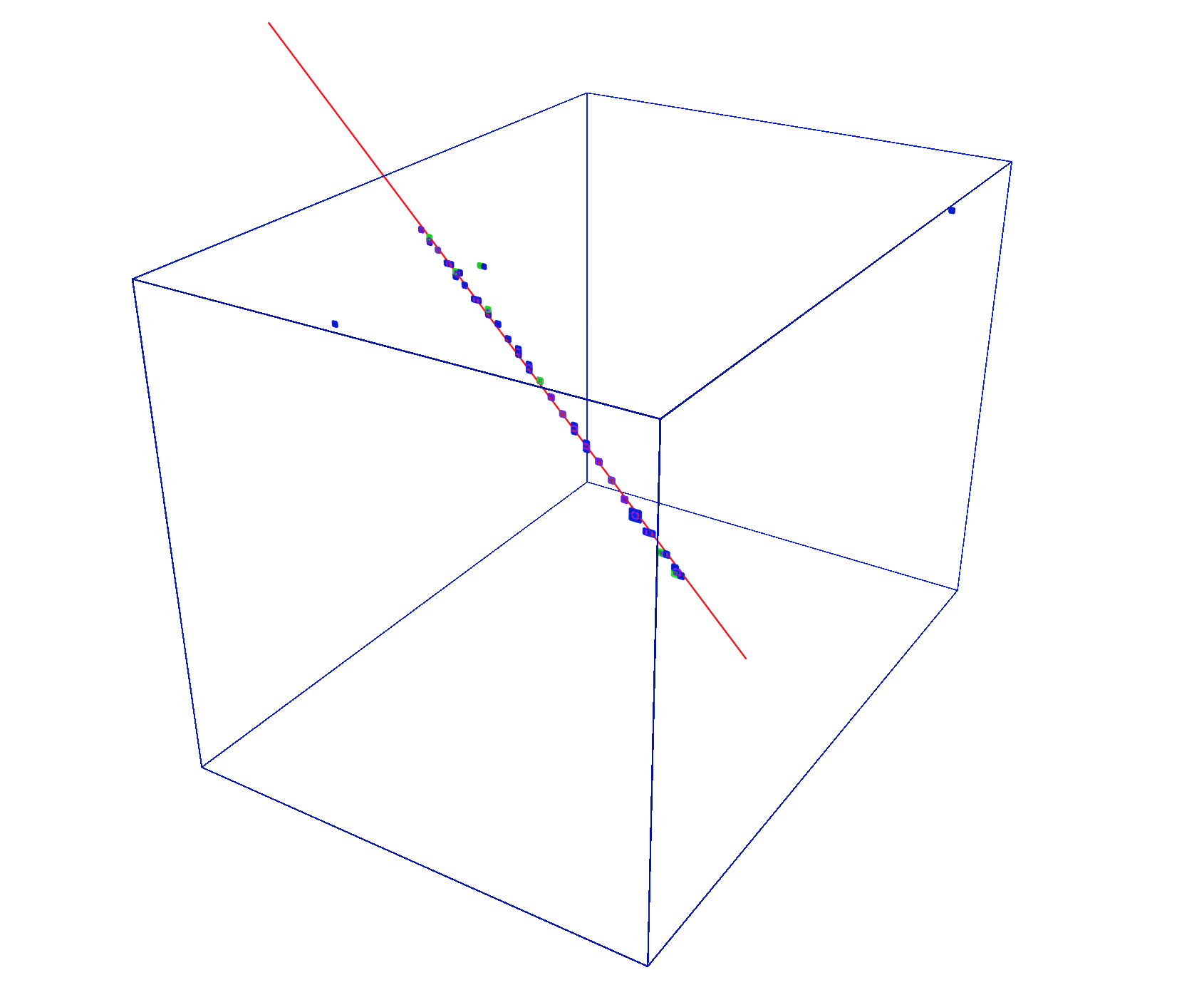}
  \caption{Example of Mip event in the SDHCAL prototype 
  (cosmic muon) }
  \label{fig:evt_cosmic}
\end{figure}

In order to estimate the performance of the GRPC, only tracks 
satisfying $\chi^2<20$ are considered for the further studies. 

\subsubsection{Efficiency and multiplicity}

The efficiency $\epsilon_{i}$ of given layer $i$ is defined as 
the probability to find at least 1 hit within  $3\unit{cm}$ of 
the reconstructed track. The considered layer "$i$" is removed 
from the track reconstruction to prevent any bias on the 
efficiency calculation. The multiplicity $\mu_{i}$ is defined as 
the mean number of hits matched on layer $i$ within $3\unit{cm}$ of 
the track intercept for tracks with at least one hit matched. 
The \fref{fig:eff_layer} and \fref{fig:mu_layer} shows the 
efficiency and multiplicity for each layer.

\begin{figure}[!h]
  \centering
  \begin{subfigure}[b]{0.45\textwidth}
    \centering
    \includegraphics[height=1.02\textwidth]{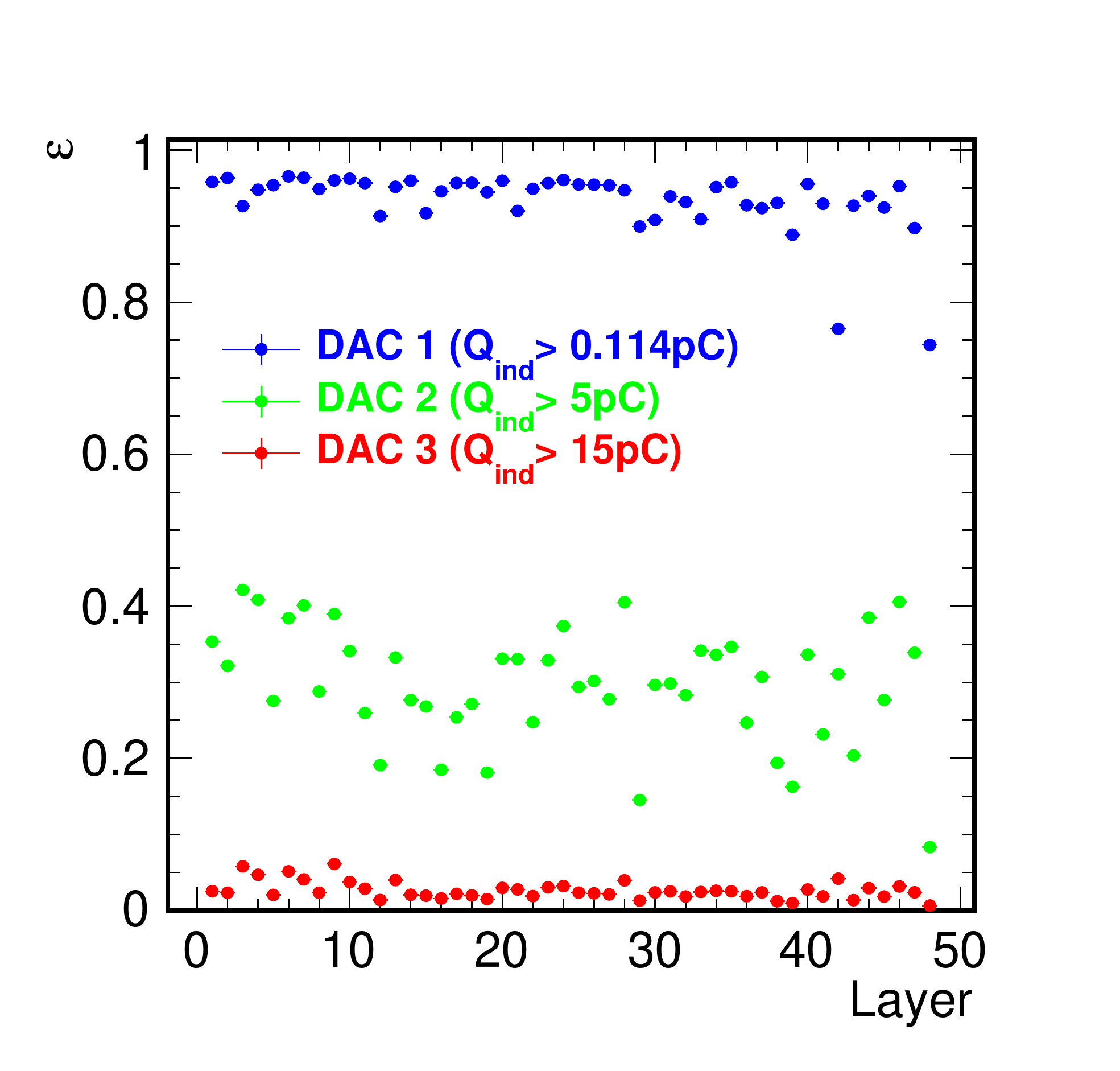}
    \caption{}
    \label{fig:eff_layer}
  \end{subfigure}
  \begin{subfigure}[b]{0.45\textwidth}
    \centering
    \includegraphics[height=1.02\textwidth]{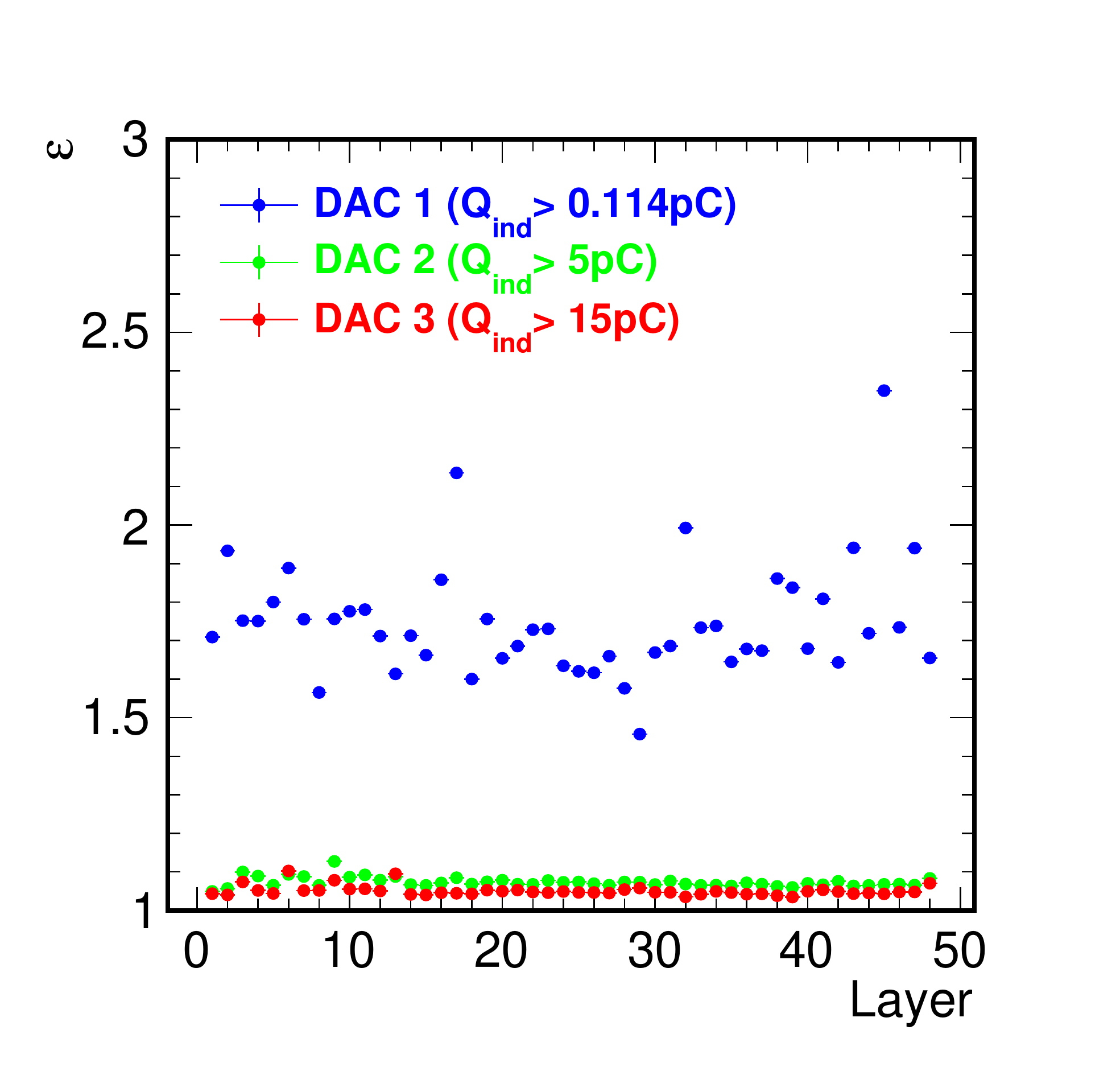}
    \caption{}	    
    \label{fig:mu_layer}
  \end{subfigure}
  
  \begin{subfigure}[b]{0.45\textwidth}
    \centering
    \includegraphics[height=1.02\textwidth]{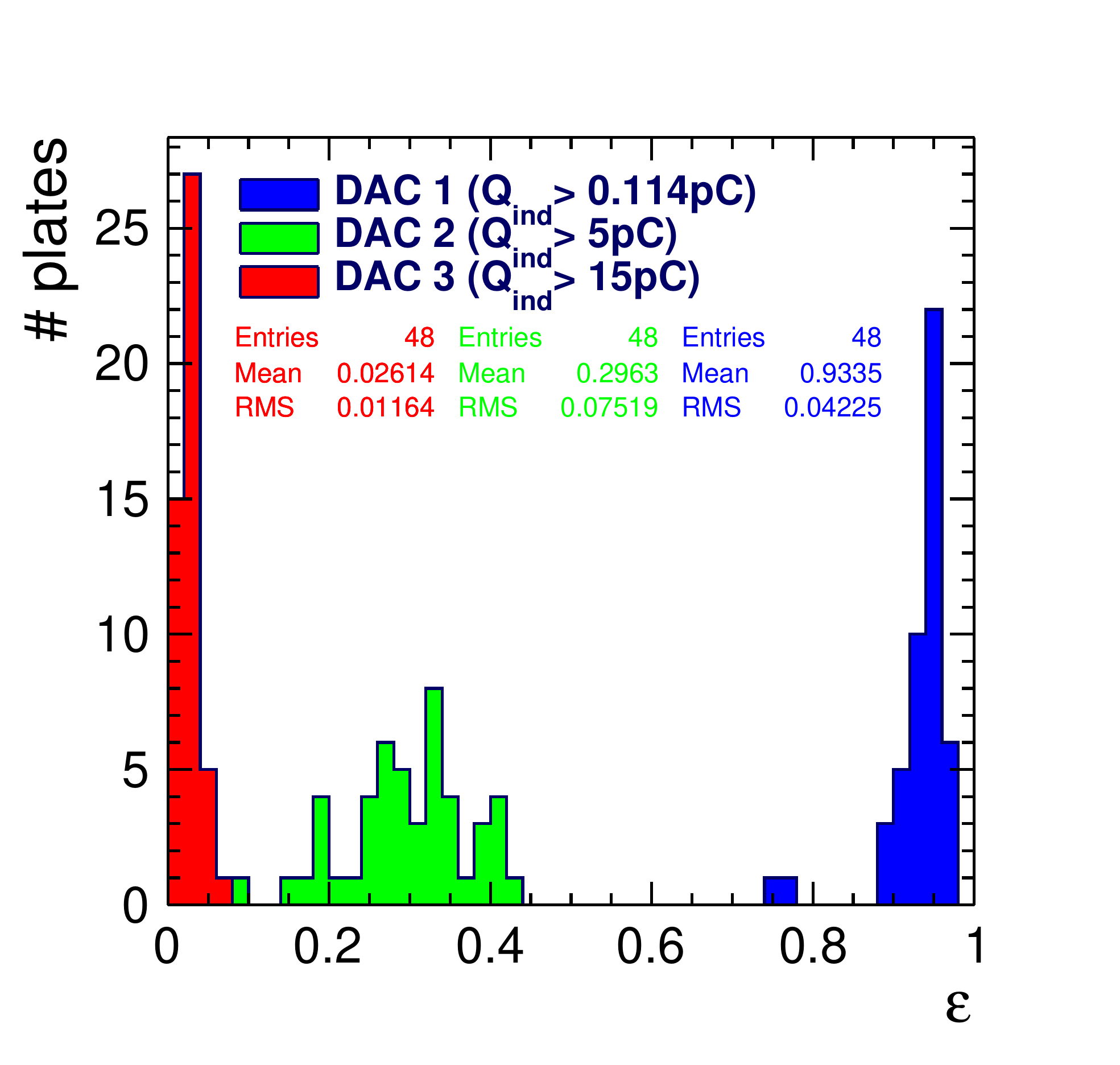}
    \caption{}
    \label{fig:eff_layer_stat}
  \end{subfigure}
  \begin{subfigure}[b]{0.45\textwidth}
    \centering
    \includegraphics[height=1.02\textwidth]{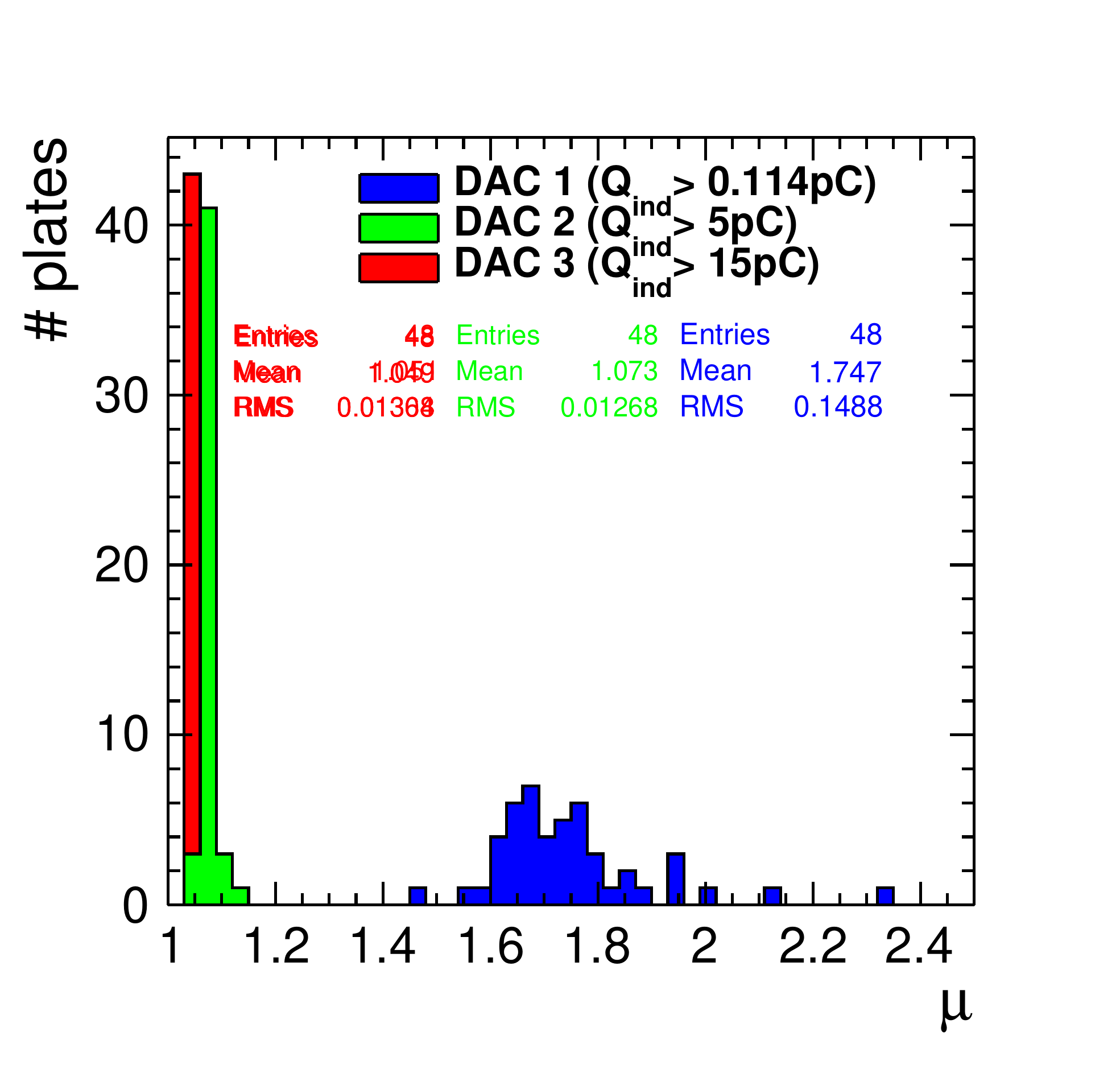}
    \caption{}
    \label{fig:mu_layer_stat}
  \end{subfigure}
  \caption{The efficiency (\subref{fig:eff_layer}) and the average multiplicity (\subref{fig:mu_layer}) for each chamber (layer) and for the three standard thresholds. (\subref{fig:eff_layer_stat}) and (\subref{fig:mu_layer_stat}) represent respectively the distributions of multiplicity and efficiency of the chambers and that for the three thresholds.}	    
\end{figure}

An efficiency of $\sim 96\%$ is observed over most of the chambers with an average multiplicity for the lowest threshold of about $1.7$. The dispersion is essentially due to the fact that no gain correction is applied to the readout channels. However using clusters rather than hits in energy resolution is expected  to reduce the effect of such dispersion.

\section{Technical performance}

To validate the SDHCAL performance, the prototype was operated with the power-pulsing mode during a beam test at the CERN SPS H2 beam line. The SPS duty cycle structure consisting of one spill with 9s time duration occurring every 45s was used.  The prototype power consumption was measured during the spill when the electronic analog part is switched on but also in-between the spills when this is switched off. During the spill a total consumption of  $2106\unit{W}$ was recorded. Of this, only $1050\unit{W}$ were consumed by the 6912 ASICs of the prototype. This consumption is equivalent to about $12  \mu\unit{W}$ per channel if the ILC duty cycle comes to be applied. The result is in a rather good agreement with the value of  $8 \mu\unit{W}$  measured using a dedicated test board on witch one HARDROC ASIC was operated in the power-pulsing mode using the foreseen ILC duty cycle scheme. The remaining part is consumed  in the buffers located on the ASUs inside the cassettes ($66\unit{W}$) and in the DIFs's components ($990\unit{W}$).  In the DIFs $750\unit{W}$ are consumed by the regulators used to reduce the voltage from $6$ to $3.5\unit{V}$ and could be eliminated in future versions of the DIF and $240\unit{W}$ are consumed in the DIFs  themselves~\footnote{This consumption could be reduced in a future version by using low-power digital parts.}. These $240\unit{W}$ are not affected by the power-pulsing mode and represent the essential of the prototype consumption during the time between the spills.   The position of the DIFs outside the detectors prevents that the heating of the DIF components and more particularly that of the regulators to influence directly the active layer performance.  Although the heat produced by the DIFs was found to have negligible effect in the case of the SDHCAL prototype, it should be evacuated  using adequate cooling system in future experiments where the SDHCAL  is expected to be placed inside the magnet. 

For what concerns the power dissipation due to the ASICs and the few buffers of the ASUs\footnote{One buffer per ASU.} the contact of the ASICs with the stainless steel cover of the cassette allows to dissipate the heat through the absorber. The simple lateral cooling system described  in the introduction is used to evacuate the absorber heat. The temperature stability of the prototype using such a system with the power-pulsing mode applied using the SPS duty cycle was checked by recording the temperature of three probes positioned inside three different cassettes.  One of these three cassettes was chosen to be in the first  part of the prototype, the second in the middle and the third in the rear.  The probes were fixed in the middle of the 1m$^2$ electronic board between the ASICs. Figure~\ref{TEMP} shows the evolution of the temperature measured by the three probes for several days. The  day-to-day as well as day and night variation of the ambiant temperature are observed but there is no clear increase due to the prototype running operation. Another important observable which reveals the stability of the active layers during the running operation is the stability of the current of each detector.  This current which is due in part \footnote{ The GRPC current includes also the contribution of  the detector leakage current.}  to the gas gain inside the detector and which increases with temperature was monitored and found to be  stable during the data taking.

\begin{figure}[h!]
  \centerline{\includegraphics[width=0.7\textwidth]{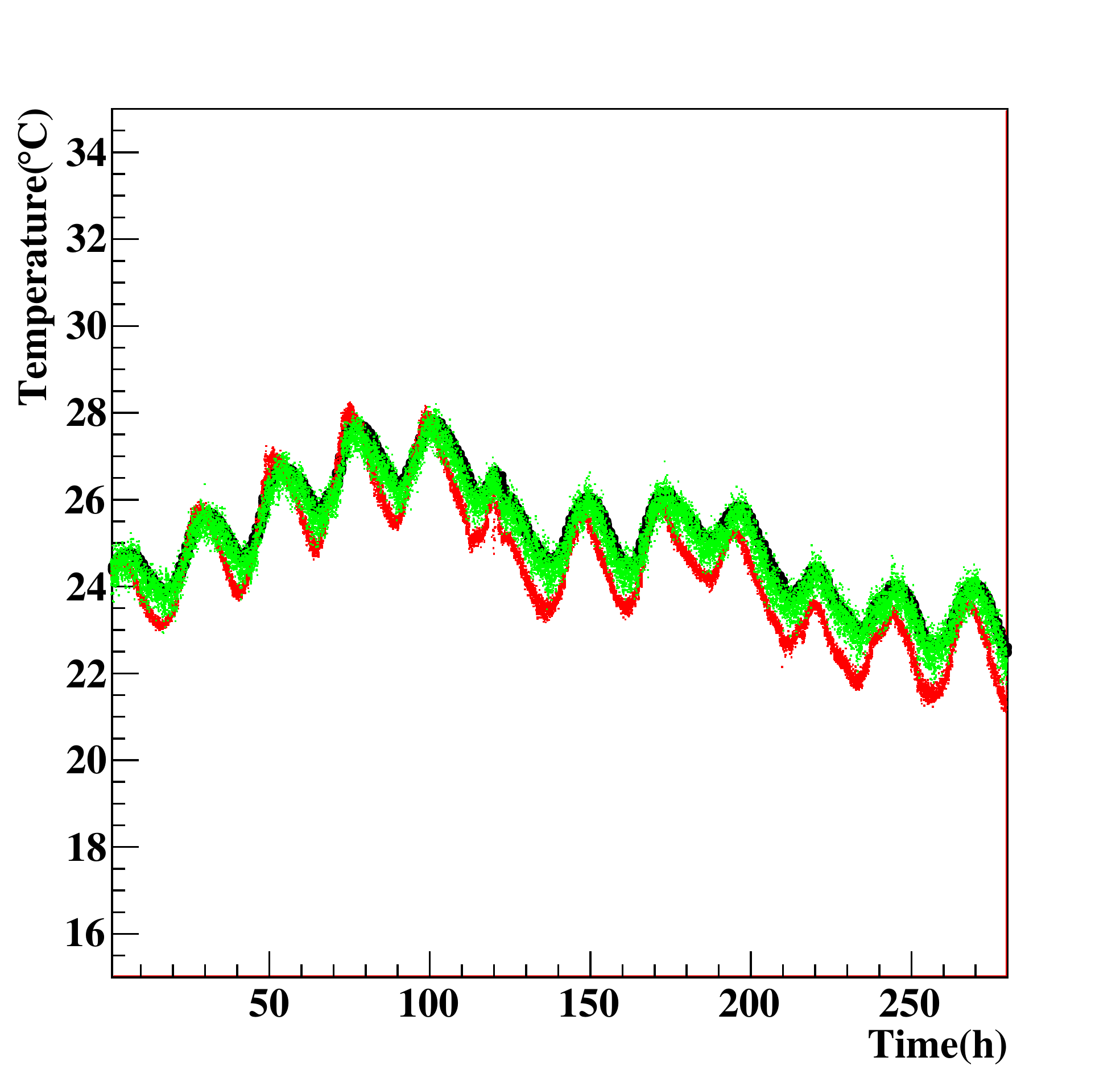}}
  \caption{Evolution of the temperature as measured by three probes placed on the electronic board inside the cassettes of three different layers; one in the front, one in the middle and the third in the rear of the prototype using a power-pulsing mode based on the SPS duty cycle.}
  \label{TEMP}
\end{figure}

No attempt was made to switch off the lateral cooling system while operating the prototype using the power-pulsing mode according to the SPS duty cycle. Indeed, a thermal study based on the simulation of the ASICs power dissipation in the prototype was performed without the lateral cooling system. A relative increase of 4.6 $^\circ$C was found in the central part of the prototype with respect to the ambient temperature. This increase of temperature would lead to more noise in the central part of the detector and thus  a degradation of the prototype data quality.  

Although one expects that running the SDHCAL prototype with the power-pulsing mode using the future ILC duty cycle will produce much less heating, the need of an appropriate cooling system should be assessed taking into account the geometry of the future SDHCAL and the constraint imposed by the global detector structure. 

The stability of the prototype mechanical structure was also checked. Muons beam as well as cosmic rays  were used to study the deformation of the  structure by measuring the relative displacement of each layer with respect to the others.  To perform this, tracks built using the  clusters left by the passage of muons in the other layers as explained in the previous section were used.  The residuals in both the horizontal (x-axis) and the vertical (y-axis) directions in each layer were obtained by comparing the expected impact parameter of the track and the barycenter of the associated cluster if any cluster is found in the layer in question. Figure~\ref{Residuals} shows the residuals $\Delta x$ and $\Delta y$ for all the layers. The residuals in the horizontal direction seem in agreement with the lateral mechanical clearance between the cassettes and the mechanical structure. The residuals in the vertical direction indicate a slight bending of few millimeters in the middle.  This  can be explained by a displacement of the plates  of the prototype due to a small incident that occurred during one of the first manipulations of the prototype as explained in~\ref{sec:meca}.

\begin{figure}[!ht]
\includegraphics[width=.48\columnwidth]{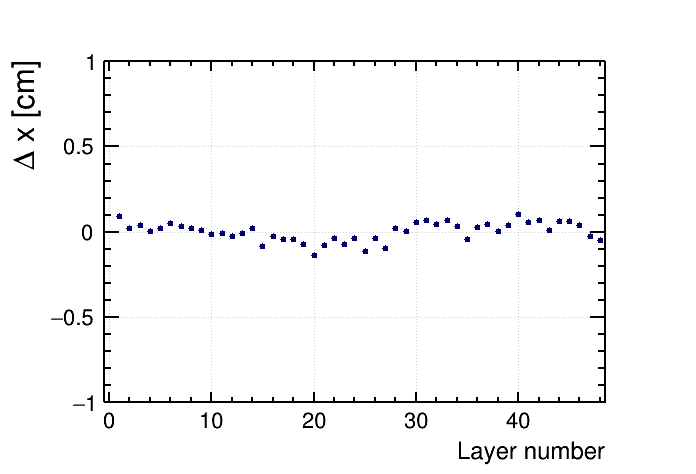}
\includegraphics[width=.48\columnwidth]{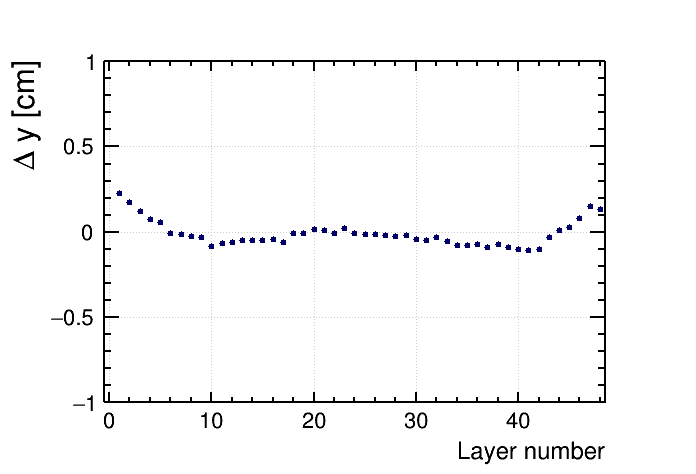}
\caption{ Relative deviation of each of the 48 layers with respect to the others in the horizontal direction (left) and in the vertical direction(right).}
\label{Residuals}
\end{figure}

\section{Conclusion}

We have conceived and built  a technological prototype as a demonstrator of the Semi-Digital Hadronic CALorimeter for the future ILC experiments.
The prototype has shown that a highly-granular calorimeter fulfilling the requirements of compactness and relatively low-power consumption HCAL with PFA capability is  achievable by using simple but efficient and robust detectors such as GRPCs.  
Preliminary results with pion beam, to be published in future papers, show that such a new generation of calorimeters can also provide very good energy measurement in addition to an excellent tracking capability. 
Additional efforts are still needed to validate completely the SDHCAL concept to equip the detectors of the future different lepton collider experiments. 
These efforts have been already initiated. They focus on improving the mechanical structure by opting for more robust and less deforming welding techniques like those using electron beams. Attempts to build mechanical structures using these techniques are ongoing in collaboration with CERN services. 
The construction of larger GRPC detectors (up to 3m$^2$) and adapting their gas distribution system to keep the same efficiency achieved in the case of 1m$^2$ detectors are also ongoing. Finally the readout electronics is being improved by replacing the daisy chain protocol by the I2C one which allows to address the different ASICs individually eliminating thus the eventual problem of neutralizing a large portion of one detector electronics if one ASIC comes to die.  Concerning this last point an updated version of the HARDROC ASIC has been already conceived and produced. The new version equipped with the I2C protocole and additional interesting features such as the zero suppression and an extended dynamic range is being tested. Preliminary results show that the new features are successful.  New electronic and acquisition boards are also being developed to cope with the new features of the HARDROC while taking advantage of the recent progress achieved within LHC update research programmes. 
\section{acknowledgment}

We would like to thank all who contributed to the construction of the prototype. We would like to acknowledge the support provided by the following funding agencies  F.R.S.-FNRS, FWO(Belgium), CNRS and ANR(France), SEIDI and CPAN (Spain).

\end{document}